\documentclass[11pt]{article}
\pdfoutput=1

\usepackage{jheppub} 

\usepackage{latexsym,amsmath,amsfonts,amssymb, tensor}
\usepackage{graphicx}
\usepackage{epsf}
\usepackage{makeidx}
\usepackage{multirow}
\usepackage[all]{xy}
\usepackage{hyperref}
\usepackage{verbatim} %for comments
\usepackage{rotating}
\usepackage{xcolor}
\usepackage{multirow}
\usepackage{bm}
\usepackage{cleveref}
\usepackage{slashed}
\usepackage[normalem]{ulem}

\usepackage{graphicx}
\usepackage{latexsym,amsmath,amsfonts,amssymb}
\usepackage{mathrsfs}
\usepackage{bbm}
\usepackage{bm}
\usepackage{subfigure}
\usepackage{paralist}
\usepackage{youngtab}
\usepackage{bclogo}
\usepackage{mathrsfs}

\usepackage{esvect} %for arrows \vv

\usepackage{tikz}
\usepackage{tikz-cd}
\usepackage{diagbox}
\usetikzlibrary{positioning}
\usetikzlibrary{calc}
\usetikzlibrary{decorations.pathreplacing,calligraphy}

\usepackage{xstring}
\usetikzlibrary{decorations.pathmorphing} 
\usetikzlibrary{decorations.markings} 
\usetikzlibrary{snakes}
\usetikzlibrary{arrows} 
\usetikzlibrary{shapes} 
\usetikzlibrary{matrix} 
\usetikzlibrary{positioning} 
\usepackage[english]{babel} 
\usepackage[autostyle]{csquotes}

\tikzstyle{brane}=[draw]
\tikzset{D7/.style={circle, draw=black, inner sep=0pt, fill=white, minimum size=3mm}}
\tikzset{hasse/.style={circle, fill,inner sep=2pt}}
\tikzset{flavor/.style={regular polygon,fill=white,regular polygon sides=4,inner sep=2.5pt, draw}}
\tikzset{gauge/.style={circle, draw,inner sep=2.5pt}}
\tikzset{gaugeb/.style={circle, draw,fill=black,inner sep=2.5pt}}
\tikzset{gauger/.style={circle, draw,fill=cyan,inner sep=2.5pt}}
\tikzset{gaugeg/.style={circle, draw,fill=red,inner sep=2.5pt}}
\tikzset{bd/.style={circle, draw=black, inner sep=0pt, fill=black, minimum size=1mm}}
\tikzset{wd/.style={circle, draw=black, inner sep=0pt, fill=white, minimum size=2mm}}
\tikzset{SUd/.style={circle, draw=black, inner sep=0pt, fill=yellow, minimum size=2mm}}
\tikzset{Dynkin/.style={circle, draw=black, inner sep=0pt, fill=white, minimum size=2mm}}
\tikzstyle{ligne}=[draw, thick] 
\tikzset{doublearrow/.style={ draw=black!75, color=black!75, thick, double distance=3pt, }}

\newcommand{\dvol}{d{\rm vol}}   %d\mathrm{vol}

\numberwithin{equation}{section}

\newcommand{\nn}{\nonumber}
\newcommand{\mat}[1]{\begin{pmatrix} #1 \end{pmatrix}}

\newcommand{\be}{\begin{equation}} 
\newcommand{\ee}{\end{equation}}
\newcommand{\bea}{\begin{equation} \begin{aligned}} \newcommand{\eea}{\end{aligned} \end{equation}}

\newcommand{\bit}{\begin{itemize}} 
\newcommand{\eit}{\end{itemize}}

\newcommand{\cA}{\mathcal{A}}
\newcommand{\cB}{\mathcal{B}}

\newcommand{\cD}{\mathcal{D}}

\newcommand{\cF}{\mathcal{F}}
\newcommand{\cG}{\mathcal{G}}
\newcommand{\cH}{\mathcal{H}}

\newcommand{\cK}{\mathcal{K}}
\newcommand{\cM}{\mathcal{M}}
\newcommand{\cN}{\mathcal{N}}

\newcommand{\bC}{\mathbb{C}}
\newcommand{\bF}{\mathbb{F}}

\newcommand{\bN}{\mathbb{N}}

\newcommand{\bP}{\mathbb{P}}

\newcommand{\Z}{\mathbb{Z}}
\newcommand{\C}{\mathbb{C}}
\newcommand{\R}{\mathbb{R}}

\renewcommand{\t}{\widetilde }
\renewcommand{\d}{\partial }

\renewcommand{\b}{\bar }

\newcommand{\alphadot}{{\dot\alpha}}
\newcommand{\betadot}{{\dot\beta}}

\newcommand{\half}{{1\over 2}}

\newcommand{\CA}{\mathcal{A}}
\newcommand{\CB}{\mathcal{B}}
\newcommand{\CC}{\mathcal{C}}
\newcommand{\CD}{\mathcal{D}}

\newcommand{\CF}{\mathcal{F}}
\newcommand{\CG}{\mathcal{G}}
\newcommand{\CH}{\mathcal{H}}
\newcommand{\CI}{\mathcal{I}}
\newcommand{\CJ}{\mathcal{J}}
\newcommand{\CK}{\mathcal{K}}
\newcommand{\CL}{\mathcal{L}}
\newcommand{\CM}{\mathcal{M}}
\newcommand{\CN}{\mathcal{N}}
\newcommand{\CO}{\mathcal{O}}
\newcommand{\CP}{\mathcal{P}}
\newcommand{\CQ}{\mathcal{Q}}

\newcommand{\CV}{\mathcal{V}}

\newcommand{\FR}{\mathfrak{R}}
\newcommand{\Fg}{\mathfrak{g}}

\newcommand{\m}{\mathfrak{m}}
\newcommand{\n}{\mathfrak{n}}

\newcommand{\h}{\widehat}

\DeclareMathOperator{\Tr}{Tr}
\DeclareMathOperator{\tr}{tr}

\newcommand{\SL}{{\mathscr L}}

\newcommand{\ov}{\over}

\newcommand{\dilog}{{\text{Li}_2}}
\newcommand{\trilog}{{\text{Li}_3}}

\newcommand{\MG}{{\mathbf X}} %M-theory singualrity
\newcommand{\amCB}{\boldsymbol{a}}

\newcommand{\Flux}{{\boldsymbol{\Pi}}}
\newcommand{\FiberOp}{\mathscr{F}}
\newcommand{\FiberOpK}{\mathscr{K}}

\newcommand{\bA}{{\boldsymbol{A}}}
\newcommand{\bB}{{\boldsymbol{B}}}
\newcommand{\bG}{{\boldsymbol{G}}}

\newcommand{\p}{\mathfrak{p}}
\renewcommand{\k}{\mathbf{k}} 
\newcommand{\mK}{\mathbf{k}}

\newcommand{\Cfib}{{\rm C}}
\newcommand{\FKK}{{\hat{\mathtt{F}}}} %{{{\rm F}^{\rm  (KK)}}}

\renewcommand{\S}{{\mathrm{S}}}

\newcommand{\Lop}{\mathbb{L}}

\newcommand{\bbeta}{{\boldsymbol{\beta}}}

\DeclareMathAlphabet{\pazocal}{OMS}{zplm}{m}{n}

\usepackage{diagbox}
\usepackage{array}
\makeatletter
\newcommand{\thickhline}{%
    \noalign {\ifnum 0=`}\fi \hrule height 1pt
    \futurelet \reserved@a \@xhline
}
\newcolumntype{"}{@{\hskip\tabcolsep\vrule width 1pt\hskip\tabcolsep}}
\makeatother

\makeatletter
\g@addto@macro{\endtabular}{\rowfont{}}% Clear row font
\makeatother
\newcommand{\rowfonttype}{}% Current row font
\newcommand{\rowfont}[1]{% Set current row font
   \gdef\rowfonttype{#1}#1%
}
\newcolumntype{L}{>{\rowfonttype}l}

\usepackage{multirow}
\usepackage{arydshln}
\begin{document}

% format
\baselineskip=18pt  % a la harvmac
\numberwithin{equation}{section}  % make eq labels (sec.num)
\allowdisplaybreaks  % allow page breaks in displayed eqs

%%%%%%%%%%%%%%%%%%%%%%%%%%%%%%%%%%%%%%%%%%%
%%%        TITLE BEGINS HERE
%%%%%%%%%%%%%%%%%%%%%%%%%%%%%%%%%%%%%%%%%%%

%% ========== title (note version) begins here ==========
%
%\vspace*{-1cm}
%\begin{center}
% {\Large\bf Title of the Document}
%\end{center}
%\vspace*{-.5cm}
%
%% ========== title (note version) ends here ==========

%% ========== title (paper version, a la harvmac) begins here ==========

\thispagestyle{empty}

\vspace*{0.8cm} 
\begin{center}
{\huge   Partition Functions and Fibering Operators
\medskip\\
 on the Coulomb Branch of 5d SCFTs}

 \vspace*{1.5cm}
Cyril Closset$^{\flat}$,  Horia Magureanu$^{\sharp}$

 \vspace*{0.7cm} 

 {\it$^{\flat}$  School of Mathematics, University of Birmingham,\\ 
Watson Building, Edgbaston, Birmingham B15 2TT, UK}\\

{\it$^{\sharp}$ Mathematical Institute, University of Oxford, \\
Andrew-Wiles Building,  Woodstock Road, Oxford, OX2 6GG, UK}\\

\end{center}

\vspace*{0.5cm}

\noindent
We study 5d $\mathcal{N}=1$ supersymmetric field theories on closed five-manifolds $\mathcal{M}_5$ which are  principal circle bundles over simply-connected K\"ahler four-manifolds, $\mathcal{M}_4$, equipped with the Donaldson-Witten twist. We propose a new approach to compute the supersymmetric partition function on $\mathcal{M}_5$ through the insertion of a fibering operator, which introduces a non-trivial fibration over $\mathcal{M}_4$, in the 4d topologically twisted field theory. We determine the  so-called Coulomb branch partition function on any such $\mathcal{M}_5$, which is conjectured to be the holomorphic `integrand' of the full partition function. We precisely match the low-energy effective field theory approach to explicit one-loop computations, and we discuss the effect of non-perturbative 5d BPS particles in this context. When   $\mathcal{M}_4$ is toric, we also reconstruct our Coulomb branch partition function by appropriately gluing Nekrasov partition functions.  As a by-product of our analysis, we provide strong new evidence for the validity of the Lockhart-Vafa formula for the five-sphere   partition function.

\newpage
%%%%%%%%%%%%%%%%%%%%%%%%%%%%%%%%%%%%%%%%%%%
%%%           TITLE ENDS HERE
%%%%%%%%%%%%%%%%%%%%%%%%%%%%%%%%%%%%%%%%%%%

\tableofcontents
%\printindex

%%%%%%%%%%%%%%%%%%%%%%%%%%%%%%%%%%%%%%%%%%%
%%%        MAIN TEXT BEGINS HERE
%%%%%%%%%%%%%%%%%%%%%%%%%%%%%%%%%%%%%%%%%%%

\section{Introduction}
 The study of supersymmetric quantum field theories on non-trivial Riemannian manifolds often opens an important window into their strongly coupled dynamics \cite{Witten:1988ze, Witten:1994ev, Nekrasov:2002qd, Pestun:2007rz}. In this work, we consider five-dimensional superconformal field theories (5d SCFTs) \cite{Seiberg:1996bd} on five-manifolds $\CM_5$ that can be constructed as circle fibrations over a K\"ahler four-manifold:
 \be\label{M5 fibration intro}
 S^1\longrightarrow \CM_5 \overset{\pi}{\longrightarrow} \CM_4~.
 \ee
 For simplicity, we will only consider principal circle bundles% 
 \footnote{By which we mean `principal $U(1)$ bundles'. The case of more general  fibrations is left for future work.}  over $\CM_4$. A particularly important example of this construction is the five-sphere (with its round metric), viewed as a circle bundle over the complex projective space:
 \be
S^1\longrightarrow  S^5 \overset{\pi}{\longrightarrow} \bP^2~.
 \ee 
The partition function of a 5d SCFT on the five-sphere, 
\be\label{FS5 intro}
F_{S^5} =  \log |{\bf Z}_{S^5}|~,
\ee
 can be defined for any 5d conformal field theory, irrespective of supersymmetry. 
It is conjecturally a good measure of the number of `degrees of freedom' of the strongly-coupled fixed point %which always decreases upon renormalisation group (RG) flow 
 \cite{Klebanov:2011gs}. Despite some heroic computations, {\it e.g.} as in \cite{Chang:2017cdx},  to compute this object exactly and efficiently in 5d SCFTs remains a challenge.

\subsection{The Donaldson-Witten twist approach}\label{subsec:DWapproach}

This paper initiates a new approach to computing the $\CM_5$ supersymmetric partition function, ${\bf Z}_{\CM_5}$, following a line of ideas which was successfully applied to  3d  $\CN=2$  theories on Seifert manifolds \cite{Closset:2017zgf} -- see also \cite{Nekrasov:2014xaa,Benini:2016hjo, Closset:2016arn,  Closset:2018ghr, Closset:2019hyt}. As our starting point, we first consider the 5d theory on a product manifold
 \be
 \CM_5 = \CM_4\times S^1~.
 \ee
 The presence of the $S^1$ factor allows us to consider the 4d $\CN=2$ Kaluza-Klein (KK) theory that one obtains by compactifying the 5d SCFT on a circle \cite{Nekrasov:1996cz}. The low-energy physics of that KK theory on its Coulomb branch (CB) is governed by a certain Seiberg-Witten geometry \cite{Seiberg:1994aj, Seiberg:1994rs, Nekrasov:1996cz, Ganor:1996pc}, like for any 4d $\CN=2$ supersymmetric field theory.%
 \footnote{The CB physics of 4d $\CN=2$ KK theories has been revisited in recent works \protect\cite{Closset:2019juk, Closset:2021lhd, Jia:2021ikh, Magureanu:2022qym, Jia:2022dra}.}
 The simplest way to put this KK theory on $\CM_4$ while preserving supersymmetry is to consider the ordinary topological twist, also known as the Donaldson-Witten (DW) twist \cite{Witten:1988ze}. 
 % On a generic four-manifold, we would preserve only one supercharge. 
  We choose $\CM_4$ to be K\"ahler so that we can preserve two supercharges \cite{Witten:1994ev} which  anticommute to a translation along the $S^1$. Supersymmetric operators must then be extended along the circle. 

The topologically twisted partition function ${\bf Z}_{\CM_4\times S^1}$, also called the twisted index \cite{Hosseini:2018uzp}, is expected to capture (generalised) K-theoretic Donaldson invariants of $\CM_4$ \cite{Nakajima:2005fg}. More precisely, those invariants should arise from the insertions of some operators $\CO$ wrapping the circle. Let us denote by
\be\label{ObsDW intro}
\langle \CO \rangle^{\rm DW}_{\CM_4\times S^1}~,
\ee
any observable in the four-dimensional topological quantum field theory (TQFT) obtained by the DW twist. The direct path integral computation of such observables is a famous and famously challenging problem \cite{Witten:1994cg, Moore:1997pc, Marino:1997gj, Marino:1998bm} which, however, is not the focus of this paper (see~\cite{Bawane:2014uka, Bershtein:2015xfa,  Korpas:2017qdo, Korpas:2019ava, Moore:2017cmm, Korpas:2019cwg, Bonelli:2020xps, Manschot:2021qqe,  Aspman:2021kfp, Korpas:2022tij, Aspman:2022sfj}
 for a lot of more recent progress, and especially~\cite{toappearKMMTZ} for the case of 4d $\CN=2$ KK theories). Instead, let us assume that we know how to compute \eqref{ObsDW intro} for any $\CO$. Then, we claim that one can compute the $\CM_5$ partition function as a particular observable in the DW theory, namely as the expectation value of a so-called {\it fibering operator}, denoted by $\FiberOp$. We write this as:
\be
{\bf Z}_{\CM_5} = \langle \FiberOp_\p \rangle^{\rm DW}_{\CM_4\times S^1}~,
\ee
where $\p$ denotes the first Chern class of the principal circle bundle \eqref{M5 fibration intro}. 
  In this approach, we should be able to compute seemingly `non-topological' 5d observables, like the $S^5$ partition function, in the four-dimensional DW theory.%
  \footnote{To prove this expectation  {\it a priori}, one should show explicitly that one can continuously deform the `DW-twist' supersymmetric background used in this paper to the `round $S^5$' supersymmetric background with vanishing $SU(2)_R$ background gauge field, and furthermore prove that this deformation in $\CQ$-exact. This would amount to an 8-supercharge equivalent to the 4-supercharge analysis of \protect\cite{Closset:2013vra}, which goes well beyond the scope of this paper. Instead, we simply observe that our computations match previous computations performed on different supersymmetric backgrounds.}
 (Similarly, in \cite{Closset:2017zgf}, the $S^3$ partition function was understood as an observable in the $A$-model, a 2d TQFT.)

Thanks to topological invariance, one can use the low-energy Seiberg-Witten description of the KK theory to compute the partition function on $\CM_4\times S^1$ and hence, upon insertion of the fibering operator, on $\CM_5$. It is useful to consider the theory at any given point on the Coulomb branch, with $\amCB$ denoting the scalars in the low-energy abelian vector multiplets.%
\footnote{We will also use $\amCB$ to denote mass terms, which arise as background vector multiplets. These will be discussed more explicitly in the main text.} One can then ask what is the `partition function' on $\CM_5$ at a fixed value of $\amCB$, and with some fixed background fluxes $\m$ for the abelian gauge fields turned on. By a slight abuse of terminology, this `off-shell' quantity will be called  the {\it CB partition function},  denoted by:
\be\label{CBZ first}
Z_{\CM_5}(\amCB)_\m~.
\ee
On any geometry with the topology of $\R^4\times S^1$, this would correspond to the DW-twist of the Seiberg-Witten theory, which is fully determined by the Seiberg-Witten prepotential, $\CF(\amCB)$. 
On a closed space, we must integrate over the full moduli space, and the partition function on $\CM_5$ will be obtained after integrating out the dynamical low-energy vector multiplets (we usually retain a dependence on background vector multiplets, which keep track of the flavour symmetry). Based on a number of previous results and conjectures in the literature (see {\it e.g.} \cite{Nekrasov:2003vi, Hosseini:2018uzp, Bonelli:2020xps}), we expect an explicit formula of the form:
\be\label{ZM5 full conj intro}
 {\bf Z}_{\CM_5}=\sum_\m\oint_\CC d\amCB \, Z_{\CM_5}(\amCB)_\m~,
\ee
where the precise form of the sum over fluxes and of the integration contour have to be determined. We hope to address this in future work. In this paper, we shall be more modest and focus on computing the integrand, namely the CB partition function $Z_{\CM_5}$. As we will see, this already entails a number of subtleties and leads to new and interesting results. 

The CB partition function \eqref{CBZ first} is a holomorphic function of $\amCB$. The schematic formula \eqref{ZM5 full conj intro} is inspired by the `holomorphic approach' to the DW twist \cite{Losev:1997tp}, and by a conjecture of Nekrasov for toric four-manifolds \cite{Nekrasov:2003vi}.
 By contrast, the Moore-Witten $u$-plane integral approach deals with a non-holomorphic integrand, which renders the vector-multiplet integration better defined. While it would be important to reconcile the two approaches,  in this paper we shall be wilfully na\"ive and ignore all $\CQ$-exact terms in the effective action. We will not discuss the contributions from Seiberg-Witten invariants, either, even though they are expected to appear prominently in the general story \cite{Witten:1994cg}.
 
  On any K\"ahler four-manifold, we have the relations:
 \be
 \chi_h \equiv 1- h^{0,1} + h^{0,2}= {\chi+\sigma\ov 4}~, \qquad\qquad b_2^+ = 1+ 2h^{0,2}~,
 \ee
 where $\chi$ and $\sigma$ are the Euler characteristic $\chi= \sum_{i=0}^4 (-1)^i b_i$ and the signature $\sigma= b_2^+- b_2^-$ of $\CM_4$, respectively, and $h^{p,q}$ are its Hodge numbers.  
 In this work, we will  further assume that the K\"ahler manifold $\CM_4$ is  simply connected (in particular, $h^{0,1}=0$), as is often done in the study of Donaldson invariants.%
 \footnote{From the point of view of 4-manifold invariants, this assumption is necessary to have any hope of classifying smooth manifolds. For us, it is just a technical restriction. See also~\protect\cite{Marino:1998rg} for a discussion of the $u$-plane approach when $\pi_1(\CM_4)\neq 0$.} This is to avoid some additional zero-mode contributing to the CB partition function, and it also simplifies the computation of certain one-loop determinants. (The case with $\pi_1(\CM_4)\neq 0$ is nonetheless rather interesting, and it is left for future work.)

\subsection{The $S^5$ partition function}
 The computation of the $S^5$ partition function of 5d SCFTs has been approached in various ways in the literature  \cite{Kallen:2012cs, Kallen:2012va, Kim:2012ava, Imamura:2012efi, Lockhart:2012vp, Kim:2012qf, Nieri:2018ghd}, mostly from the perspective of the 5d $\CN=1$ supersymmetric gauge theories that may arise as infrared phases -- see {\it e.g.} the reviews \cite{Qiu:2016dyj, Pasquetti:2016dyl}. 
 Conjecturally, the full partition function takes the form:
\be\label{ZS5 Nek intro}
{\bf Z}_{S^5} = \int d\sigma \prod_{l=1}^3 Z_{\C^2\times S^1}(i\sigma^{(l)}, \epsilon_1^{(l)},  \epsilon_2^{(l)})~,
\ee
 where the integral is over the 5d CB parameters, and the integrand is a product of three K-theoretic Nekrasov partition functions (including the classical and perturbative contributions), with the $\Omega$-deformation parameters $\epsilon_1, \epsilon_2$ which become supersymmetric squashing parameters on $S^5$. It was further conjectured by Lockhart and Vafa \cite{Lockhart:2012vp} that the $S^5$ partition function can also be written as:
 \be\label{ZS5 BPS intro}
 {\bf Z}_{S^5} = \int d\sigma \prod_{\alpha \; {\rm BPS}} Z^{(\alpha)}_{S^5, {\rm BPS}}(\sigma)~,
 \ee
 schematically,  where the integrand is a product over all 5d BPS states -- that is, the electrically-charged particles on the 5d Coulomb branch. The present paper corroborates the Lockhart-Vafa formula \eqref{ZS5 BPS intro}. We also revisit and clarify the  factorisation formula \eqref{ZS5 Nek intro}, and we generalise it to other five-manifolds.

Using the DW-twist approach outlined above, we find the following CB partition function on $S^5$:
\be\label{ZS5 intro}
Z_{S^5}(\amCB)= \bA(\amCB)^3\, \bB(\amCB) \, \FiberOp(\amCB)^\half~.
\ee
with the fibering operator:
\be\label{FiberOp form 1 intro}
\FiberOp(\amCB)= \exp\left(- 4 \pi i \left(\CF(\amCB)- \amCB {\d \CF(\amCB)\ov \d \amCB}+\half \amCB^2 {\d^2 \CF(\amCB)\ov \d \amCB^2} \right)\right)~.
\ee
 Here, $\bA(\amCB)$ and $\bB(\amCB)$ are the standard gravitational couplings in the DW theory   \cite{Witten:1995gf}, and $\CF(\amCB)$ is the full effective prepotential of the 4d $\CN=2$ KK theory. The formula \eqref{FiberOp form 1 intro} is our general result for the fibering operator. 
Note that we could set the gauge fluxes to zero ($\m=0$) in \eqref{ZS5 intro} because all fluxes on $\bP^2$ trivialise when lifted to $S^5$. In writing \eqref{ZS5 intro} we also assumed, for simplicity, that the KK theory can be consistently coupled to $\CM_4=\bP^2$, which is only true if a certain condition on the BPS spectrum is satisfied (roughly speaking, there should be no hypermultiplets). We will explain how to lift this artificial assumption momentarily.

We are then naturally led to the following conjecture for the round $S^5$ partition function of any 5d SCFT:
\be\label{ZS5 conj intro}
{\bf Z}_{S^5} = \int_{i\R} d\amCB \, Z_{S^5}(\amCB)~,
\ee
where the integration contour is over the imaginary axis for the CB variables $\amCB$.%
\footnote{In general, we expect that the contour should be slightly deformed so that the integral converges. This is the analogue of the `$\sigma$-contour' discussed in \protect\cite{Closset:2017zgf}.  There will also be a representation of ${\bf Z}_{S^5}$ as in \protect\eqref{ZM5 full conj intro} that involves a sum over $\bP^2$ fluxes. This and the relations between different formulas for the full partition function will be discussed elsewhere.} This matches many previous computations whenever a comparison is possible, including the $\epsilon_1, \epsilon_2\rightarrow 0$ limit of \eqref{ZS5 Nek intro} with $\amCB=i \sigma$. In our approach, the  Lockhart-Vafa  factorisation \eqref{ZS5 BPS intro} into BPS states is completely equivalent to the following expansion of the effective prepotential of the 4d $\CN=2$ KK theory:
\be\label{sum CF intro}
\CF(\amCB) = \CF_{\rm cl}(\amCB)-{1\ov (2\pi i)^3}\sum_{\alpha \; {\rm BPS}}  d^{(\alpha)}\; \trilog(Q^{(\alpha)})~,
\ee
with $Q= e^{2\pi i \amCB}$, schematically.  An analogous expansion must also hold for $\log \bA$ and $\log \bB$. 
 Whenever the 5d SCFT can be engineered at a Calabi-Yau threefold singularity in M-theory, the sum in \eqref{sum CF intro} is a sum over wrapped M2-branes near the resolved singularity, with $Q$ the exponentiated K\"ahler parameters, and the numerical constants $d^{(\alpha)}$ are then essentially the (refined) Gopakumar-Vafa invariants of the resolved threefold \cite{Gopakumar:1998ii, Gopakumar:1998jq, Hollowood:2003cv, Iqbal:2007ii}. The term $\CF_{\rm cl}(\amCB)$ in \eqref{sum CF intro} denotes the `classical' terms, which are supergravity contributions in M-theory; for our purposes in this paper, we will essentially ignore these terms.

\subsection{Three ways to obtain the CB partition functions}
We can compute the CB partition function on $\CM_5$ using three complementary methods, which all yield the same answer:
\begin{itemize}
\item[(1)] {\bf One-loop determinants.} In this approach, we first study ordinary 5d $\CN=1$ gauge theories on our supersymmetric background $\CM_5$. This is in keeping with standard supersymmetric localisation computations  \cite{Qiu:2016dyj}, and it would seem to only capture the `perturbative' part of the full 5d SCFT partition function. However, if we assume the validity of the Lockhart-Vafa factorisation: 
\be
Z_{\CM_5}(\amCB)_\m=\prod_{\alpha \; {\rm BPS}}  Z^{(\alpha)}_{\CM_5, {\rm BPS}}(\amCB)_\m~,
\ee
on any $\CM_5$, then each factor $Z^{(\alpha)}_{\CM_5, {\rm BPS}}$ can be computed by a simple generalisation of the one-loop determinant computation for a free hypermultiplet. We pay particular attention to regularising the one-loop results in a way that fully respects five-dimensional gauge invariance (and hence, in general, breaks parity), following \cite{Closset:2018bjz}. 

\item[(2)] {\bf Low-energy effective action.} In this approach, which forms the core of the paper, we study the low-energy effective couplings on the Coulomb branch. This is the TQFT logic summarised in section~\ref{subsec:DWapproach} above. The fibering operator arises as a background flux insertion for the $U(1)_{\rm KK}$ symmetry (that is, the conserved momentum along the circle). This general method  does not require us to assume the Lockhart-Vafa factorisation, or to assume anything about the 5d SCFT, and it works for any (simply connected) K\"ahler manifold $\CM_4$. (A somewhat similar approach was taken in \cite{Crichigno:2018adf, Jain:2021sdp, Jain:2022avc, Santilli:2020uht} for geometries that are more closely related to the 3d setup of \cite{Closset:2017zgf}.)

\item[(3)] {\bf Nekrasov partition function gluing.} Our third approach is only valid for $\CM_5$ a circle bundle over $\CM_4$ a toric manifold. Then, we can construct $\CM_5$ as a toric gluing  of $\chi(\CM_4)$ distinct patches  $\C^2\times S^1$, with the CB partition function obtained by gluing Nekrasov partition functions $Z_{\C^2\times S^1}$ -- this approach has been discussed extensively in the literature, see {\it e.g.} \cite{Lockhart:2012vp, Kim:2012qf, Imamura:2012efi,  Qiu:2013aga, Qiu:2014oqa, Qiu:2015rwp}.%
\footnote{The gluing approach is more general and can also be applied beyond the topological twist by gluing `topological' and `anti-topological' Nekrasov partition functions, as first discussed by Pestun for $S^4$~\protect\cite{Pestun:2007rz} and later generalised in various directions \cite{Nieri:2013yra, Festuccia:2016gul, Festuccia:2018rew, Festuccia:2019akm, Festuccia:2020yff}.}
 We revisit the case of the five-sphere, and we propose a generalisation of \eqref{ZS5 Nek intro} to any such $\CM_5$, which reads:
\be
    {\bf Z}_{\cM_5} = \sum_{\mathfrak{n}_l} \oint d\hspace{0.7pt}\amCB\prod_{l=1}^{\chi(\CM_4)} Z_{\bC^2\times S^1} \left({\amCB+\tau_1^{(l)}\mathfrak{n}_l + \tau_2^{(l)}\mathfrak{n}_{l+1} \ov \gamma^{(l)}}, {\tau_1^{(l)} \ov \gamma^{(l)}}, {\tau_2^{(l)} \ov \gamma^{(l)}}\right)~.
\ee
The quantities in the arguments will be explained in section~\ref{sec:NekGlue}. (In particular, here $\tau_i= \beta \epsilon_i$ denotes the $\Omega$-deformation parameters, with $\beta$ the radius of the $S^1$ fiber.) In the non-equivariant limit, $\tau_1, \tau_2\rightarrow 0$, we recover the CB partition function for the DW twist.

\end{itemize}

\noindent
An important subtlety, which we will address in detail, arises because the DW twist is actually not well-defined for every 5d SCFT. This is true whenever $\CM_4$ is not spin and the twisted theory contains fields that transform as spinors. For instance, this is the case for the simplest theory, a free hypermultiplet on a non-spin manifold like $\bP^2$. In general one needs to consider (background) spin$^c$ connections instead of ordinary $U(1)$ connections \cite{Witten:1994cg, Labastida:1997rg, Manschot:2021qqe}. In physics language, this simply means that we should consider abelian gauge fields $A$ with some half-integer-quantised fluxes:
\be\label{spinc intro}
{1\ov 2\pi}\int_{\S} F = \m +\varepsilon \k~, \qquad \varepsilon \in \half \Z~.
\ee
with $F=dA$. Here $\k$ denotes the first Chern class of the canonical line bundle $\CK$ over $\CM_4$, and the `ordinary' fluxes $\m$ are integer-quantised. Recall that a K\"ahler manifold $\CM_4$ is spin if and only if the `square root' $\CK^\half$ exists. The partition function will ultimately depend on the choice of $\varepsilon$ in \eqref{spinc intro}. We call the corresponding supersymmetric background the {\it extended DW twist}.%
\footnote{Slightly more general choices of spin$^c$ connections are possible (see {\it e.g.} \protect\cite{Manschot:2021qqe}), but we will restrict ourselves to the quantisation condition \protect\eqref{spinc intro} for simplicity.} The allowed values of the extended DW-twist parameter $\varepsilon$ depend on the theory. Moreover, requiring that the 5d theory can be defined on any $\CM_4$ (with a consistent choice of $\varepsilon$) actually imposes the existence of an interesting `spin/charge relation'  \cite{Seiberg:2016rsg, Cordova:2018acb}, which is a non-trivial constraint on the 5d BPS spectrum.%
\footnote{It would be interesting to know whether any 5d SCFT necessarily satisfies this spin/charge relation. We will only show that it holds explicitly in some rank-1 theories.}

For $\CM_5$ a principal circle bundle of first Chern class $\p$ over $\CM_4$, the full CB partition function  takes the form:
\be \label{master formula}
    Z_{\cM_5}(\amCB; \varepsilon)_\m = \bA(\amCB)^{\chi}\bB(\amCB)^{\sigma}\bG(\amCB;\varepsilon)^{2\chi+3\sigma} \Flux(\amCB)^{\half (\m+2\varepsilon \mK, \m)} \FiberOpK(\amCB)^{(\p, \m+ \varepsilon \mK)}\,  \FiberOp(\amCB)^{\half (\p, \p)}~.
\ee
Here, $\chi$ and $\sigma$ are the Euler characteristic and signature of $\CM_4$, respectively. The first three couplings, $\bA$, $\bB$ and $\bG$, are effective gravitational couplings (including $\bG$, the coupling to the spin$^c$ connections), the fourth coupling, $\Flux$, is dubbed the `flux operator', which inserts some abelian (background) flux on $\CM_4$, and
\be
 \h \FiberOp_\p(\amCB; \varepsilon)_\m \equiv  \FiberOpK(\amCB)^{(\p, \m+ \varepsilon \mK)}\,  \FiberOp(\amCB)^{\half (\p, \p)}~,
\ee
gives us the fibering operator, which is a flux operator for the $U(1)_{\rm KK}$ symmetry. The relation \eqref{ZS5 intro} is a simple instance of this master formula, for $S^5$ with $\m=0$ and $\varepsilon=0$. 

Given the very general result \eqref{master formula}, the next step will be to compute the full partition function by performing the integration over the Coulomb branch of the 4d $\CN=2$ KK theory, with the hope of deriving \eqref{ZM5 full conj intro} as well as our conjecture \eqref{ZS5 conj intro}. For theories of rank one, such computations would build upon recent analyses of the Coulomb branch geometry \cite{Aspman:2021vhs, Closset:2021lhd, Magureanu:2022qym}. This is left for future work. 

\medskip

\noindent
This paper is organised as follows. In section~\ref{sec:DW twist}, we discuss the topological twist on  $\CM_4$, setting up our conventions. In section~\ref{sec:5dDW}, we define and study a class of supersymmetric backgrounds for 5d $\CN=1$ theories on $\CM_5$. In section~\ref{sec:oneloopdet}, we compute the relevant one-loop determinants on $\CM_5$. In section~\ref{sec:fluxFiberingOps}, we derive the master formula \eqref{master formula} using the low-energy effective action in curved space. Finally, in section~\ref{sec:NekGlue}, we study the gluing of Nekrasov partition functions for $\CM_4$ a toric surface. Our geometry and supersymmetry conventions, as well as some useful additional  results, are collected in various appendices.

\section{Topological twist on a K\"ahler surface}\label{sec:DW twist}
In this section, we review the  topological twist of four-dimensional $\CN=2$ supersymmetric gauge theories \cite{Witten:1988ze, Witten:1994ev}. We focus on K\"ahler four-manifolds (also known as  K\"ahler surfaces), on which we can preserve two supercharges, and we discuss the `twisted variables' that are most useful in that context. By a slight abuse of terminology, we call this the Donaldson-Witten (DW) twist. We will particularly insist on the `extended DW twist' of the hypermultiplet. 

\subsection{K\"ahler surfaces and the $\CN=2$ topological twist}
Consider a K\"ahler four-manifold, $\CM_4$, viewed as a hermitian manifold $(\CM_4, J, g)$ whose complex structure is covariantly constant, 
$\nabla_\mu {J^{\nu}}_\rho=0$. 
Let us use some local complex coordinates $(z^i)= (z^1, z^2)$, in terms of which the K\"ahler metric $g$ reads:  
\be\label{K metric M4}
ds^2 = 2 g_{i\b j} dz^i d\b z^{\b j}~, \qquad \qquad g_{i\b j}={ \d^2 K \ov \d z^i \d\b z^{\b j}}~,
\ee
where $K$ is the K\"ahler potential. Our geometric conventions are spelled out in appendix~\ref{app:4dconventions}.

We are interested in 4d $\CN=2$ quantum field theories in Euclidean space-time with an exact $SU(2)_R$ $R$-symmetry. 
 Let us first recall the standard definition of the topological twist, as originally given by Witten~\cite{Witten:1988ze}. We start with a theory with a global symmetry that includes the Euclidean rotation ${\rm Spin}(4)\cong SU(2)_l\times SU(2)_r$ and the $R$-symmetry, and we relabel the spins of fields according to the new `twisted spin',
 \be
 SU(2)_l\times SU(2)_D~, \qquad\qquad SU(2)_D \equiv {\rm diag}(SU(2)_r\times SU(2)_R)~.
 \ee
Alternatively, the DW twist can be understood as a supergravity background ({\it i.e.} a rigid limit of some 4d $\CN=2$ supergravity), consisting  of a metric  \eqref{K metric M4} and of a background gauge field ${\bf A}^{(R)}$ for the $R$-symmetry,  preserving some fraction of the flat-space supersymmetry (see {\it e.g.} \cite{Karlhede:1988ax, Festuccia:2011ws, Klare:2013dka}).  One preserves a right-chiral supersymmetry:
\be
\delta_{\t\xi} \equiv \t \xi^I_\alphadot \t Q_I^\alphadot~,
\ee
on any background $(\CM_4, g, {\bf A}^{(R)})$ that admits a covariantly-constant spinor $\t\xi_I$:
\be\label{KSE 4d gen}
 D_\mu \t\xi_I \equiv \left(\nabla_\mu {\delta_I}^J- i{ ({\bf A}^{(R)}_\mu)_I}^J \right)\t\xi_J=0~.
\ee
Here and in the following, $I=1,2$ are $SU(2)_R$ indices.%
\footnote{We usually keep the $SU(2)_R$ indices explicit, while suppressing the spinor indices $\alpha, \alphadot$.}
Such a background exists on any Riemannian four-manifold: one obtains a solution to \eqref{KSE 4d gen} by identifying the $SU(2)_R$ connection with the spin connection \cite{Witten:1988ze}. In our conventions, we  have:
 \be\label{top sol gen}
 { ({\bf A}^{(R)}_\mu)_I}^J= {i\ov 2} \omega_{\mu a b} {(\t\sigma^{ab})^\alphadot}_\betadot\,  {\delta^J}_\alphadot \, {\delta_I}^\betadot~, \qquad
 \quad({ \t \xi^\alphadot}_I)=({\delta^\alphadot}_I)= \mat{1 & 0 \\ 0 &1}~.
 \ee
More invariantly, the Killing spinor is a section of a complex vector bundle:
 \be
\t\xi \in \Gamma[S_+ \otimes E_R]~,
\ee
where $E_R$ is a rank-2 $SU(2)_R$ vector bundle.  The topological twist \eqref{top sol gen} consists in choosing $E_R \cong S_+$, in which case $S_+ \otimes E_R$ decomposes as a direct sum
\be\label{SER exp}
S_+ \otimes E_R \cong \CO\oplus \Omega^+~,
\ee
where $\Omega^+$ is the rank-3 vector bundle of self-dual $2$-forms.%
\footnote{At the level of ${\rm Spin}(4)$ representations, we have $(0, \half)\otimes (0, \half) = (0,0)\oplus (0,1)$.} 
  Then our Killing spinor $\t\xi$ is simply the constant section of the trivial line bundle $\CO$. It is also important to note that the topological twist is defined on any four-manifold, irrespective of whether it is a spin manifold, because the bundle \eqref{SER exp} is well-defined even when $S_+$ is not.  

We are interested in  K\"ahler surfaces, in which case we actually preserve  two distinct supersymmetries: 
\be\label{susy on Kahler}
\delta_1 \equiv \t \xi_{(1) \alphadot}^I \t Q_I^\alphadot~,\qquad\qquad\quad
\delta_2 \equiv \t \xi_{(2) \alphadot }^I\t Q_I^\alphadot~.
\ee
The Levi-Civita connection on a K\"ahler manifold has reduced holonomy $U(2) \cong SU(2)_l \times U(1)_r \subset SU(2)_l \times SU(2)_r$, and we then only need to `twist' $U(1)_r$ by turning on a non-trivial gauge field for the $R$-symmetry subgroup $U(1)_R \subset SU(2)_R$. 
In the complex frame basis $(e^1, e^2)$, the spin connection reads:
 \be
 \half \omega_{\mu ab} \t\sigma^{ab} = -(\omega_{\mu 1\b 1}+ \omega_{\mu 2\b 2}) \tau^3~, \quad \qquad \tau^3\equiv \mat{1 & 0 \\ 0 & -1}~. 
 \ee
 By choosing the background $SU(2)_R$ gauge field:
 \be
 { ({\bf A}^{(R)}_\mu)_I}^J = \sum_{{\bf a}=1}^3 A_\mu^{\bf a} {(\tau^{\bf a})_I}^J= A^{\bf 3}_\mu {(\tau^3)_I}^J~,
 \qquad 
 A_\mu^{\bf 3}  dx^\mu= -i  \left( \omega_{\mu 1 \b 1} + \omega_{\mu 2\b 2}\right)~,
 \ee
with $\tau^{\bf a}$   the Pauli matrices, we preserve the two Killing spinors:
 \be\label{top sol Kahler}
( \t \xi_{(1) I}^\alphadot)=(\delta^{\alphadot {\dot 1}} \delta_{I 1})= \mat{1 & 0 \\ 0 &0}~, \qquad\qquad
( \t \xi_{(2) I}^\alphadot)=  (\delta^{\alphadot {\dot 2}} \delta_{I 2}) = \mat{0 & 0 \\ 0 &1}~.
 \ee
Note that the solution $\t\xi_I$ in \eqref{top sol gen} is the sum of these two Killing spinors, 
\be\label{txi gen sum}
\t\xi= \t\xi_{(1)}+\t\xi_{(2)}~.
\ee
 Correspondingly, we preserve the flat-space supercharges $\t Q_{\dot 2}^2$ and $\t Q_{\dot 1}^1$ on any K\"ahler manifold, while on a generic four-manifold we only preserve their sum, $\t Q_{\dot 1}^1+ \t Q_{\dot 2}^2$.

Let us describe this  Donaldson-Witten twist more covariantly. On any K\"ahler surface $\CM_4$, the spin bundle ${\bf S} \equiv S_- \oplus S_+$ formally decomposes as:
\be\label{spin bndl 4d gen}
S_- \cong \CK^{\half} \otimes \Omega^{0,1}~,\qquad \qquad S_+ \cong \CK^{\half} \oplus \CK^{-\half}~,
\ee 
with $\CK$ the canonical line bundle. Here, $\CM_4$ is spin if and only if the `square-root' $\CK^\half$ actually exists. Recall that the second Stiefel-Withney class of a complex surface $\CM_4$ is related to its first Chern class, namely
\be
w_2(\CM_4)\cong c_1(\CK) \mod 2~.
\ee
 Let us choose an $SU(2)_R$ vector bundle of the form
\be\label{ER def}
E_R = L_R^{-1} \oplus L_R~,
\ee
for $L_R$ some $U(1)_R$ line bundle. The Killing spinors \eqref{top sol Kahler} are really sections
\be\label{KS as sections 4d}
\t\xi_{(1)} \in \Gamma[S_+ \otimes L_R]~, \qquad \qquad 
\t\xi_{(2)} \in \Gamma[S_+ \otimes L_R^{-1}]~,
\ee
and the DW twist amounts to the formal identification
\be\label{LR top twist}
L_R \cong \CK^{-\half}~.
\ee
In general,  $\CM_4$ is not spin and therefore $\CK^\half$ does not exist, but the bundles $S_+ \otimes L_R^{\pm 1}$ are nonetheless well-defined spin$^c$ bundles. We will further comment on this point in section~\ref{subec: 4d hyper} below,  
 where we discuss the topological twist of the hypermultiplet.

Note also that the Killing spinors \eqref{KS as sections 4d} are precisely the ones that give rise to a single curved-space supercharge for $\CN=1$ supersymmetric field theories on $\CM_4$ a K\"ahler manifold  \cite{Witten:1994ev, Dumitrescu:2012ha}. The two distinct $\CN=1$ subalgebras correspond to $\t Q^{I=1}$ and  $\t Q^{I=2}$ for the spinors $\t\xi_{(1)}$ and $\t\xi_{(2)}$, respectively.

\paragraph{Spinor bilinears and K\"ahler structure.} Given the Killing spinors introduced so far,  one can construct well-defined two-forms on $\CM_4$. First of all, given any solution $\t\xi$ to the Killing spinor equation \eqref{KSE 4d gen}, we can define the $SU(2)_R$-neutral anti-self-dual two-form:
\be\label{CJ bilinear 4d}
\CJ_{\mu\nu}[\t\xi] \equiv - 2i  { {\t \xi}^{\dagger I} \t \sigma_{\mu\nu} \t \xi_I\ov |\t\xi|^2}~, \qquad \quad 
|\t\xi|^2\equiv  {\t \xi}^{\dagger I}  \t \xi_I~,
\ee
where the sum over repeated indices is understood. For the Killing spinor \eqref{top sol gen} on a general four-manifold, the bilinear \eqref{CJ bilinear 4d} identically vanishes. On the other hand, from the Killing spinors  \eqref{top sol Kahler}, we obtain:
 \be\label{J from spinors 4d}
J_{\mu\nu} \equiv  \CJ_{\mu\nu}[\t\xi_{(1)}]   =  -  \CJ_{\mu\nu}[\t\xi_{(2)}]~,
 \ee
 which satisfies:
\be
{J^\mu}_\nu{J^\nu}_\rho = - {\delta^\mu}_\rho~, \qquad \nabla_\mu J_{\nu\rho}=0~.
 \ee
 Thus, \eqref{J from spinors 4d} gives us the complex structure (and the associated K\"ahler form) of the hermitian K\"ahler manifold $\CM_4$. In this way, one can show that there are two linearly independent solutions to \eqref{KSE 4d gen} if and only $\CM_4$ is K\"ahler \cite{Dumitrescu:2012ha}. 
 Given the two Killing spinors \eqref{KS as sections 4d}, we may also write down the bilinears:
\bea
p_{(1)}^{2,0} \equiv \t\xi_{(1)} \sigma_{\mu\nu} \t\xi_{(1)}\, dx^\mu \wedge dx^\nu 
\;  &\; \in \, \Gamma[\CK \otimes L_R^2]~, \cr
p_{(2)}^{0,2} \equiv \t\xi_{(2)} \sigma_{\mu\nu} \t\xi_{(2)} \, dx^\mu \wedge dx^\nu  &\; \in \, \Gamma[\CK^{-1} \otimes L_R^{-2}]~. \cr
\eea
In the frame basis, we have:
\be
p_{(1)}^{2,0}=  - e^1 \wedge e^2~, \qquad 
p_{(2)}^{0,2} = - e^{\b1} \wedge e^{\b 2}~,
\ee
for the solutions \eqref{top sol Kahler}.
These are nowhere-vanishing sections of line bundles, therefore the corresponding line bundles are trivial.  
This is another way to see that \eqref{LR top twist} must hold, or more precisely $L_R^{2} \cong \CK^{-1}$ if $\CM_4$ is not spin. This topological twist is therefore also an `$\CN=1$ holomorphic twist' (see {\it e.g.}~\cite{Dumitrescu:2012ha, Closset:2014uda}) for either of the two $\CN=1$ subalgebras mentioned above.

\subsection{The vector multiplet on $\CM_4$}
Let us consider the $\CN=2$ vector multiplet $\CV$, in the adjoint representation of some Lie algebra $\Fg= {\rm Lie}(G)$, on $\CM_4$  a K\"ahler manifold. It consists of a gauge field $A_\mu$, two sets of gauginos $\lambda^I$, and a triplet of auxiliary scalar fields $D_{IJ}=D_{JI}$: 
  \be 
 \CV = \left(A_\mu, \phi, \t\phi, \lambda^I,\t \lambda_I, D_{IJ}\right)~.
 \ee
 Its off-shell supersymmetry transformations on $\R^4$ are reviewed in Appendix~\ref{App:subsec:4dSUSY}.
The gauge connection $A=A_\mu dx^\mu$ is well defined on any four-manifold. After the topological twist, the gauginos are also well-defined on $\CM_4$. The left-chiral gauginos $\lambda=(\lambda^I)$ are sections
\be\label{lambda twisting}
\lambda\in \Gamma[S_-\otimes E_R] \cong \Gamma[\Omega^{0,1} \oplus (\CK\otimes \Omega^{0,1})]\cong  \Gamma[\Omega^{0,1} \oplus \Omega^{1,0}]~.
\ee
We use the Hodge star operator to map  $(2,1)$-forms (the sections of  $\CK\otimes \Omega^{1,0}$) to $(1,0)$-forms, according to $\omega^{1,0}= \star\, \omega^{2,1}$. Similarly, the  right-chiral gauginos $\t\lambda=(\t\lambda_I)$ are sections
\be\label{tilde lambda twisting}
\t\lambda\in \Gamma[S_+\otimes \b E_R] \cong \Gamma[ \CO \oplus \CO\oplus \CK \oplus \CK^{-1}]~.
\ee
These $(p,q)$-forms can be constructed explicitly from the ordinary (flat-space) spinors, by contracting the gauginos with the Killing spinors \eqref{top sol Kahler} to form $SU(2)_R$-neutral tensors. For the left-chiral gauginos,  the holomorphic and anti-holomorphic 1-forms in \eqref{lambda twisting} are given by: 
\be
\Lambda^{1,0}\equiv \t \xi_{(1) I} \t \sigma_\mu \lambda^I \, dx^\mu~, \qquad\quad
\Lambda^{0,1}\equiv \t \xi_{(2) I} \t \sigma_\mu \lambda^I \, dx^\mu~.
\ee
  Similarly, starting from the right-chiral gauginos $\t\lambda=(\t \lambda_I)$, we define the two scalars
 \be
 \t \Lambda^{0,0}_{(1)} \equiv \t\xi_{(1)}^I \t\lambda_I~, \qquad \quad
 \t \Lambda^{0,0}_{(2)} \equiv \t\xi_{(2)}^I \t\lambda_I~,
 \ee
and the holomorphic and anti-holomorphic two-forms:
\bea
&\t \Lambda^{2,0} \equiv \left( \t\xi^I_{(1)} \t\sigma_{\mu\nu} \t\lambda_I + {i\ov 2} \t\xi^I_{(1)} \t\lambda_I \, J_{\mu\nu} \right) dx^\mu \wedge dx^\nu = \half \t \Lambda^{2,0}_{\mu\nu} dx^\mu \wedge dx^\nu~, \cr
&\t \Lambda^{0,2} \equiv \left( \t\xi^I_{(2)} \t\sigma_{\mu\nu} \t\lambda_I - {i\ov 2} \t\xi^I_{(2)} \t\lambda_I \, J_{\mu\nu} \right) dx^\mu \wedge dx^\nu = \half \t \Lambda^{0,2}_{\mu\nu} dx^\mu \wedge dx^\nu~,
\eea
in agreement with \eqref{tilde lambda twisting}.
 In our choice of local frame basis, we have:
\bea
& \Lambda^{1,0}= -\lambda_1^1 \, e^1 + \lambda_2^1 \,e^2~, \qquad\qquad
&&\Lambda^{0,1}=   \lambda_2^2 \,e^{\b 1} + \lambda_1^2 \,e^{\b 2}~, \cr
& \t \Lambda^{0,0}_{(1)}= \t\lambda^{\dot 2}_2~, \qquad &&   \t \Lambda^{0,0}_{(2)}= \t\lambda^{\dot 1}_1 
~, \cr
&   \t\Lambda^{2,0} = - \t\lambda^{\dot 1}_2\, e^1 \wedge e^2~, \qquad &&   \t\Lambda^{0,2} = \t\lambda^{\dot 2}_1 \, e^{\b 1} \wedge e^{\b 2}~. \cr
\eea
Let us also define the $SU(2)_R$-neutral auxiliary fields:
\bea \label{4d CD0}
&\CD^{2,0} \equiv - i  D_{IJ}\, \xi_{(1)}^I \t\sigma_{\mu\nu} \t\xi^J_{(1)} \, dx^\mu\wedge dx^\nu &&= - i D_{22} p_{(1)}^{2,0}~,\cr
&\CD^{0,2} \equiv  i  D_{IJ}\, \xi_{(2)}^I \t\sigma_{\mu\nu} \t\xi^J_{(2)} \, dx^\mu\wedge dx^\nu &&=- i D_{11} p_{(2)}^{0,2}~, \cr
& \CD^{0,0} \equiv   \half J^{\mu\nu} \left(- i  D_{IJ}\, \xi_{(1)}^I \t\sigma_{\mu\nu} \t\xi^J_{(2)} + F_{\mu\nu}\right) &&=D_{12} +   \h F~,
\eea
with
\be
  \h F \equiv  \half J^{\mu\nu} F_{\mu\nu} =  2i(F_{1\b 1} + F_{2 \b 2})~.
\ee
The flat-space supersymmetry transformations of the vector multiplet are written explicitly in Appendix~\ref{App:subsec:4dSUSY} -- see equation \eqref{susy vec offshell Neq2}. Using the twisted variables, one obtains the following curved-space supersymmetry transformations under the two supercharges \eqref{susy on Kahler}:
\bea\label{susy vec twisted}
& \delta_1 \phi =0~, && \delta_2 \phi= 0~, \cr
& \delta_1 \t\phi =\sqrt2 \t\Lambda_{(1)}^{0,0}~, \; && \delta_2\t \phi= \sqrt2 \t\Lambda_{(2)}^{0,0}~, \cr
& \delta_1 A   =- i \Lambda^{1,0}~, \qquad && \delta_2  A   = - i \Lambda^{0,1}~, \cr
& \delta_1 \Lambda^{1,0}   =0~, \qquad && \delta_2 \Lambda^{1,0}    = 2 i \sqrt2 \d_A \phi~, \cr
& \delta_1  \Lambda^{0,1}  =2i \sqrt2 \b \d_A \phi~, \qquad && \delta_2  \Lambda^{0,1}   = 0~, \cr
& \delta_1  \t \Lambda_{(1)}^{0,0}  =0~, \qquad && \delta_2 \t \Lambda_{(1)}^{0,0}    = i \CD^{0,0} - i [\t\phi, \phi]~, \cr
& \delta_1   \t \Lambda_{(2)}^{0,0} =- i \CD^{0,0} - i [\t\phi, \phi]~, \qquad && \delta_2  \t \Lambda_{(2)}^{0,0}   = 0~, \cr
& \delta_1  \t\Lambda^{2,0}  =\CD^{2,0}~, \qquad && \delta_2     \t\Lambda^{2,0}  = 4 F^{2,0}~, \cr
& \delta_1    \t\Lambda^{0,2}  =4 F^{0,2}~, \qquad && \delta_2     \t\Lambda^{0,2}  = \CD^{0,2}~, \cr
& \delta_1   \CD^{0,0} =\sqrt2 [\phi, \t\Lambda_{(1)}^{0,0}]~, \qquad && \delta_2   \CD^{0,0}   =- \sqrt2 [\phi, \t\Lambda_{(2)}^{0,0}]~, \cr
& \delta_1 \CD^{2,0}   =0~, \; && \delta_2   \CD^{2,0}  = 4 i\d_A \Lambda^{1,0}+ 2 i \sqrt2 [\phi, \t\Lambda^{2,0}]~, \cr
& \delta_1   \CD^{0,2}  =4 i \b\d_A \Lambda^{0,1}+ 2 i \sqrt2 [\phi, \t\Lambda^{0,2}]~, \; && \delta_2   \CD^{0,2}   = 0~.
\eea
Here,  $\d_A$ and $\b \d_A$ denote the Dolbeault operators twisted by the gauge field $A=A_\mu dx^\mu$:
\be
d_A  = d -i A= \d_A + \b \d_A~.
\ee
Moreover, $F^{2,0}$ and $F^{0,2}$ denote the $(2,0)$ and $(0,2)$ projection of the field strength $F= \half F_{\mu\nu} dx^\mu\wedge dx^\nu$. Note that we have:
\be\label{rel Dolb square 4d}
\d_A^2= - i F^{2,0}\wedge~, \qquad\qquad
\b\d_A^2= - i F^{0,2}\wedge~, \qquad\qquad
\{\d_A, \b\d_A\}= - i F^{1,1}\wedge~.
\ee
 It is straightforward to check that the  transformations \eqref{susy vec twisted} realize the supersymmetry algebra:
\be \label{4d vec susy closure}
\delta_1^2  =0~, \qquad\qquad \delta_2^2  =0~, \qquad\qquad \{ \delta_1, \delta_2\} = 2  \sqrt2 \delta_{g(\phi)}~,
\ee
where $\delta_{g(\phi)}$ is a gauge transformation with parameter $\phi$. In particular, we have $\delta_{g(\phi)} A = d_A \phi= d\phi + i [\varphi, A]$ for the gauge field, and $\delta_{g(\phi)}\varphi = i [\phi, \varphi]$
for any field $\varphi$  transforming in the adjoint representation of $\Fg$.

\noindent 
{\bf Supersymmetric Lagrangian.} The 4d $\CN = 2$ SYM Lagrangian (given by equation \eqref{Neq2 SYM} in flat-space)  can be written compactly as
\be    \label{4d SYM lagrangian}
 \SL_{\rm SYM}  = {1\ov g^2}  \delta_1 \delta_2 \tr\left({1\ov 8}  \star\left(  \t \Lambda^{2,0} \wedge \t \Lambda^{0,2}\right) - i {\sqrt{2}\ov 4}  \t\phi \left( \CD^{0,0} - 2 \h F\right)  \right)  - {1\ov 2 g^2}\star \tr\left(F\wedge F\right)~.
\ee
%with the overall trace over the gauge indices left implicit.
Note that it is ``mostly'' $\CQ$-exact. The bosonic terms read:
\bea   
& \SL_{\rm SYM}\big|_{\rm bos}  &=&\; {1\ov g^2}  \tr\Bigg(
 \star \left(2 F^{2,0}\wedge F^{0,2}- \half  F\wedge F\right) -\t\phi D_\mu D^\mu\phi\\
&&&\qquad  -{1\ov 8}\star (\CD^{2,0}\wedge \CD^{0,2}) - \half\CD^{0,0}\left(\CD^{0,0}-2 \h F\right)
\Bigg)~.
\eea
On supersymmetric configurations, the $\CQ$-exact terms in \eqref{4d SYM lagrangian} evaluate to zero. Defining the `instanton number' -- more precisely, (minus) the second Chern class of any holomorphic vector bundle associated to a principal $G$ bundle --  as
\be
k =- {1\ov 16\pi ^2}\int_{\CM_4} \tr\left(F\wedge F\right)~,
\ee
and adding the topological coupling
\be
S_{\rm top}= i  {\theta \ov 16\pi^2}\int \tr\left(F\wedge F\right) =-i \theta k~,
\ee
to the Lagrangian,  any supersymmetric vector-multiplet configuration is weighted by a factor:
\be
e^{-S_{\rm SYM}- S_{\rm top}}= e^{2\pi i \tau k}~,
\ee
where we defined the holomorphic gauge coupling as
\be
\tau= {\theta\ov 2\pi} +  {4\pi i \ov g^2}~.
\ee
In particular, the classical saddles are Yang-Mills instantons%
\footnote{More precisely, {\it anti-instantons}, satisfying the anti-self-duality condition $F=- \star F$. We  can call them `instantons' since self-dual instantons do not play a role in Donaldson-Witten theory.} 
and they contribute in this way.

%%%%%%%%%%%%%%%%%%%%%%%%%%%%%%%%%%%%%%%%%%%

\subsection{The hypermultiplet on $\CM_4$}\label{subec: 4d hyper}
Consider a hypermultiplet $\CH$ charged under a Lie group $G$. When considered as part of a larger gauge theory, $G$ will include both the gauge group, with its dynamical gauge fields, and the flavor symmetry group, with its background gauge fields. On-shell, this 4d $\CN=2$ multiplet consists of two complex scalars, $q^I$, forming a doublet of $SU(2)_R$, and of two $SU(2)_R$-neutral Dirac fermions:
\be
\CH= \left( q^I~,\, \t q_I~,\, \eta~,\, \t\eta~,\, \chi~,\, \t\chi \right)~.
\ee
In addition, we will need to introduce some auxiliary fields in order to realize the two supersymmetries $\delta_1$ and $\delta_2$ off-shell.%
\footnote{It is well-known that one cannot realize the full flat-space $\CN=2$ supersymmetry off-shell with a finite number of auxiliary fields, but there is no problem with realizing these two particular supercharges off-shell. }
The fields $q^I$, $\eta$ and $\t \chi$ transform in some representation $\FR$ of the gauge algebra $\Fg$, and the fields $\t q_I$, $\chi$ and $\t \eta$ transform in the conjugate representation  $\b \FR$.
After the topological twist, the scalars $q^I$, $\t q_I$ become right-chiral spinors, which are therefore not well-defined unless $\CM_4$ is a spin manifold \cite{Alvarez:1994ii}. For charged hypermultiplets, this issue can be remedied by introducing a spin$^c$ structure \cite{Witten:1994cg, Hyun:1995mb, Labastida:2005zz}. Such a structure exists on any oriented closed four-manifold, but it is important to emphasise that this is an additional choice that we make when considering hypermultiplets. We thus call this an `extended DW twist'.

Without too much loss of generality, let us  consider $\CH$ charged under some gauge group $\t G = U(1) \times G$, where the $U(1)$ gauge field is really a spin$^c$ connection. It is associated with a line bundle $\CL_0$ such that $\CL_0^\half \otimes S_+$ is well-defined. On a K\"ahler manifold, with the spin bundle  formally given by \eqref{spin bndl 4d gen}, we will choose:
 \be
 \CL_0 \cong \CK^{-1}~.
 \ee
 We insist on the fact that this is a somewhat arbitrary choice, however natural it appears on a K\"ahler manifold. For our purposes, it will be important to consider the more general case:
 \be\label{CL0 extended twist}
\CL_0 \cong \CK^{2\varepsilon}~, 
 \ee
 with $\varepsilon\in \half \Z$ a free parameter. Roughly speaking, the extended topological twist is simply a choice of $\varepsilon$ for each hypermultiplet in a theory; this must be done in a consistent way, as we will discuss further in later sections. 
  
 \medskip
 \noindent
 The ordinary DW twist of the hypermultiplet scalars gives us the right-chiral spinors
 \be
 {\bf q}\equiv \t \xi_I q^I~, \qquad \qquad
 \t {\bf q} \equiv  \epsilon^{IJ}\t\xi_I \t q_J~,
 \ee
 where we are using the DW Killing spinor \eqref{txi gen sum}. On an arbitrary K\"ahler manifold, the extended topological twist exists when these spinors are further valued in $\CL_0^{\half}$, namely:
 \be
  {\bf q}\in \Gamma(S_+\otimes \CK^{\varepsilon})~, \qquad\qquad
   \t {\bf q} \in \Gamma(S_+\otimes \CK^{-\varepsilon})~.
 \ee
 Note that ${\bf q}$ being valued in   $\CK^{\varepsilon}$ means that the charged conjugate scalar $\t {\bf q}$ is valued in $\CK^{-\varepsilon}$. 
 In the rest of this section, we  will set $\varepsilon=-\half$. Reinstating a general $\varepsilon$ will simply correspond to having the twisted hypermultiplet, as described below, also valued in a line bundle $\CK^{\varepsilon+\half}$. 
 Thus, setting $\varepsilon=-\half$,  the scalars become $(p,q)$-forms:
  \be
  {\bf q}\in \Gamma(\CO \oplus \CK^{-1})~, \qquad\qquad
   \t {\bf q} \in \Gamma(\CK\oplus \CO)~.
 \ee
  All fields are also valued in the appropriate vector bundles $E_\FR$ or $E_{\b\FR}$ determined by the representation $\FR$ of $G$ -- we omitted this from the notation to avoid clutter.
 To perform the explicit change of basis between flat-space and twisted variables, it is convenient to introduce the spinors
 \be
(\epsilon_+^\alphadot)=\mat{1\\ 0} \in\Gamma\left(\CK^\half\right)~, \qquad \qquad
(\epsilon_-^\alphadot)=\mat{0\\ -1} \in\Gamma\left(\CK^{-\half}\right)~, 
 \ee
 which are, formally, sections of $\CK^{\pm\half}$, as indicated. For any right-chiral spinor $\t\psi^{\dot\alpha}$, let us  define the contractions
 \bea
& {\bf C}^{0,2}(\t\psi)&=&\, \left( \t\epsilon_- \t\sigma_{\mu\nu} \t\psi -i\, \t\epsilon_-\t\psi\, J_{\mu\nu} \right) dx^\mu \wedge dx^\nu&=&\, \t\psi^{\dot 2} e^{\b1} \wedge e^{\b 2}~,\\
& {\bf C}^{0,2}(\t\psi)&=&\, -\left( \t\epsilon_+ \t\sigma_{\mu\nu} \t\psi +i\, \t\epsilon_+\t\psi\, J_{\mu\nu} \right) dx^\mu \wedge dx^\nu&=&\, \t\psi^{\dot 1} e^{1} \wedge e^{2}~.\\
 \eea
The `twisted scalars' for this extended topological twist are then 
 \bea
& Q^{0,0} &=&\, \epsilon_- {\bf q}&=&\,  q^1~,     \qquad\qquad &&  \t Q^{0,0} &=& \, - \epsilon_+ {\bf q}&=&\, \t q_1~,    \\
&Q^{0,2} &=&\,  {\bf C}^{0,2}(\t {\bf q})   &=&\,q^2 \, e^{\b1} \wedge e^{\b 2}~,   \qquad\qquad
&&  \t Q^{2,0} &=&\, {\bf C}^{2,0}(\t {\bf q}) &=&\,\t q_2\,  e^1 \wedge e^2~.
\eea
We also have the two Dirac spinors:
\be\label{spinc sections}
 \Psi= \mat{\eta_\alpha \cr \t\chi^\alphadot} \in \Gamma[S \otimes \CK^{-\half}]~, \qquad\quad
\t \Psi= \mat{\chi_\alpha \cr \t\eta^\alphadot} \in \Gamma[S \otimes \CK^{\half}]~,
\ee
which are sections of spin$^c$ bundles as explained above.  They can be conveniently decomposed into $(p,q)$-forms:
\be
 \Psi^{0, \bullet}=  (\eta^{0,1}, \t\chi^{0,0},  \t\chi^{0,2})\in \Omega^{0, \bullet}~,\qquad\quad
 \t \Psi^{\bullet,0}= (\chi^{1,0}, \t\eta^{0,0},\t\eta^{2,0} )\in \Omega^{ \bullet,0}~.
\ee
For instance, the spinor $\chi$ is a section of
$S_- \otimes \CK^\half \cong \Omega^{0,1}\otimes \CK \cong \Omega^{2,1}\cong \Omega^{1,0}$, 
where we find it convenient to use $\chi^{1,0} \equiv \star \chi^{2,1}$. The explicit change of variables is given by
\bea
&\eta^{0,1}  =\t \epsilon_- \t\sigma_\mu \eta \, dx^\mu~, \qquad\qquad  && \chi^{1,0}= \t\epsilon_+ \t\sigma_\mu \chi \, dx^\mu~, \\
& \t\chi^{0,0}= \t \epsilon_- \t\chi~, \qquad && \t \eta^{0,0}=\t\epsilon_+ \t\eta~, \\
& \t \chi^{0,2}=  {\bf C}^{0,2}(\t \chi)~, \qquad &&\t\eta^{0,0}=   {\bf C}^{2,0}(\t \eta)~.
\eea
We also need to introduce the auxiliary one-forms $h^{0,1}$ and $\t h^{1,0}$ in order to close the curved-space supersymmetry algebra off-shell. In fact, under the two supersymmetries $\delta_1, \delta_2$, the hypermultiplet splits into two off-shell multiplets (coupled to the vector multiplet):
\be\label{CH twisted content}
\CH \cong (Q^{0,0} , Q^{0,2},  \Psi^{0, \bullet}, h^{0,1}) \oplus ( \t Q^{0,0},  \t Q^{2,0}, \t \Psi^{\bullet,0} , \t h^{1,0})~,
\ee
which consist of purely anti-holomorphic and holomorphic forms, respectively.
 The supersymmetry transformations read:
\bea    \label{4d SUSY hyper part 1}
& \delta_1 Q^{0,0} =0~, && \delta_2  Q^{0,0} = \sqrt2 \t\chi^{0,0}~, \cr
& \delta_1   Q^{0,2}=\sqrt2 \t\chi^{0,2}~, && \delta_2   Q^{0,2}= 0~, \cr
& \delta_1  \eta^{0,1}=  2 i \sqrt2 \b\d_A Q^{0,0}+ h^{0,1}~,\quad && \delta_2   \eta^{0,1}= i \sqrt2 \star \left(\d_A Q^{0,2}\right)~, \cr
& \delta_1  \t\chi^{0,0}=2 i \phi Q^{0,0}~, && \delta_2   \t\chi^{0,0}= 0~, \cr
& \delta_1  \t\chi^{0,2}=0~, && \delta_2   \t\chi^{0,2}= 2 i \phi Q^{0,2}~, \cr
& \delta_1  h^{0,1}=0~, && \delta_2   h^{0,1}= X^{0,1}~, \cr
\eea
and:
\bea     \label{4d SUSY hyper part 2}
& \delta_1 \t Q^{0,0} =-\sqrt2 \t\eta^{0,0}~, && \delta_2  \t Q^{0,0} = 0~, \cr
& \delta_1   \t Q^{2,0}=0~, && \delta_2  \t  Q^{2,0}= \sqrt2 \t\eta^{2,0}~, \cr
& \delta_1  \chi^{1,0}=   i \sqrt2 \star \left(\b \d_A \t Q^{2,0}\right) ~,\quad && \delta_2   \chi^{1,0}=2 i \sqrt2 \d_A \t Q^{0,0}+ \t h^{1,0}~, \cr
& \delta_1  \t\eta^{0,0}=0~, && \delta_2   \t\eta^{0,0}= 2 i \t Q^{0,0} \phi~, \cr
& \delta_1  \t\eta^{2,0}=-2 i \t Q^{2,0}\phi ~, && \delta_2   \t\eta^{2,0}= 0~, \cr
& \delta_1\t  h^{1,0}= \t X^{1,0}~, && \delta_2  \t h^{1,0}=0~, \cr
\eea
with:
\bea
& X^{0,1} & \equiv&\; \; - 4 i \b\d_A \t \chi^{0,0}-2 i \star(\d_A \t \chi^{0,2})+ 2 i \sqrt2 \Lambda^{0,1} Q^{0,0} 
+  i \sqrt2 \star(\Lambda^{1,0} \wedge Q^{0,2})\\
&&&\; \;+ 2i \sqrt2 \phi \eta^{0,1}~, \cr
& \t X^{1,0} & \equiv &\; \; 4 i \d_A \t \eta^{0,0}-2 i \star(\b\d_A \t \eta^{2,0})- 2 i \sqrt2 \t Q^{0,0} \Lambda^{1,0}
-  i \sqrt2 \star(\t Q^{2,0} \wedge \Lambda^{0,1} )\\
&&&\; \;- 2i \sqrt2 \chi^{1,0} \phi ~.  
\eea
In the above expressions, the covariant derivatives act as
\be \label{convention covariant derivative}
    D_{\mu} \, f = \partial_{\mu} \, f - i A_{\mu} \, f~, \qquad \quad D_{\mu}\,\t f = \partial_{\mu}\,\t f + i\,\t f\,A_{\mu}~,
\ee
on any fields $f=(Q, \t\chi, \eta, h)$ and $\t f=(\t Q, \t \eta, \chi, \t h)$ in the gauge representation $\FR$ and $\b \FR$, respectively. The equations of motion for the fermions are $X^{0,1}=0$ and $\t X^{1,0}=0$. Note that we have:
\bea
& \delta_1 X^{0,1} = 2 i\sqrt2 \phi h^{0,1}~, \qquad &&  \delta_2 X^{0,1} =  0~,\\
& \delta_1 \t X^{1,0} = 0~, \qquad &&  \delta_2\t X^{1,0} =  -2 i\sqrt2\, \t h^{1,0} \phi~.
\eea
These transformations reproduce the supersymmetry algebra \eqref{4d vec susy closure}. In particular, we have
\be
    \{\delta_1,\delta_2\} \, f = 2i\sqrt{2} \, \phi \, f~, \qquad \quad \{\delta_1,\delta_2\}\, \t f = -2i\sqrt{2} \, \t f \,\phi~.
\ee
The hypermultiplet Lagrangian on $\CM_4$ can be obtained by starting from the flat-space Lagrangian \eqref{H kin 4d}, writing it in twisted variables, and adding in the auxiliary fields  to ensure off-shell supersymmetry. The important fact is that it is $\CQ_{DW}$-exact. We find
\bea    \label{4d hypermultiplet lagrangian}
    \SL_\CH & = {1\ov 4} (\delta_1+\delta_2) \star\Bigg( \t h^{1,0}\wedge\star \, \eta^{0,1} - 2i\sqrt{2} \, \chi^{1,0} \wedge \star \, \b \partial_A Q^{0,0} + i\sqrt{2} \, \chi^{1,0}\wedge\partial_A Q^{0,2} \\
    & \qquad -{i\ov 4} \, \t \eta^{0,0} \, \t \phi \, Q^{0,0} \, \dvol  + i \, \t \eta^{2,0} \wedge \t \phi \, Q^{0,2} - i \, \t Q^{2,0} \wedge \t \Lambda^{0,2} \, Q^{0,0} \\
    & \qquad  + {i\ov 4} \, \t Q^{0,0} \, \t\Lambda^{0,0}_{(1)}\, Q^{0,0} \, \dvol  + i \, \t Q^{2,0}\wedge \t \Lambda^{0,0}_{(2)} \, Q^{0,2} + i \, \t Q^{0,0} \, \t \Lambda^{2,0}\wedge Q^{0,2}\Bigg)~,
\eea
with $\dvol = {1\ov4} e^1\wedge e^2 \wedge e^{\b 1}\wedge e^{\b 2}= \star 1$. We will discuss the corresponding kinetic terms in detail in section~\ref{subec:1loop 4d}.

%%%%%%%%%%%%%%%%%%%%%%%%%%%%%%%
\section{Five-dimensional DW twist on circle bundles}\label{sec:5dDW}
In this section, we uplift the topological twist of the previous section to a supersymmetric background for 5d $\CN=1$ supersymmetric field theories on any five-manifold $\CM_5$ which is a principal circle bundle over a K\"ahler surface,
\be
S^1\longrightarrow \CM_5 \overset{\pi}{\longrightarrow} \CM_4~.
\ee
Supersymmetric backgrounds of similar geometries were discussed by many authors -- see {\it e.g.} \cite{Kallen:2012cs, Hosomichi:2012ek, Kallen:2012va, Qiu:2013pta, Imamura:2014ima, Alday:2014bta, Alday:2015lta}. Our approach here is limited to a supersymmetric background that reduces to the (extended) topological twist on $\CM_4$.

\subsection{Circle-bundle geometries and the 5d Killing spinor equation}\label{subsec:M5 geom}

Let the five-manifold $\CM_5$ be a principal circle bundle over a K\"ahler four-manifold $\mathcal{M}_4$. This fibration is fully determined by the first Chern class:
\be
c_1(\mathcal{L}_{\rm KK}) =\frac{1}{2\pi}\FKK \in H^2(\mathcal{M}_4, \mathbb{Z})~ , 
\ee
where the `defining line bundle' $\CL_{\rm KK}$ is the complex line bundle associated to the $S^1$ bundle. We define the Chern numbers $\p_k$ by:
\be \label{ps for CL0}
 c_1(\mathcal{L}_{\rm KK})  = \sum_k \p_k [S_k]~, 
\end{equation}
with the 2-cycles $\S_k \subset \CM_4$ forming a basis of $H_2(\CM_4, \Z)$, and $[S_k]\in H^2(\CM_4, \Z)$ their Poincar\'e duals. We  have:
\be\label{Ikl def}
   {\bf I}_{kl}=  \int_{\S_k} [\S_l]= \S_k\cdot \S_l~,
\ee
 the intersection numbers on $\CM_4$. 
Given a K\"ahler metric \eqref{K metric M4} on $\CM_4$ with local coordinates $(z^1, z^2)$,  we choose the  five-dimensional metric
\be\label{dsM5}
    ds^2(\mathcal{M}_5) = ds^2(\mathcal{M}_4)+ \eta^2~, \qquad \eta\equiv \beta (d\psi+  \Cfib)~,
\ee
with the fiber coordinate $\psi$ subject to the identification $\psi \sim \psi + 2\pi$, and the connection $\Cfib$ on $\CM_4$ such that:
\be
    d\Cfib= \FKK= 2\pi\, c_1(\CL_{\rm KK})~.
\ee
In \eqref{dsM5} we also introduced $\beta$, the radius of the circle fiber. 
It has been shown in \cite{Alday:2015lta} that theories with $\mathcal{N}=1$ supersymmetry can be defined on five-manifolds that admit such a metric. The existence of curved-space supersymmetry is related to the existence of a transversely holomorphic foliation (THF) structure defined by the one-form $\eta$, similarly to the three-dimensional geometries studied in \cite{Closset:2012ru, Closset:2017zgf}.

By assumption, since we have a fibration structure, the metric \eqref{dsM5} admits a Killing vector $K$ with dual one-form given by $\eta$, $K^M \equiv \eta^M$, namely:
\be
K= {1\ov \beta}\d_\psi~.
\ee
 Note that we have $d \eta =\beta  \FKK = 2 \pi \beta  \sum_k \p_i [\S_k]$ and 
$\nabla_M \eta_N+ \nabla_N \eta_M=0$, 
which both follow from the relation:
\be
    \nabla_M \eta_N = \frac{\beta}{2} \FKK_{MN}~.
\ee
Here $\nabla_M$ is the 5d Levi-Civita connection.  
We would like to construct a supersymmetric background on $\CM_5$ which is the uplift of the four-dimensional DW twist. In particular, such a background should admit two five-dimensional Killing spinors $\zeta_{(i)}^I$, for $i=1,2$,  related to the four-dimensional Killing spinors \eqref{top sol Kahler} by 
\be\label{KS explicit 5d}
\zeta_{(i) I} = \mat{ 0\\  \t \xi_{(i)I}}~,
\ee
with $\t \xi^{\dot \alpha}_{(1) I} = \delta^{\dot \alpha \dot 1}\delta_{I1}$ and $\t \xi^{\dot \alpha}_{(2) I} = \delta^{\dot \alpha \dot 2}\delta_{I2}$.
Once we posit the Killing spinors \eqref{KS explicit 5d}, we must reconstruct the 5d Killing spinor equations that they satisfy. 
The trivial uplift of the 4d Killing spinor equation,
\be \label{FalseKillingSpinorEqn}
    \left(\nabla_{M}\delta\indices{_I^J} -i(A_{M}^{R})\indices{_I^J} \right) \zeta_J = 0~,
\ee
 only holds for the trivial circle fibration. This is related to the fact that the connection $\nabla_M$ does not preserve the decomposition of tensors into vertical and horizontal components with respect to the fibration, since 
\be
    \nabla_{M} \eta_{N} \neq 0~,
\ee
unless $\p_i=0$. To correct this, we  can simply introduce a new connection that preserves the fibration structure. The price to pay is that such a connection will have non-zero torsion.

%%%%%%%%%%%%%%%%%%%%%%%%%%%%%%%%%%%%%%
\subsubsection{Transversely holomorphic foliation and adapted connection} 

The five-dimensional manifolds that we are considering here are fibrations with a K\"ahler base -- in particular, they are transversely holomorphic foliations (THF) with an adapted metric. A one-dimensional foliation structure on the five-manifold $\CM_5$ is generated by a nowhere-vanishing vector field $K^M = g^{MN}\eta_N$. The foliation is transversely holomorphic if there exists a tangent bundle endomorphism $\Phi$ -- {\it i.e.} a two-tensor ${\Phi^M}_N$ -- whose restriction to the kernel of the one-form $\eta$ gives an integrable complex structure,
\be\label{prop Phi to J}
\Phi\Big|_{\ker(\eta)}= J~.
\ee
In particular, we have the relation
\be\label{Phi2 rel}
    \Phi\indices{^{M}_{N}}\Phi\indices{^{N}_{P}} = - \delta\indices{^{M}_{P}} + K^{M}\eta_{P}~.
\ee
As shown in \cite{Alday:2015lta}, the existence of one supercharge on a five-manifold%
\footnote{Together with a few other assumptions about the type of theories considered.} 
implies the existence of a THF. In the present case, we have two supercharges and the foliation is actually a fibration. We further restrict ourselves to the case when the holomorphic base manifold is K\"ahler, as required by the DW twist with two supercharges. 

 Focussing then on the class of fibered five-manifolds introduced above, with the adapted metric \eqref{dsM5}, it is convenient to introduce a modified connection $\hat{\nabla}$ that preserves the THF and fibration structure,
\be\label{Nabetaphi}
    \hat{\nabla}_{M}g_{NP} = 0~,\qquad \hat{\nabla}_{M}\eta_{N} = 0~, \qquad \hat{\nabla}_{M}\Phi\indices{^{N}_{P}}=0~.
\ee
This connection can be expressed in terms of the Levi-Civita connection as
\be
    \hat{\Gamma}^{P}_{MN} = \Gamma^{P}_{MN} + K^{P}_{MN}~,
\ee
where $K$ is the cotorsion tensor. In terms of the circle-bundle curvature $\FKK_{MN}$, this becomes:
\be
    K_{PMN} = \frac{\beta}{2}\left( \eta_{P}\FKK_{MN} - \eta_{N}\FKK_{MP} - \eta_{M}\FKK_{NP} \right)~.
\ee
The torsion tensor of the modified connection reads
\be
    T\indices{^{P}_{MN}} = K\indices{^{P}_{MN}} - K\indices{^{P}_{NM}} = \beta\eta^{P}\FKK_{MN}~.
\ee
From here on, we will denote by $\hat{D}_{M}$ the covariant derivative with respect to the modified connection (which is also $SU(2)_R$- and gauge-covariant, as the case may be). For instance, for a scalar field $\phi$ we have
\be
    \left[\hat{D}_{M}, \hat{D}_{N}\right] \phi = -T\indices{^{P}_{MN}}\hat{D}_{P}\phi~.
\ee
%%%%%%%%%%%%%%%%%%%%%%%%%%%%%%%%%%%%%%

\subsubsection{Killing spinor equation and spinor bilinears} 
Given the adapted connection $\hat{D}_M$ on $\CM_5$, we choose the  Killing spinor equation
\be\label{5d KSE final}
    \hat{D}_{M} \zeta_{I} \equiv \left(\hat{\nabla}_{M} \delta\indices{^I_J} - i({\bf A}^{(R)}_M)\indices{^I_J} \right) \zeta_J = 0~,
\ee
with ${\bf A}^{(R)}_M$ the $SU(2)_{R}$ background gauge field. One can easily check that the 5d Killing spinors \eqref{KS explicit 5d} are indeed solutions to \eqref{5d KSE final}, once we take ${\bf A}^{(R)}_M$ to be the pull-back of the corresponding DW-twist connection on $\CM_4$. 
 In fact, we only need to turn on a $U(1)_{R} \subset SU(2)_{R}$ background, as in the four-dimensional case. As a result, we can introduce the spinors:
\be
    \zeta_{(1)} = \zeta_{(1)\,I=1}~, \qquad \quad \zeta_{(2)} = \zeta_{(2)\,I=2}~,
\ee
 for which we have:
\be\label{KSE 5d}
\left(\hat \nabla_M- i A_M^{(R)} \right)\zeta_{(1)}=0~,\qquad \qquad
\left(\hat \nabla_M+ i A_M^{(R)} \right)\zeta_{(2)}=0~,
\ee
with the $U(1)_R$ gauge field
\be
 A^{(R)} =- i \left(\omega_{\mu 1 \b 1} +\omega_{\mu 2 \b 2} \right) dx^\mu\big|_{\CM_4}~.
\ee
%This 5d DW twist can be formulated more covariantly, as in 4d. 

\medskip
\noindent
{\bf THF from spinors.} Given   two distinct nowhere-vanishing solutions to \eqref{5d KSE final}, we can reconstruct THF tensors. We define the one-form
\bea
  \eta_{M} = - \frac{1}{ |\zeta_{(1)}|^2} \zeta^{\dagger\, I}_{(1)}\gamma_{M}\zeta_{(1) \, I} = - \frac{1}{ |\zeta_{(2)}|^2} \zeta^{\dagger \, I}_{(2)}\gamma_{M}\zeta_{(2)\, I}~,
\eea
with the Hermitian conjugate defined as for four-dimensional spinors \cite{Dumitrescu:2012ha, Dumitrescu:2012at}.  Additionally, similarly to the three-dimensional analysis of \cite{Closset:2012ru}, the quantity defined as:
\be
    K^{M} = \zeta_{(1)I}\gamma^{M}\zeta_{(2)}^I = - \zeta_{(1)}\gamma^{M}\zeta_{(2)}~,
\ee
is a non-vanishing Killing vector, whose orbits define a foliation of $\mathcal{M}_5$. Note that $K^{M} = \eta^{M}$ for our choice of metric. For future reference, let us also define the scalar
\be\label{def kappa 5d}
\kappa\equiv  \zeta_{(1)}^I\zeta_{(2) \, I}~.
\ee
Note that, when plugging in \eqref{KS explicit 5d}, we have $K^M = \delta^{M5}$ and $\kappa=1$. 
Finally, we define a two-tensor
\be
   {\Phi^{MN}} = \frac{i\zeta^{\dagger}_{(1)}{\gamma^{MN}}\zeta_{(1)}}{|\zeta_{(1)}|^2}~, 
\ee
with $\gamma^{MN}\equiv \half [\gamma^M, \gamma^N]$, 
which satisfies \eqref{prop Phi to J} and \eqref{Phi2 rel}.   The Killing spinor equation \eqref{5d KSE final} also implies \eqref{Nabetaphi}.

\medskip
\noindent
{\bf The 5d DW twist.}  The five-dimensional uplift of the DW twist on $\CM_4$ can be formulated a little bit more covariantly. 
 To do this, it is useful to consider the two-forms
\bea\label{2forms5dDW}
&\CP_{(1)}\equiv  \zeta_{(1)} \Sigma_{MN}  \zeta_{(1)} \, dx^M \wedge dx^N = i e^1 \wedge e^2~, \cr
&\CP_{(2)}\equiv  \zeta_{(2)} \Sigma_{MN}  \zeta_{(2)} \, dx^M \wedge dx^N = i e^{\b1} \wedge e^{\b 2}~, \cr
\eea
where we use the complex frame \eqref{5d frame}. 
Let us define the canonical line bundle $\CK_{\CM_5}$ on $\CM_5$ as the pull-back of the canonical line bundle $\CK_{\CM_4}$ on the K\"ahler manifold $\CM_4$, using the fibration structure $\pi : \CM_5 \rightarrow \CM_4$,  namely:
\be\label{def K in 5d}
\CK_{\CM_5} = \pi^\ast \CK_{\CM_4}~.
\ee 
Since the spinors $\zeta_{(1)}$ and   $\zeta_{(2)}$ have $U(1)_R$ charges $+1$ and $-1$, respectively, the two forms \eqref{2forms5dDW} are nowhere-vanishing sections 
\be
\CP_{(1)} \in \Gamma[\CK \otimes L_R^2]~, \qquad \qquad
\CP_{(2)} \in \Gamma[\b\CK \otimes L_R^{-2}]~,
\ee 
with $L_R$ the $U(1)_R$ bundle on $\CM_5$. 
We, therefore, have the 5d uplift of the DW twist,
 \be
L_R \cong \CK^{-\half}~,
\ee
literally as in \eqref{LR top twist} but now written in terms of line bundles on $\CM_5$. 
As before, $\CK^\half$ will not be well-defined unless $\CM_4$ is spin, but the Killing spinors are  well-defined sections of appropriate spin$^c$ bundles nonetheless. 

%%%
\subsubsection{$(p,q)$-forms and twisted Dolbeault operators on $\CM_5$}\label{subsec:twisted Dolbeault 5d}
The 5d uplift of the DW twist remains independent of the choice of K\"ahler metric on $\CM_4$. To make this property manifest, we express all fields in terms of differential forms, exactly like in 4d. On $\CM_5$, differential forms can be further decomposed into horizontal and vertical forms ({\it i.e.} along the base $\CM_4$ and the circle fiber, respectively). 
This can be done explicitly by using the projectors:
\bea
{\Pi^M}_N = & \half\left({\delta^M}_N- i {\Phi^M}_N- K^M \eta_N\right)~,\\
{{\b\Pi}^M}_{\phantom{M}N} = &  \half\left({\delta^M}_N+i {\Phi^M}_N- K^M \eta_N\right)~,\\
{\Theta^M}_N = & K^M \eta_N~.
\eea
Any $k$-form on $\CM_5$ decomposes into $(p,q|\ell)$-forms, with $p+q+\ell=k$. Here, $\ell$ denotes the form degree along the fiber. By abuse of notation, a five-dimensional $(p,q|0)$-form is called $(p,q)$-form, denoted by $\omega^{p,q}$. Any $(p,q|1)$-form can be written as
\be
\omega^{(p,q|1)}= \omega^{p,q} \wedge \eta~.
\ee
For instance, for any one form $\omega=\omega_M dx^M$, we have
\be\label{1form 5d dec}
 \omega= \omega^{1,0}+ \omega^{0,1}+ \omega_5 \eta=  \omega^{1,0}_i dz^i+ \omega^{0,1}_{\b i}d\b z^{\b i}+ \omega_5 \eta~,
\ee
where:
\be
  \omega_M {\Pi^M}_N =  \omega^{1,0}_N~, \qquad
   \omega_M {{\b\Pi}^M}_{\phantom{M}N} =   \omega^{0,1}_N~,\qquad 
    \omega_5 \equiv K^M \omega_M~.
\ee
In particular, the vertical component is defined by contracting with $K$.
For future reference, we also note that any $2$-form $F$ decomposes as:
\be\label{F 2form gen decomp}
F= F^{2,0}+ F^{0,2}+ F^{1,1}+ F^{1,0}\wedge \eta+F^{0,1}\wedge \eta~.
\ee
In particular, $(2,0)$-forms are sections of the 5d canonical line bundle \eqref{def K in 5d}. 

\medskip
\noindent
{\bf Dolbeault operators on $\CM_5$.} The differential  $d: \Omega^k \rightarrow \Omega^{k+1}$ on $\CM_5$ decomposes as
\be
d= \d + \b \d + \h \d_5~,
\ee
where $\d$ and $\b\d$ denote the twisted Dolbeault operators: 
\be\label{def d twisted 5d}
\d : \Omega^{(p,q|\ell)} \rightarrow \Omega^{(p+1,q|\ell)}~, \qquad \qquad
\b\d : \Omega^{(p,q|\ell)} \rightarrow \Omega^{(p,q+1|\ell)}~,
\ee
and $\h \d_5:  \Omega^{(p,q|\ell)} \rightarrow \Omega^{(p,q|\ell+1)}$ is given by:%
\footnote{Note that the operator $\d_5$ does not change the form degree. We denote this way the Lie derivative along $K$, which is equal to $K^M\d_M$ on forms because $\iota_K \omega=0$ for any horizontal form, and moreover $\CL_K \eta=0$ because $K^M= \eta^M$ is a Killing vector.}
\be
\h \d_5 \equiv \eta \wedge \d_5~, \qquad\qquad \d_5 \equiv  \CL_K = K^M \d_M~.
\ee
In terms of the local coordinates $(x^M)=(z^i, \b z^{\b i}, \psi)$, the twisted Dolbeault operators are given explicitly by:
\be
\d= dz^i \wedge (\d_i - \Cfib_i \d_\psi)~, \qquad \qquad
\b\d= d\b z^{\b i} \wedge (\d_{\b i} - \Cfib_{\b i} \d_\psi)~, 
\ee
where $\Cfib_i$ and $\Cfib_{\b i}$ and the holomorphic and anti-holomorphic component of the connection $\Cfib$ introduced in \eqref{dsM5}. Whenever the fibration is non-trivial, the twisted Dolbeault operators are not nilpotent. Instead, they satisfy the relations:
\be
\d^2= - \beta \FKK^{2,0}\wedge \d_5~, \qquad \quad
\b\d^2= - \beta \FKK^{0,2}\wedge \d_5~,\qquad\quad
\{\d, \b\d\}= - \beta \FKK^{1,1}\wedge \d_5~.
\ee
Of course, they reduce to the ordinary Dolbeault operators on $\CM_4$ upon dimensional reduction along the fiber. We also have that
\be
\{\d+\b\d, \h \d_5\} = \beta \FKK\wedge \d_5~, \qquad\qquad  \h \d_5^2=0~,
\ee
where $\beta\FKK=d\eta$. Note that $\FKK$ is a horizontal 2-form on $\CM_5$.

\subsubsection{Background fluxes on $\CM_5$}\label{subsub:introFlux}
Let us consider supersymmetry-preserving background fluxes for gauge fields on $\CM_5$. Equivalently, we  consider line bundles $L_{\CM_5}$ with first Chern class
\be
c_1(L_{\CM_5}) \in H^2(\CM_5, \Z)~.
\ee
The supersymmetry-preserving line bundles are pull-back of holomorphic line bundles on $\CM_4$:
\be
L_{\CM_5} = \pi^\ast L_{\CM_4}~.
\ee 
 Given our assumption that $\CM_4$ is simply connected, the Gysin sequence implies the following simple relation between the second cohomologies of $\CM_4$ and $\CM_5$:
\be\label{relH2M4M5}
H^2(\CM_5, \Z) = {\rm coker} \Big(c_1(\CL_{\rm KK}) \, : \, H^0(\CM_4, \Z)\rightarrow  H^2(\CM_4, \Z)\Big)~.
\ee
Let us introduce the notation $\m$ for the abelian flux on $\CM_4$:
\be
c_1(L_{\CM_4})= \sum_k \m_k [S_k]~,
\ee
as in \eqref{ps for CL0}.%
\footnote{Assuming that $\CM_4$ is simply connected, all 2-cycles in $\CM_5$ are inherited from 2-cycles in $\CM_4$. More generally, the same would remain true of supersymmetry-preserving fluxes.} 
 The relation  \eqref{relH2M4M5} means that we can write any five-dimensional flux, denoted by $\m_{\rm 5d}\in H^2(\CM_5,\Z)$, as:
 \be
\m_{\rm 5d} = \m \mod \p~.
\ee
  One important example is the lens space $S^5/\Z_p$ obtained as a degree-$p$ fibration over $\mathbb{P}^2$ (hence $\p=p$), in which case we have $\m_{\rm 5d} \in \Z_p$, with $p=1$ corresponding to the five-sphere. 
  
  For future reference, let us introduce the intersection pairing.  Given two line bundles $L$ and $L'$ on $\CM_4$ with fluxes $\m$ and $\m'$, respectively, we define:
\be\label{defInterPairing}
(\m, \m')= \int_{\CM_4} c_1(L)\wedge c_1(L')= \sum_{k, l} \m_k {\bf I}_{kl} \m_l ~,
\ee
with ${\bf I}_{kl}$ defined in \eqref{Ikl def}.
%%%

\subsection{The 5d $\CN=1$ vector multiplet on $\CM_5$}
Let us now consider the simplest supersymmetry multiplets on $\CM_5$. (See appendix \ref{app:5dconventions} for a review of the standard flat-space results.)
The 5d vector multiplet contains a gauge field $A_M$, a real scalar $\sigma$, an $SU(2)_R$ doublet of gauginos, $\Lambda_I$, transforming as a Majorana-Weyl spinor, and an $SU(2)_R$ triplet of auxiliary scalars ${ D}_{IJ}$, with the flat-space supersymmetry transformations reviewed in appendix~\ref{app:5dsusy}.   
On our curved-space background $\CM_5$, the supersymmetry transformations read:
\bea\label{vec5d susy curved}
& \delta A_M &&=\;  i \zeta_I \gamma_M \Lambda^I~, \cr
& \delta \sigma&&=\; - \zeta_I \Lambda^I~, \cr
& \delta \Lambda_I &&=\;- i \Sigma^{MN} \zeta_I\, \left( F_{MN} - i \beta \sigma  \FKK_{MN}\right) + i \gamma^M \zeta_I \hat D_M \sigma - i { D}_{IJ} \zeta^J~, \cr
& \delta { D}_{IJ} &&=\;  \zeta_I \gamma^M \hat D_M \Lambda_J+  \zeta_J \gamma^M \hat D_M \Lambda_I + \zeta_I [\sigma, \Lambda_J] +\zeta_J [\sigma, \Lambda_I]~.
\eea
Note that the difference from the flat-space algebra arises due to the expression for the field strength, which, when written in terms of the new covariant derivative, reads:
\be\label{FMN full}
    F_{MN} = \hat{\nabla}_{M} A_N - \hat{\nabla}_N A_M - i[A_M, A_N] + \beta \eta^P \FKK_{MN} A_P~.
\ee
The curved-space supersymmetry algebra on $\CM_5$ reads:
\be\label{5d susy alg}
\delta_1^2=0~, \qquad \delta_2^2=0~,\qquad
\{\delta_1, \delta_2\}= -2 i \CL_K^{(A)} + 2 \kappa \delta_{g(\sigma)}~,
\ee
where $\CL_K^{(A)}$ is the gauge-covariant Lie derivative along $K^M$, $\kappa$ is defined in \eqref{def kappa 5d}, and $\delta_{g(\sigma)}$ denotes a gauge transformation with parameter $\sigma$. 
The supersymmetry algebra \eqref{5d susy alg}  reproduces \eqref{4d vec susy closure} upon dimensional reduction along the fiber direction. 
It is equivalent to the flat-space algebra \eqref{susy transformations vect 5d flat} with the derivatives $D_M$ replaced with $\hat D_M$.

Upon topological twisting, the various fields become differential forms on $\cM_5$, exactly like in 4d. We decompose them into $(p,q|\ell)$-forms, following the notation introduced in section~\ref{subsec:twisted Dolbeault 5d}.  In particular, the decomposition \eqref{1form 5d dec} holds for the 5d connection $A\equiv A_M dx^M$, and the field-strength 2-form \eqref{FMN full} decomposes as in  \eqref{F 2form gen decomp}.
 We then write  the supersymmetry variations in terms of the twisted Dolbeault operator \eqref{def d twisted 5d}, which need to be further twisted by the gauge fields:
\be
\d_A \equiv \d- i A^{1,0}~, \qquad\qquad
\b\d_A  \equiv\b \d- i A^{0,1}~,\qquad\qquad
\d_{5,A}  \equiv \d_5- i A_5~,
\ee
to preserve gauge covariance. 
Note that they satisfy:
\bea\label{rel Dolb square 5d}
\d_A^2=& - i F^{2,0}\wedge - \beta \FKK^{2,0}\wedge \d_5 ~, \\
\b\d_A^2=& - i F^{0,2}\wedge - \beta \FKK^{0,2}\wedge \d_5~, \\
\{\d_A, \b\d_A\}=& - i F^{1,1}\wedge - \beta \FKK^{1,1}\wedge \d_5~,
\eea
generalising \eqref{rel Dolb square 4d}.
We then have the vector-multiplet supersymmetry variations:
\bea\label{5d susy vec twisted}
& \delta_1 \sigma = \t\Lambda_{(1)}^{0,0}~, && \delta_2 \sigma  = \t\Lambda_{(2)}^{0,0}~, \cr
& \delta_1 A   =- i \Lambda^{1,0}+i  \t\Lambda_{(1)}^{0,0}\eta~, \qquad && \delta_2  A   = - i \Lambda^{0,1}+i \t\Lambda_{(2)}^{0,0}\eta~, \cr
& \delta_1 \Lambda^{1,0}   =0~, \qquad && \delta_2 \Lambda^{1,0}    = 2 i \d_A \sigma - 2 F^{1,0}~, \cr
& \delta_1  \Lambda^{0,1}  =2i \b \d_A \sigma - 2F^{0,1}~, \qquad && \delta_2  \Lambda^{0,1}   = 0~, \cr
& \delta_1  \t \Lambda_{(1)}^{0,0}  =0~, \qquad && \delta_2 \t \Lambda_{(1)}^{0,0}    = i \h\CD^{0,0} - i\d_{5,A} \sigma~, \cr
& \delta_1   \t \Lambda_{(2)}^{0,0} =- i \h\CD^{0,0} - i \d_{5,A} \sigma~, \qquad && \delta_2  \t \Lambda_{(2)}^{0,0}   = 0~, \cr
& \delta_1  \t\Lambda^{2,0}  =\CD^{2,0}~, \qquad\qquad && \delta_2     \t\Lambda^{2,0}  = 4 (F^{2,0}-i\beta \sigma \FKK^{2,0})~, \cr
& \delta_1    \t\Lambda^{0,2}  =4 (F^{0,2}-i\beta \sigma \FKK^{0,2})~,\qquad \qquad && \delta_2     \t\Lambda^{0,2}  = \CD^{0,2}~, \cr
& \delta_1  \h \CD^{0,0} =  [\sigma, \t\Lambda_{(1)}^{0,0}]- \d_{5, A}  \t\Lambda_{(1)}^{0,0}~, \qquad && \delta_2  \h \CD^{0,0}   =-   [\sigma, \t\Lambda_{(2)}^{0,0}]+ \d_{5, A}\t\Lambda_{(2)}^{0,0}~, \cr
\eea
and:
\bea    \label{5d susy vec twisted - part 2}
& \delta_1 \CD^{2,0}   =0~, \qquad\qquad \qquad\qquad  \delta_2   \CD^{2,0}  = 4 i\d_A \Lambda^{1,0}+ 2i[\sigma,\t\Lambda^{2,0}] - 2 i \d_{5,A} \t\Lambda^{2,0} ~,\cr
& \delta_1   \CD^{0,2}  =4 i \b\d_A \Lambda^{0,1} + 2i [\sigma, \t\Lambda^{0,2}]- 2 i \d_{5,A}\t\Lambda^{0,2}~,\qquad \qquad\qquad \qquad \delta_2   \CD^{0,2}   = 0~.
\eea
The scalar appearing in $\h\CD^{0,0}$ in \eqref{5d susy vec twisted} is related to the 4d scalar $\CD^{0,0}$ defined in \eqref{4d CD0} by:
\be
    \h \CD^{0,0} \equiv  \CD^{0,0} + 2\beta \sigma \FKK^{0,0}~,
\ee
where $\FKK^{0,0}$ is defined as $\FKK^{0,0} \equiv {1\ov 4}\Phi^{MN} \FKK_{MN}$. 
To check the closure of  \eqref{5d susy vec twisted}-\eqref{5d susy vec twisted - part 2}, it is useful to first compute the supersymmetry variations of the field strength:
\bea
    &  \delta_1   F^{2,0}  = -i\d_A \Lambda^{1,0} + i\beta \FKK^{2,0} \t\Lambda^{0,0}_{(1)}~,  \;  \qquad  && \delta_2 F^{2,0}   =i\beta\FKK^{2,0}\t\Lambda^{0,0}_{(2)}~,  \\
    &  \delta_1   F^{0,2}  = i\beta \FKK^{0,2} \t\Lambda^{0,0}_{(1)}~,  \;  \qquad  && \delta_2 F^{0,2}   =-i\b \d_A \Lambda^{0,1} + i\beta\FKK^{0,2}\t\Lambda^{0,0}_{(2)}~,  \\
    &  \delta_1   F^{1,0}  = i\d_A\Lambda^{0,0}_{(1)} + i \d_{5,A}\Lambda^{1,0}~,  \;  \qquad  && \delta_2 F^{1,0}   = i\d_A\Lambda^{0,0}_{(2)}~,  \\
    &  \delta_1   F^{0,1}  = i\b\d_A\Lambda^{0,0}_{(1)}~,  \;  \qquad  && \delta_2 F^{0,1}  =  i\b\d_A \Lambda^{0,0}_{(2)} + i \d_{5,A}\Lambda^{0,1}~. 
\eea
It is then straightforward to check that $\delta_1^2 = 0 = \delta_2^2$, while  $\{\delta_1, \delta_2 \} f = 2i[\sigma,f] - 2i\d_{5,A}f$ for any of the fields $f$ in the vector multiplet.%
\footnote{This also holds for the field-strength $F$ upon using the Bianchi identity.}
 The twisted vector multiplet therefore realises the supersymmetry algebra \eqref{5d susy alg}, namely:
\be\label{5d SUSY algebra twist}
\delta_1^2 = 0~, \qquad
\delta_2^2 = 0~, \qquad
    \{\delta_1, \delta_2 \}  = -2i \CL_{K}+ 2\delta_{g(\sigma + i\, \iota_K A)}~,
\ee
where $\CL_{K}$ is the usual Lie derivative along $K$, $\iota_K A = K^M A_M = A_5$ is the contraction with the vector field $K$, while $\delta_{g(\epsilon)}$ is the gauge transformation with parameter $\epsilon$ introduced in \eqref{4d vec susy closure}. 
Finally, one can check that the 5d $\CN=1$ SYM Lagrangian on $\CM_5$ is `almost' $\CQ$-exact, similarly to the 4d Lagrangian \eqref{4d SYM lagrangian}.

%%%%%%%%%%%%%%%%%%%%%%%%%%5

\subsection{The 5d $\CN=1$ hypermultiplet on $\CM_5$}

The 5d $\CN=1$ hypermultiplet consists of an $SU(2)_R$ doublet of complex scalar fields $q^I$ and of a Dirac spinor $\Psi, \t \Psi$. (See appendix~\ref{app:5dconventions} for our 5d conventions.) As in the case of the 4d $\CN = 2$ hypermultiplet, we can realise the two supercharges of the DW twist off-shell by introducing some appropriate auxiliary fields. This is explained in appendix~\ref{app:5d hyper} in the case of flat-space supersymmetry. On our  curved space background, one simply replaces the derivatives $D_M$ in \eqref{susy h off shell R5 -1}-\eqref{susy h off shell R5 -2} with the torsionfull adapted connection $\hat D_M$.

The hypermultiplet can be recast in twisted variables, exactly as in 4d.  The field content is formally the same as in \eqref{CH twisted content}, with the $(p,q)$-forms being now interpreted as forms on $\CM_5$, following the discussion of section~\ref{subsec:twisted Dolbeault 5d}. The supersymmetry  variations read:
 \bea
& \delta_1 Q^{0,0} =0~, && \delta_2  Q^{0,0} = \sqrt2 \, \t\chi^{\,0,0}~, \cr
& \delta_1   Q^{0,2}=\sqrt2 \, \t\chi^{\,0,2}~, && \delta_2   Q^{0,2}= 0~, \cr
& \delta_1  \eta^{0,1}=  2 i \sqrt2\, \b\d_A Q^{0,0}+ h^{0,1}~,\quad\qquad && \delta_2   \eta^{0,1}= i \sqrt2 \star \left(\d_A Q^{0,2}\right)~, \cr
& \delta_1  \t\chi^{\,0,0}=i\sqrt{2} (\sigma-\d_{5,A}) Q^{0,0}~, && \delta_2   \t\chi^{\,0,0}= 0~, \cr
& \delta_1  \t\chi^{\,0,2}=0~, && \delta_2   \t\chi^{\,0,2}=  i\sqrt{2} (\sigma-\d_{5,A}) Q^{0,2}~, \cr
& \delta_1  h^{0,1}=0~, && \delta_2   h^{0,1}= X^{0,1}~, \cr
\eea
and:
\bea
& \delta_1 \t Q^{0,0} =-\sqrt2 \, \t\eta^{\,0,0}~, && \delta_2  \t Q^{0,0} = 0~, \cr
& \delta_1   \t Q^{2,0}=0~, && \delta_2  \t  Q^{2,0}= \sqrt2 \, \t\eta^{\,2,0}~, \cr
& \delta_1  \chi^{1,0}=   i \sqrt2 \star \left(\b \d_A \t Q^{2,0}\right) ~,\quad && \delta_2   \chi^{1,0}=2 i \sqrt2 \,\d_A \t Q^{0,0}+ \t h^{1,0}~, \cr
& \delta_1  \t\eta^{\,0,0}=0~, && \delta_2   \t\eta^{\,0,0}= i\sqrt{2} \left(\t Q^{0,0} \sigma + \d_{5,A} \t Q^{0,0}\right)~, \cr
& \delta_1  \t\eta^{\,2,0}=-i \sqrt{2}\left( \t Q^{2,0} \sigma + \d_{5,A}\t Q^{2,0}\right)~,\qquad && \delta_2   \t\eta^{\,2,0}= 0~, \cr
& \delta_1\t  h^{1,0}= \t X^{1,0}~, && \delta_2  \t h^{1,0}=0~, \cr
\eea
where we defined:
\bea
& X^{0,1} & \equiv&\; \; - 4 i\, \b\d_A \t \chi^{\,0,0}-2 i \star(\d_A \t \chi^{\,0,2})+ 2 i \sqrt2 \, \Lambda^{0,1} Q^{0,0} 
+  i \sqrt2 \star(\Lambda^{1,0} \wedge Q^{0,2})\\
&&&\; \;+ 2i (\sigma-\d_{5,A}) \eta^{0,1}~, \cr
& \t X^{1,0} & \equiv &\; \; 4 i \, \d_A \t \eta^{\,0,0}-2 i \star(\b\d_A \t \eta^{\,2,0})- 2 i \sqrt2 \, \t Q^{0,0} \Lambda^{1,0}
-  i \sqrt2 \star(\t Q^{2,0} \wedge \Lambda^{0,1} )\\
&&&\; \;- 2i \left(\chi^{1,0} \sigma + \d_{5,A} \chi^{1,0}\right) ~.  
\eea
This naturally generalises \eqref{4d SUSY hyper part 1}-\eqref{4d SUSY hyper part 2}. 
Note that the four-dimensional scalar  is replaced by a differential operator:
\be
\phi  \rightarrow {1\ov \sqrt2}(\sigma \mp \d_{5,A} )~,
\ee
when acting on a field in the representation $\FR$ or $\b\FR$, respectively.
 Let us also point out that the Hodge star operator used above is in fact the Hodge dual on $\cM_4$, obtained from the 5d Hodge dual by the contraction
    $\star \equiv \iota_K \star_{5}$. 
One can easily check that the supersymmetry algebra \eqref{5d SUSY algebra twist} is satisfied. 
The kinetic Lagrangian is again $\CQ$-exact. 
The five-dimensional uplift of \eqref{4d hypermultiplet lagrangian}  reads:
\bea   
    \SL_H & = {1\ov 4}\star (\delta_1+\delta_2)\Bigg( \t h^{1,0}\wedge\star \, \eta^{0,1} - 2i\sqrt{2} \, \chi^{1,0} \wedge \star \, \b \d_A Q^{0,0} + i\sqrt{2} \, \chi^{1,0}\wedge\d_A Q^{0,2} \\
    &  -{i\sqrt2 \ov 8} \, \t \eta^{\,0,0} (\sigma + \d_{5,A}) \, Q^{0,0} \,\dvol  + {i\sqrt2 \ov 2} \, \t \eta^{\,2,0} \wedge (\sigma+\d_{5,A}) \, Q^{0,2} - i \, \t Q^{2,0} \wedge \t \Lambda^{0,2} \, Q^{0,0} \\
    &   + {i\ov 4} \, \t Q^{0,0} \, \t\Lambda^{0,0}_{(1)}\, Q^{0,0} \, \dvol  + i \, \t Q^{2,0}\wedge \t \Lambda^{0,0}_{(2)} \, Q^{0,2} + i \, \t Q^{0,0} \, \t \Lambda^{2,0}\wedge Q^{0,2}\Bigg)~.
\eea

%%%%%%%%%%%%%%%%%%%%%%%%%%%%%%%%%%%%%%%%%%%%%%

\section{One-loop determinants on $\cM_5$}\label{sec:oneloopdet}
In this section, we compute one-loop determinants on $\CM_5$. We start by considering the contribution of a free hypermultiplet in 4d, before obtaining the 5d result by  summing over the KK modes. We then generalise this result to derive the one-loop contribution of any 5d BPS particle running along the circle fiber.  

%%%%%%%%%%%
\subsection{Free hypermultiplet on $\cM_4$}\label{subec:1loop 4d}
Consider the 4d $\CN=2$ hypermultiplet on $\CM_4$ with the extended topological twist discussed in section~\ref{subec: 4d hyper}.   We couple the hypermultiplet to a background vector multiplet, which should preserve our two supercharges on $\CM_4$. Hence, we impose:
\bea \label{SUSYbackground1}
    \CD^{2,0} = 0~, \qquad \CD^{0,2} = 0, \qquad F^{0,2} = 0~, \qquad F^{2,0} = 0~,
\eea
and:
\be \label{SUSYbackground2}
    \partial_A \phi = 0~, \qquad \b \partial_A \phi = 0~, \qquad \CD^{0,0} =0~, \qquad [ \t \phi, \phi] = 0~.
\ee
These conditions are obtained by imposing that the gaugino variations in  \eqref{susy vec twisted} vanish. Note that, in particular, the background gauge field must be the connection of a holomorphic vector bundle. 
Given the hypermultiplet Lagrangian \eqref{4d hypermultiplet lagrangian}, let us extract the quadratic fermionic and bosonic fluctuations around this supersymmetric background. They can be expressed
in the following compact form:
\be \label{Quadratic fluctuations compact}
    \SL_{\rm bos} = \star \left( \t Q \wedge \star\Delta_{\rm bos} Q\right)~, \qquad \quad \SL_{\rm fer} = \star \left( \t \Psi \wedge \star \Delta_{\rm fer} \Psi\right)~.
\ee
Here the bosonic fields are written as $\t Q = (\t Q^{0,0}~, \t Q^{2,0})$ and $Q = (Q^{0,0}~, Q^{0,2})^T$, while the fermionic fields are given by $\t \Psi = (\chi^{1,0}~, \t \eta^{\,0,0}~, \t \eta^{\,2,0})$ and $\Psi = (\eta^{0,1}~, \t \chi^{\,0,0}~, \t \chi^{\,0,2})^T$. The bosonic kinetic operator reads:
\be \label{bosonic kinetic operator}
    \Delta_{\rm bos} = \left( \begin{matrix}
     -2\star \partial_A \star \b \partial_A + 2\phi \t\phi & 0~ \\ 0 & -{1\ov 2}\star \b\partial_A \star  \partial_A + {1\ov 2}\phi \t\phi ~
    \end{matrix} \right)~,
\ee
and the fermionic operator takes the form
\bea    \label{fermionic kinetic operator}
    \Delta_{\rm fer} = \left( \begin{matrix}
     -{i\sqrt{2}\ov 2}\,\phi \quad& i\,\b\partial_A \quad& ~{i\ov 2}\star \partial_A~ \\
     ~-i\star \partial_A\star~ \quad& i\sqrt{2}\,\t\phi \quad& ~0~ \\
     -{i\ov 2}\star \b \partial_A \quad& 0 \quad& ~-{i\sqrt{2}\ov 4}\,\t\phi~
    \end{matrix}\right)~.
\eea
(Note that $\star^2 = \pm 1$, depending on whether the form is of even or odd degree.) 
The corresponding Gaussian integration will give the exact answer for the hypermultiplet in this supersymmetric background, by a  standard scaling argument,  because the kinetic Lagrangian \eqref{4d hypermultiplet lagrangian} is $\CQ$-exact.  The  partition function of the 4d hypermultiplet on $\CM_4$ is then given by:
\be\label{ZM4 detFdetB}
Z_{\CM_4}^\CH = {\det   \Delta_{\rm fer} \ov \det   \Delta_{\rm bos} }~.
\ee
Supersymmetry leads to huge cancellations between fermionic and bosonic fluctuations, so that \eqref{ZM4 detFdetB} can be  further simplified. With the DW twist on $\CM_4$, the boson-fermion degeneracy is almost perfect -- we discuss the pairing more explicitly in appendix~\ref{app:hyperModePairing}. 
 Here, we follow a standard argument (see {\it e.g.} \cite{Pestun:2007rz, Closset:2013sxa}) which gives us the final answer almost immediately. First, we note that the bosonic and fermionic kinetic operators are related by:
\bea\label{detferbos rel M4}
    \Delta_{\rm fer}
    \left( \begin{matrix} 1 \quad& -2i \b \partial_A \quad& i\star \partial_A  \\
    0 \quad& -i\sqrt{2} \phi \quad& 0 \\
    0 \quad& 0 \quad& i\sqrt{2}\phi \end{matrix} \right) =  \left( \begin{matrix} -{i\sqrt{2} \ov 2}\phi \quad & 0 \quad 0~ \\
   \begin{matrix} -i \star \partial_A \star  \\
    {i\ov 2} \star \b \partial_A \end{matrix} & \Delta_{\rm bos} \end{matrix}\right)~,
\eea
which is a consequence of supersymmetry. It will be convenient to introduce the quantity
\be
\Lop = -i\sqrt{2}\phi~,
\ee
viewed as an operator on our bosonic and fermionic $\FR$-valued forms. 
 Taking the determinant of both sides of \eqref{detferbos rel M4}, we obtain:
\be \label{1-loop det hyper reduction M4}
Z_{\CM_4}^\CH = {\det   \Delta_{\rm fer} \ov \det   \Delta_{\rm bos} } = { \det(\Lop^{0,1}) \ov \det(\Lop^{0,0}) \det(\Lop^{0,2})}  ~,
\ee
with the superscript indicating the type of $(p,q)$-forms that $\Lop: \Omega^{p,q} \rightarrow \Omega^{p,q}$ acts upon.  Now, recall that the Dolbeault operators $\b \partial:\Omega^{p,q-1} \rightarrow \Omega^{p,q}$ and $\partial:\Omega^{p-1,q} \rightarrow \Omega^{p,q}$ have adjoints $\b\partial^*:\Omega^{p,q} \rightarrow \Omega^{p,q-1}$ and $\partial^*: \Omega^{p,q} \rightarrow \Omega^{p-1,q}$, respectively, defined as:
%\footnote{Up to some signs, which are not important for this discussion.}
\be
    \b\partial^* =- \star\partial\star~, \qquad \quad \partial^* = - \star \b\partial \star~,
\ee
and similarly for the gauge-covariant generalisation. 
Then, one can check that: 
\bea
    \ker(\b\partial_A) = \ker(\star \partial_A \star \b\partial_A)~, \qquad \quad \ker(\b\partial_A^*) = \ker( \b\partial_A\star \partial_A\star )~,\\
    \ker(\partial_A) = \ker(\star \b\partial_A \star \partial_A)~, \qquad \quad \ker(\partial_A^*) = \ker( \partial_A\star \b\partial_A\star )~.
\eea
As a result, the non-zero eigenvalues of $\b \partial_A^* \b\partial_A$ and those of $\b\partial_A\b\partial_A^*$ are in one-to-one correspondence, and similarly for the operators in the second line. Note that these operators do not change the degree of the differential form they act upon. Moreover, they clearly commute with the operators $\Lop$ introduced above and, as a result, the eigenvalues of $\Lop$ that lie outside these kernels will cancel in the one-loop determinant. Thus, we have:
\be\label{OneLoopDetFullOps}
  Z_{\CM_4}^\CH  = {\det_{\ker(\b\partial_A^*)\oplus \ker(\b\partial_A)}(\Lop^{0,1}) \ov \det_{\ker(\b\partial_A)}(\Lop^{0,0}) \det_{\ker(\b\partial^\ast_A)}(\Lop^{0,2})} ~.
\ee
That is, we are restricting attention to the zero modes:
\bea    \label{hyper zero modes M4}
    \b\partial_A Q^{0,0} = 0~, \qquad \quad \b\partial_A^\ast (\star Q^{0,2}) = 0~, \qquad\quad
    \b\partial_A^* \eta^{0,1} = 0~, \qquad \quad \b\partial_A \eta^{0,1} = 0~,
\eea
using the fact that $\star : \Omega^{0,2}\rightarrow \Omega^{0,2}$. 
Up to an irrelevant numerical factor, we find that:
\be  
     Z_{\CM_4}^\CH   = \left(-i\sqrt{2}\phi\right)^{-\CI}~,
\ee
with $\CI$ the net number of zero-modes \eqref{hyper zero modes M4} contributing. It is given by:
\bea\label{CI for det}
&\CI=&& \dim \ker(\b\d_A :E^{0,0}\rightarrow E^{0,1})+\dim \ker(\b\d_A^\ast :E^{0,2}\rightarrow E^{0,1})\\
&&& - \dim \ker(\b\d_A^\ast :E^{0,1}\rightarrow E^{0,0})-\dim \ker(\b\d_A :E^{0,1}\rightarrow E^{0,2})~,
\eea
where we denoted by $E^{0,q}\equiv \Omega^{0,q}\otimes E$ the space of $(0,q)$-forms valued in the gauge bundle $E$ with connection $A$. Let ${\rm ind}(\b\partial_A)$ denote the index of the Dolbeault complex twisted by $E$:
\be
0 \longrightarrow  \Omega^{0,0}\otimes E\overset{\b\d_A}{\longrightarrow}  \Omega^{0,1}\otimes E \overset{\b\d_A}{\longrightarrow} \Omega^{0,2}\otimes E \longrightarrow 0~.
\ee
Formally, one finds:
\be
\CI= {\rm ind}(\b\partial_A)- \dim( \Omega^{0,1}\otimes E)~.
\ee
By the assumption that   $\CM_4$ is simply connected,  however, we have $\dim( \Omega^{0,1}\otimes E)=0$ and thus $\CI= {\rm ind}(\b\partial_A)$. In summary, the one-loop determinant of the hypermultiplet on a simply-connected K\"ahler manifold reads:
\be \label{1-loop hyper M4}
     Z_{\CM_4}^\CH   = \left(-i\sqrt{2}\phi\right)^{-{\rm ind}(\b\partial_A)}~.
\ee
 Some useful facts about the twisted Dolbeault complex index are reviewed in appendix~\ref{app:index}.

%%%%%%%%%%%%
\subsection{Five-dimensional uplift of the hypermultiplet}

Let us now compute the partition function of a charged 5d $\CN=1$ hypermultiplet on $\CM_5$. The simplest way to do this is to expand the 5d fields into 4d modes, by KK reduction along the $S^1$ fiber, and use the 4d  result \eqref{1-loop hyper M4} for the 4d modes of fixed KK charge. 
Due to the non-trivial fibration structure on $\CM_5$, the supersymmetric background for the 5d vector multiplet is slightly more complicated than the 4d background  \eqref{SUSYbackground1}-\eqref{SUSYbackground2}. We have the conditions:
\bea    \label{5d SUSYbackground1}
    \cD^{2,0} = 0~, \qquad \cD^{0,2} = 0~, \qquad F^{2,0} = i\beta\sigma \FKK^{2,0}~, \qquad F^{0,2} = i\beta \sigma \FKK^{0,2}~,
\eea
together with:
\bea
    i\d_A \sigma = F^{1,0}~, \qquad i\b \d_A \sigma = F^{0,1}~, \qquad \h \CD^{0,0} = \d_{5,A} \sigma = 0~,
\eea
in terms of the 5d twisted Dolbeault operators. 
 On the four-dimensional supersymmetric background \eqref{SUSYbackground1}, the condition \eqref{SUSYbackground1} implied  $\d_A^2=0$ and $\b \d_A^2=0$, which simplified the computation. This is no longer the case on $\CM_5$ if the fibration is non-trivial. 
 %For instance, we have $\d_A^2 = -i\left(F^{2,0} - i \beta \FKK^{2,0}\, \d_{5,A}\right)$, which does not vanish. 
  The kinetic operators for the bosonic and fermionic  fluctuations around the 5d background read:
  \bea
    \Delta_{\rm bos} =& \left( \begin{matrix}
     -2\star \d_A \star \b \d_A + \left(\sigma \sigma - \d_{5,A}\d_{5,A}\right) & *\d_A \d_A ~ \\ -\star \b\d_A \b\d_A & -{1\ov 2}\star \b\d_A \star  \d_A + {1\ov 4}\left(\sigma \sigma - \d_{5,A}\d_{5,A}\right) ~
    \end{matrix} \right)~,\\
    \Delta_{\rm fer} =& \left( \begin{matrix}
     -{i\ov 2}\left(\sigma - \d_{5,A}\right) & i\,\b\d_A & ~{i\ov 2}\star \d_A~ \\
     ~-i\star \d_A\star~ & i \left(\sigma + \d_{5,A}\right) & ~0~ \\
     -{i\ov 2}\star \b \d_A & 0 & ~-{i\ov 4}\left(\sigma + \d_{5,A}\right)~
    \end{matrix}\right)~.
\eea
They are related as follows:
\bea
    \Delta_{\rm fer}
    \left( \begin{matrix} 1 & -2i \b \d_A & i\star \d_A  \\
    0 & -i\left(\sigma-\d_{5,A}\right) & 0 \\
    0 & 0 & i\left(\sigma-\d_{5,A}\right) \end{matrix} \right) =  \left( \begin{matrix} -{i\ov 2}\left(\sigma-\d_{5,A}\right) \quad & 0 \quad 0~ \\
   \begin{matrix} -i \star \d_A \star  \\
    {i\ov 2} \star \b \partial_A \end{matrix} & \Delta_{\rm bos} \end{matrix}\right) ~.
\eea
As a result, the one-loop determinant reduces to:
\be\label{ZM5 from L}
   Z_{\CM_5}^\CH=  {\det(\Delta_{\rm fer}) \ov \det(\Delta_{\rm bos}) } = { \det(\Lop^{(0,1)}) \ov \det(\Lop^{(0,0)}) \det(\Lop^{(0,2)})}~, \qquad \quad \Lop = i\left(\sigma - \d_{5,A}\right)~,
\ee
giving us  the five-dimensional uplift of \eqref{1-loop det hyper reduction M4}. 
%Note that the operator $\Lop$ is defined  only up to an irrelevant constant prefactor. 
 Using a similar argument to the one given above, the only modes that contribute to \eqref{ZM5 from L} are the zero modes of the twisted Dolbeault operator on $\CM_5$.
 To evaluate this explicitly, we expand the 5d fields $\varphi$ into 4d KK modes, as a Fourier decomposition along the $S^1$ fiber:
  \be
 \varphi(z, \b z, \psi)=\sum_{n\in \Z} \varphi_{(n)}(z, \b z, \psi)~, \qquad\quad \varphi_{(n)}(z, \b z, \psi)\equiv  e^{- i n \psi}  \varphi_n(z, \b z)~.
 \ee
We then have:
\be
    \Lop \varphi_n = \lambda_n \varphi_n~, \qquad \quad \lambda_n = i\beta(\sigma + i A_5) + n~.
\ee
From the 4d point of view, each KK mode $\varphi^{p,q}$ of a given $(p,q)$-degree is a section of a bundle
\be
\Omega^{p,q}\otimes V_n~, \qquad \qquad V_n =  E_\FR \otimes (\CL_{\rm KK})^n~,
\ee
where $E_\FR$ is the gauge bundle and $\CL_{\rm KK}$ is the defining line bundle introduced in section~\ref{subsec:M5 geom}. The 5d hypermultiplet partition function then takes the form:
\be
   Z_{\CM_5}^\CH = \prod_{n\in \mathbb{Z}} \lambda_n ^{-{\rm ind}(\b\d_{V_n})}~,
\ee
where ${\rm ind}(\b\d_{V_n})$ is the index of the 4d Dolbeault complex twisted by the vector bundle $V_n$.  This infinite product needs to be properly regularised,  as we discuss next. 

%%%%%%%%%%%%
\subsection{Regularisation: summing up the KK tower}\label{sec:hyperOneloopReg}
For our purposes, we will only consider abelian gauge bundles, by choosing a maximal torus of the (background or dynamical) gauge group. Then, without loss of generality, we can consider the hypermultiplet coupled to a single $U(1)$ gauge field with background flux $\m$, as discussed in section~\ref{subsub:introFlux}.   The complex scalar in the effective 4d $\CN=2$ vector multiplet is denoted by
 \be
 a\equiv  i \beta(\sigma + i A_5)~,
 \ee
 with the identification $a\sim a+1$ under a $U(1)$ large gauge transformation. 
 The 5d hypermultiplet partition function is given formally by the infinite product:
\be
Z^\CH_{\CM_5} = \prod_{n\in \Z} \left({1\ov a+ n}\right)^{{\rm ind}(\b\d_{V_{n, \varepsilon}})}~,
\ee
in terms of the index of the $V_{n, \varepsilon}$-twisted Dolbeault complex. Here, we take $V_{n, \varepsilon}$ to be the line bundle:
\be\label{Vndelta hyper}
V_{n, \varepsilon} \cong \CK^{\varepsilon+\half} \otimes L \otimes (\CL_{\rm KK})^n~,
\ee
where the $L$ connection is the background $U(1)$ gauge field   with flux $\m$, and $\varepsilon$ indexes our choice of extended twist for the hypermultiplet, as discussed around \eqref{CL0 extended twist}.  
The canonical choice on a generic K\"ahler base $\CM_4$ is $\varepsilon=-\half$, while if $\CM_4$ is spin it is also natural to choose $\varepsilon=0$. 
Note that:
\be
c_1(V_{n, \varepsilon}) =\sum_l\left( \left(\varepsilon+\half\right)\,  \k_l + \m_l + n\,  \p_l\right) [\S_l]~, 
\ee
where $\k$ denotes the first Chern class of the canonical line bundle on $\CM_4$,
\be\label{def c1k}
 c_1(\CK)= \sum_l \k_l [S_l]~,
\ee
and $\p$ was defined in \eqref{ps for CL0}. For simplicity of notation, we can absorb the $(\varepsilon+\half) \k$ term into $\m$, effectively setting $\varepsilon=-\half$ in what follows. 
Using the index theorem (see appendix~\ref{app:index}), we find:
\be\label{indVn 5d}
{\rm ind}(\b\d_{V_{n}})  = \int_{\CM_4} {\rm Td}(T\CM_4) \wedge {\rm ch}(V_n) = \chi_h+ \half (\m + n\, \p - \k,  \m +  n\,\p)~,
\ee
with $\chi_h=   {\chi+ \sigma\ov 4}$ the holomorphic Euler characteristic, 
and with the intersection pairing $(-, -)$ on $\CM_4$ as defined in \eqref{defInterPairing}.  

\medskip
\noindent
{\bf Regularisation of the result.}
Given \eqref{indVn 5d},  the infinite product to be regularised takes the explicit form:
\be\label{ZM5 as product}
Z^\CH_{\CM_5}(a)_\m = \prod_{n\in \Z} \left({1\ov a+ n}\right)^{ \chi_h+ \half (\m + n\, \p - \k,  \m +  n\,\p) }~.
\ee
The notation $Z^\CH_{\CM_5}(a)_\m$ makes the dependence  on $a$ and $\m$ manifest.
It is convenient to factor \eqref{ZM5 as product} as follows: 
\be \label{Pert part funct}
Z^\CH_{\CM_5}(a)_\m =  \Flux^\CH(a)^{\chi_h+\half(\m-\k,\m)} \, \FiberOpK^\CH(a)^{(\m- \half  \k,\p)}\,
\FiberOp^\CH(a)^{\half (\p, \p)}~.
\ee
Here, we formally defined the following functions in terms of divergent products:
\be\label{H K and F hyper infinite prod}
\Flux^\CH(a)\equiv \prod_{n\in \Z}{1\ov  a + n}~, \qquad  
\FiberOpK^\CH(a) \equiv \prod_{n\in \Z}\left({1\ov a+ n}\right)^n~,\qquad
\FiberOp^\CH(a) \equiv \prod_{n\in \Z}\left({1\ov a+ n}\right)^{n^2}~.
\ee
These formal products give us  information on the analytic structure of the corresponding meromorphic functions, with poles or zeros at $a\in \Z$. Namely, $\Flux^\CH$ has poles of order $1$ at any integer $a \in \Z$, $\FiberOpK^\CH$ has poles of order $n$ at $a=n$ for every negative integer $n$ (and zeros at the positive integers), and $\FiberOp^\CH$ has poles of order $n^2$ at $a=n$ for any integer $n$.
 Following the discussion in \cite{Closset:2017zgf, Closset:2018ghr, Closset:2018bjz}, we choose the gauge-invariant regularisation, also known as the `$U(1)_{-\half}$ quantisation'. We then find:
\bea\label{H K and F hyper}
\Flux^\CH(a)= &\; {1\ov 1-e^{2\pi i a}}~,\\
\FiberOpK^\CH(a)=&\; \exp\left( {1\ov 2\pi i} \dilog(e^{2\pi i a})+a  \log(1-e^{2\pi i a}) \right)~,\\
\FiberOp^\CH(a) =&\; \exp\left(-{1\ov 2\pi^2} \trilog(e^{2\pi i a}) -{a\ov \pi i} \dilog(e^{2\pi i a})- a^2 \log(1-e^{2\pi i a}) \right)~.
\eea
Despite the appearance of polylogarithms, these functions are meromorphic in $a$, with the poles mentioned above. They also have simple transformation properties under large gauge transformations, $a\sim a+1$, with $\Flux^\CH(a+1)=\Flux^\CH(a)$ and
\be 
\FiberOpK^\CH(a+1)= \Flux^\CH(a)^{-1} \FiberOpK^\CH(a)~, \qquad\quad
\FiberOp^\CH(a+1)= \Flux^\CH(a) \FiberOpK^\CH(a)^{-2}  \FiberOp^\CH(a)~.
\ee
Using these relations, we can check that the partition function is gauge invariant. Whenever the circle is non-trivially fibered over $\CM_4$, a large gauge transformation amounts to the simultaneous shift $(a, \m)\rightarrow (a+1, \m+\p)$. More invariantly, this corresponds to tensoring the $U(1)$ line bundle with the defining line bundle, $L \rightarrow L \otimes \CL_{\rm KK}$. 
We indeed find that:
\be\label{LGTZH}
Z^\CH_{\CM_5}(a+1)_{\m + \p}= Z^\CH_{\CM_5}(a)_{\m}~,
\ee
as expected. 

%%%%%%%%%%%%%
\medskip

\noindent \paragraph{Example: Trivial fibrations.} Let us first  consider the case $\cM_5 = \cM_4 \times S^1$. Since $\p=0$, the partition function takes the simple form:
\be \label{Z Hyper M4xS1}
    Z_{\cM_4 \times S^1}^{\cH}(a)_\m =\Flux^\CH(a)^{\chi_h+ \half (\m-\k, \m)}= \left( {1\ov 1-e^{2\pi i a}}\right)^{\chi_h+ \half (\m-\k, \m)}~,
\ee
for $\varepsilon=-\half$. This agrees with previous results \cite{Hosseini:2018uzp}, up to some differences in conventions.%
\footnote{The most important difference is in the choice of regularisation. As emphasised in \protect\cite{Closset:2018bjz}, our choice is singled out by requiring gauge invariance under large gauge transformations.}
  For a more general choice of extended DW twist, we find:
\be \label{Z Hyper M4xS1 gen eps}
    Z_{\cM_4 \times S^1}^{\cH}(a; \varepsilon)_\m =\left( {1\ov 1-e^{2\pi i a}}\right)^{-{\sigma\ov 8}+{\varepsilon^2\ov 2}(2\chi+3\sigma) + \half (\m+2 \varepsilon \k, \m)}~,
\ee
where we used the relation $(\k,\k)= 2\chi +3\sigma$. (See appendix~\ref{app:index}.)

%%%%%%%%%%%%%
\medskip

\noindent \paragraph{Example: The five-sphere $S^5$.}  Let us now consider the example of $S^5$, fibered over $\mathbb{P}^2$, setting $\varepsilon=-\half$ for simplicity. We have $\chi=3$, $\sigma=1$ and ${\bf k}=-3$, and thus \eqref{Z Hyper M4xS1} gives us:
\be
Z^\CH_{\bP^2\times S^1}(a)_\m = \left( {1\ov 1-e^{2\pi i a}}\right)^{1+\half \m(\m+3)}~,
\ee
for any flux $\m\in \Z$. The $S^5$ is obtained by a fibration with $\p=1$, so that:
\be
Z^\CH_{\CM_5}(a)_\m = Z^\CH_{\bP^2\times S^1}(a)_\m   \, \FiberOpK^\CH(a)^{\m- {3\ov 2} }\,
\FiberOp^\CH(a)^{\half}~,
\ee
and we can set $\m=0$ by a large gauge transformation \eqref{LGTZH}, reflecting the fact that $H^2(S^5){=}0$. Hence we find:
\be
Z^\CH_{\CM_5}(a) =  \exp\left( - {1\ov 4 \pi^2} \trilog(e^{2\pi i a}) - {2a-3 \ov 4\pi i} \dilog(e^{2\pi i a})- {a^2-3a+2\ov 2} \log(1-e^{2\pi i a})\right)~. 
\ee
This is in good agreement with previous results \cite{Kallen:2012va}.

\medskip
\noindent \paragraph{Example: The five-manifold $T^{p_1,p_2}$.} As another example, consider the fibration over $\CM_4=\bF_0=\bP^1\times \bP^1$, with $\p=(p_1,p_2)$, which is sometimes called $T^{p_1, p_2}$.%
\footnote{In particular, for $p_1=p_2=1$,  $T^{1,1}$ famously admits a Sasaki-Einstein metric \protect\cite{Candelas:1989js}.} (We take the two $\bP^1$ factors as our basis curves, $\bF_0\cong \S_1\times \S_2$.) Then we have $\chi=4$, $\sigma=0$ and $\k= (-2,-2)$, hence:
\be
Z^\CH_{\bP^2\times S^1}(a)_\m =  \left( {1\ov 1-e^{2\pi i a}}\right)^{4 \varepsilon^2- 2 \varepsilon \m_1 \m_2 + \m_1\m_2}~,
\ee 
for any flux $\m=(\m_1, \m_2)$, and keeping an arbitrary $\varepsilon$. (Since $\bF_0$ is spin, we can choose it as we like, including choosing the DW twist value $\varepsilon=0$.) We then have:
\be
Z^\CH_{T^{p_1,p_2}}(a)_\m =  Z^\CH_{\bP^2\times S^1}(a)_\m  \, \FiberOpK^\CH(a)^{(p_1\m_2+p_2\m_1)-2 \varepsilon (\m_1+\m_2)}\,
\FiberOp^\CH(a)^{p_1 p_2}~.
\ee
For $T^{1,1}$ with $\varepsilon=0$, for instance, this gives:
\bea
&Z^\CH_{T^{1,1}}(a)_\m &=&\;\; \exp\Bigg( - {1\ov 2 \pi^2} \trilog(e^{2\pi i a}) - {2a-\m_1-\m_2 \ov 4\pi i} \dilog(e^{2\pi i a})\\
&&&\qquad\qquad - (a^2- (\m_1+\m_2)a- \m_1 \m_2) \log(1-e^{2\pi i a})\Bigg)~.
\eea
Note that we have the gauge equivalence $(a, \m_1, \m_2)\sim (a+1, \m_1+1, \m_2+1)$. Using \eqref{relH2M4M5}, one can check that $H^2(T^{1,1},\Z)\cong \Z$.

%%%%%%%%%%%%
\subsection{Higher-spin particles on $\CM_5$}\label{subsec:HSM5}
By a small generalisation of the above computation, one can also capture the contribution of higher-spin states. Such electrically-charged states generally appear on the real Coulomb branch of 5d SCFTs. For instance, when we have an infrared non-abelian gauge theory phase, the W-bosons give spin-one states. More generally, 5d BPS particles of arbitrary spin can contribute. Following the approach of \cite{Gopakumar:1998ii, Gopakumar:1998jq, Lockhart:2012vp}, we expect that, in the topologically-twisted theory, they contribute to the partition function on the Coulomb branch as KK towers of 4d off-shell hypermultiplets of $SU(2)_l\times SU(2)_r$ spin $(j_l, j_r)$. In the 5d interpretation, $(j_l,j_r)$ is the representation under the little group of the massive particle.

Let us first recall some elementary properties of the half-BPS massive  representations of the 5d  $\CN=1$ supersymmetry algebra. The BPS states saturate the BPS mass bound with $M=Z_{\rm 5d}$, where we take the fifth direction to be time, with $P_M=(0,0,0,0,-M)$. Such states are annihilated by the `right-chiral' supercharges (in the 4d notation), and the supersymmetry algebra \eqref{5d susy alg R5}, after Wick rotation, is realised as:
\be
\{ Q_\alpha^I, Q_\beta^J\} = -4 i M \epsilon^{IJ}\epsilon_{\alpha\beta}~, \qquad \qquad  \t Q^\alphadot_I=0~.
\ee
 Picking the supercharges $Q_\alpha^{I=1}$ and $Q_\alpha^{I=2}$ (of $R$-charge $R=\pm 1$, under $U(1)_R\subset SU(2)_R$, respectively) as the creation and annihilation operators, we obtain the supermultiplet:
\be\label{jljr states}
\left(j_l, j_r; \half\right)^{(-1)^{2j_l+2j_r}} \; \oplus \left(j_l+\half, j_r;  0\right)^{-(-1)^{2j_l+2j_r}} \oplus  \left(j_l-\half, j_r;  0\right)^{-(-1)^{2j_l+2j_r}} ~,
\ee
for any spin $(j_l, j_r)$ for the `ground state' -- this is for $j_l>0$, while for $j_l=0$ there is no third summand in \eqref{jljr states}.  Here $(j_l, j_r, s)^{\pm}$ denotes a 5d massive state of spin $(j_l,j_r)$ and of $SU(2)_R$ `isospin' $s$, with the superscript $\pm$ corresponding to bosons and fermions, respectively. The statistics is determined by the spin-statistics theorem.

Now, consider the standard DW twist of the multiplet \eqref{jljr states}. We obtain states of twisted $SU(2)_l\times SU(2)_D$ spins:
\be
\left(j_l, j_r\pm \half \right)^{(-1)^{2j_l+2j_r}}\; \oplus\,  \left(j_l\pm \half, j_r\right)^{-(-1)^{2j_l+2j_r}}~,
\ee
which are most conveniently written as:
\be
\left[ \left(0,\half\right)^{(-1)^{2j_l+2j_r}}\oplus \left(\half,0\right)^{-(-1)^{2j_l+2j_r}} \right] \otimes \left(j_l, j_r\right)~.
\ee
 The states in the bracket give us a standard massive hypermultiplet in the twisted theory (up to a choice of statistics), and we simply need to tensor by the general spin $(j_l, j_r)$.

For $(j_l,j_r)=(0,0)$, we recover the standard hypermultiplet.  As a first non-trivial example, it is interesting to consider the massive vector multiplet after the DW twist. Such massive vectors appear as W-bosons through the Higgs mechanism on the CB, for instance. In this case, we can compute their contribution explicitly, as a one-loop computation, by considering the gauge-fixed SYM Lagrangian. This is discussed in some detail in appendix~\ref{appendix: vec one-loop}. After the topological twist, we have the multiplet:
\be
\left[ \left(0,\half\right)^{-}\oplus \left(\half,0\right)^{+} \right] \otimes \left(0, \half\right) =  \left(0,1\right)^- \oplus \left(0,0\right)^- \oplus \left(\half, \half\right)^+~,
\ee
corresponding to  the on-shell gauginos and the massive vector, respectively. Thus the massive vector multiplet corresponds to $(j_l, j_r)= (0, \half)$.

 \medskip
\noindent
{\bf Extended topological twist at higher spin.}  
Consider a massive particle of spin $(j_l, j_r)$ charged under some abelian gauge symmetry $\prod_K U(1)_K$ with charges $q_K\in \Z$, after the standard DW twist. The corresponding KK tower of fields on $\CM_4$ are valued in the bundles:
\be\label{full jljr bundle}
\left[S_- \oplus S_+ \right]\otimes \CK^{\varepsilon} \otimes \bigotimes_K (L_{K})^{q_K} \otimes S^{2j_l}(S_-) \otimes S^{2j_r}(S_+) \otimes  (\CL_{\rm KK})^n~,
\ee
where $L_{K}$ are  $U(1)_K$ bundles (to be discussed further in  section~\ref{subsec:fluxop} below), and $S^k(E)$ denotes the symmetrised product of $k$ copies of the bundle $E$. 
When the K\"ahler manifold $\CM_4$ is not spin, the extended twist parameter $\varepsilon$ cannot be zero unless $2j_l+2j_r$ is odd. In general, we need to choose $\varepsilon$ so that:
\be
 \varepsilon + j_l+j_r +\half \in \Z~, 
\ee
which ensures that the bundle \eqref{full jljr bundle} is well-defined. This generalises the discussion of  section~\ref{subec: 4d hyper}. 
In a given 5d theory, there might be any number of massive particles of various spins that will contribute in this way, and the $\varepsilon$ parameters for each cannot be chosen independently. 
 We will come back to this important point in section~\ref{subsubsec:spincharge} below.

 \medskip
\noindent
{\bf Partition function at spin $(j_l, j_r)$.} 
 We can now generalise the previous results for the partition function of a hypermultiplet with spin $(j_l, j_r)$. It is determined in terms of the  Dolbeault complex twisted by the KK tower of `higher-spin' bundles: 
\be\label{def Vnetc}
V_{n, \varepsilon; (j_l, j_r)}= \CK^{\half}\otimes \bigotimes_K \left(\CK^{\varepsilon^K} \otimes L_{K}\right)^{q_K}\otimes S^{2j_l}(S_-) \otimes S^{2j_r}(S_+) \otimes  (\CL_{\rm KK})^n~.
\ee
The computation of that index is discussed in appendix~\ref{app:index}.  Using the notation $\m \equiv q_K \m^K$ and $\varepsilon= q_K \varepsilon^K$, one finds: 
 \bea\label{indexjljrfull}
&{\rm ind}(\b\d_{V_{n, \varepsilon; (j_l, j_r)}})  = (2j_l+1)(2j_r+1)\Bigg[-{\sigma\ov 8} +\half \varepsilon^2 (2\chi+3\sigma)-{2\ov 3} j_l(j_l+1)\chi \\
&\qquad\qquad\qquad\qquad  + {j_l(j_l+1)+j_r(j_r+1)\ov 6} (2\chi+3\sigma)+\half (\m+n \p +2 \varepsilon \k, \m + n \p)\Bigg]~.
\eea
Then, using the building blocks \eqref{H K and F hyper} and the notation $a=q_K a^K$, we obtain:
\bea\label{ZM5jljr}
&Z_{\CM_5}^{(j_l,j_r)}(a)_\m &=&\; \Flux^\CH(a)^{c_\CA  \chi+c_\CB  \sigma + c_0 \left[ \half \varepsilon^2 (2\chi + 3\sigma) + \half(\m+2\varepsilon \k, \m)\right]}  \\
&&&\qquad
\times \FiberOpK^\CH(a) ^{c_0 (\m+\varepsilon  \k,\p)}\,
\FiberOp^\CH(a)^{\half c_0  (\p, \p)}~,
\eea
in terms of the following spin-dependent numbers:
\bea\label{cAB0 coefs 1}
& c_\CA = (-1)^{2j_l+2j_r} (2j_l+1)(2j_r+1)  {j_r(j_r+1)- j_l(j_l+1) \ov 3}~, \\
& c_\CB = (-1)^{2j_l+2j_r} (2j_l+1)(2j_r+1)\left(-{1\ov 8}+{j_l(j_l+1)+ j_r(j_r+1) \ov 2}\right)~, \\
& c_0 =(-1)^{2j_l+2j_r} (2j_l+1)(2j_r+1)~,
\eea
which are independent of the geometry.

%%%%%%%%%%%%%%%%%%%%%%%%%%%%%%%%%%%%%%%%%%%%%%%%%%%%%%%%%%
\section{Flux and fibering operators on the Coulomb branch}\label{sec:fluxFiberingOps}

In this section, we consider the low-energy effective action of a 5d $\CN=1$ field theory compactified on a circle. The 5d theories we have in mind are 5d SCFTs, but the following infrared approach is independent of the exact UV completion. We consider the effective 4d $\CN=2$ KK theory compactified on $\CM_4$, at arbitrary fixed values of the extended Coulomb branch vector multiplets. As explained in the introduction, this is a crucial step towards a systematic computation of the $U$-plane integral.

\subsection{KK theory on $\CM_4\times S^1$: flux operators and gravitational couplings}\label{subsec:fluxop}
Consider any 4d $\CN=2$ theory on $\CM_4$. For definiteness, let us assume it is a KK theory, so that we have a scale $\beta^{-1}$ set by the inverse radius of the circle. We wish to study the Coulomb branch of this theory, where the low-energy degrees of freedom are $r$ 4d $\CN=2$ abelian vector multiplets -- $r$ is the `rank' of the 5d theory, by definition. We denote by $a^i$ the scalars in the $U(1)^r$ vector multiplets, which are related to the 5d $\CN=1$ abelian vector multiplets as:
\be
a^i = i \beta\left( \sigma^i+ i A^i_5\right)~, \qquad \qquad i=1, \cdots, r~.
\ee
Note that $a^i$ is dimensionless, in our conventions. Furthermore, large-gauge transformations along the fifth direction give us the periodicity $a^i \sim a^i+1$, $\forall i$. We also consider background vector multiplets for some maximal torus of the flavour symmetry group,
$U(1)^{r_F}\subset G_F$, 
where $r_F$ denotes the rank of the flavour group. The corresponding background scalars are simply complex masses, denoted by:
\be
\mu^\alpha = i \beta\left( m^\alpha+ i A^\alpha_{F,5}\right)~, \qquad \qquad \alpha=1, \cdots, r_F~,
\ee
with the identification $\mu^\alpha \sim \mu^\alpha+1$.
The total space of values for $(a^i, \mu^\alpha)$ is called the extended Coulomb branch, of dimension $r+r_F$. It is convenient to introduce the notation:%
\footnote{{\it Beware the indices:} In this section and the next, the indices $I, J, \cdots$ run over the gauge and flavor maximal torus, while $i, j, \cdots$ are gauge indices and $\alpha, \beta$ are flavor indices. This is distinct from the conventions in other sections, for instance $z^i$ denoted holomorphic coordinates on $\CM_4$, and $\alpha, \beta$ are also 4d left-chiral spinor indices;  no confusion is likely there. Note also that $I, J$ were previsouly used as $SU(2)_R$ indices, but we are now dealing with DW-twisted fields which are $SU(2)_R$-neutral, therefore this notation switch should cause no confusion.}
\be
(\amCB^I)= (a^i, \mu^\alpha)~, \qquad \qquad I=(i, \alpha)~,
\ee
which treats dynamical and background vector multiplets democratically. We will furthermore assume that the vector multiplets are the only massless degrees of freedom at generic points on the (extended) CB.%
\footnote{More generally, there could be additional massless hypermultiplets, giving us a so-called enhanced CB. We will not consider this possibility in this paper.}

The low-energy 4d $\CN=2$ effective field theory in flat space is then governed by the effective prepotential, denoted by $\CF(a, \mu)$. We define $\CF(\amCB)$ for the KK theory to be dimensionless (it is related to the usual 4d prepotential, $\CF_{\rm 4d}$,  by $\CF= \beta^2 \CF_{\rm 4d}$). The flat-space Lagrangian can be coupled to the DW-twist background on $\CM_4$. Its key property is that it is `almost' $\CQ$-exact, similarly to \eqref{4d SYM lagrangian}. Discarding the $\CQ$-exact pieces, we are left with the following topological action, which is well-defined on any $\CM_4$ \cite{Losev:1997tp}:
\bea\label{Sflat gen 4d}
&S_{\rm flat} &=&\; {i \ov 4 \pi} \int_{\CM_4} \Bigg(F^I \wedge F^J {\d^2 \CF(\amCB)\ov \d \amCB^I \d \amCB^J} -{i\ov 2} F^I \wedge \Lambda^J\wedge \Lambda^K {\d^3 \CF(\amCB)\ov \d \amCB^I \d \amCB^J  \d \amCB^K}\\
&&&\qquad\qquad\qquad\qquad\qquad\qquad  -{1\ov 48}  \Lambda^I \wedge \Lambda^J\wedge \Lambda^K\wedge \Lambda^L {\d^4 \CF(\amCB)\ov \d \amCB^I \d \amCB^J  \d \amCB^K \d \amCB^L} \Bigg)~,
%+ \delta\left(\cdots \right)~,
\eea
 where the sum over repeated indices is understood.
Here $F= dA$ for an abelian gauge field, and we also introduced the one-form $\Lambda= \Lambda^{1,0}+ \Lambda^{0,1}$, in the notation of \eqref{susy vec twisted}. Formally, \eqref{Sflat gen 4d} can be viewed as the fourth descendant, $\int_{\CM_4} \CO^{(4)}$, with respect to the DW supercharge $\delta= \delta_1+\delta_2$, of the $0$-form:
\be
\CO^{(0)}=- {2i \ov \pi} \CF(\amCB)~,
\ee 
where we used the descent relations $\delta\CO^{(n)}=d \CO^{(n-1)}$ with the supersymmetry variations:
\be
\delta a=0~, \qquad\qquad \delta \Lambda= 2 d a~,\qquad \qquad \delta F= - i d\Lambda~,
\ee
for an abelian vector multiplet, with $a= i\sqrt2 \phi$. The fermionic terms in \eqref{Sflat gen 4d} only depend on the one-form $\Lambda$. Correspondingly, they will only affect the low-energy physics on $\CM_4$ if the $\Lambda$ fields have zero-modes, which is to say if $H^1(\CM_4, \R)$ is non-trivial. Since we have assumed that   $H^1(\CM_4, \R)=0$ -- {\it i.e.} $b_1=0$ -- in this paper, we can ignore the effect of these fermionic couplings in the following. We hope to address the more general case in future work.

 Let us now consider any background gauge field configuration for the $U(1)_I$ symmetries, assuming it preserves our two supercharges. We denote the corresponding fluxes on $\CM_4$ by
\be
    c_1(F^I) = {1\ov 2 \pi}   F^I= \sum_k \m^I_k [\S_k]~.
\ee
 Recall that we denote the intersection pairing on $H_2(\CM_4, \Z)$ by $(-, -)$, so that we have:
\be\label{def inter pairing}
(\m^I, \m^J) =    {1\ov 4\pi^2} \int_{\CM_4} F^I \wedge F^J = \sum_{k,l}\m_k^I  {\bf I}_{kl} \m_l^J~,
\ee
with ${\bf I}_{kl}$ as in \eqref{Ikl def}.
At any generic point on the Coulomb branch, taking $\amCB^I$ to be constant, the action \eqref{Sflat gen 4d}  evaluates to:
\be\label{S flux def}
S_{\rm flux} \equiv  S_{\rm flat}\Big|_{\rm CB}=     \pi i (\m^I, \m^J)    {\d^2 \CF(\amCB) \ov \d \amCB^I \d \amCB^J}~.
\ee
In addition to \eqref{Sflat gen 4d}, the infrared theory compactified on $\CM_4$ is governed by well-studied gravitational couplings \cite{Witten:1995gf, Moore:1997pc, Losev:1997tp, Marino:1998bm, Shapere:2008zf}. 
Up to $\CQ$-exact terms and away from Seiberg-Witten singularities,  the topologically-twisted Coulomb-branch theory takes the simple form:
\be\label{STFT 5d}
S_{{\rm TFT}} =  S_{\rm flat}+ S_{\rm grav}~.
\ee
The second term in \eqref{STFT 5d} consists of couplings to the background metric:
\bea
&S_{\rm grav} &=&\; {i\ov 64\pi}   \int d^4 x\sqrt{g} \, \epsilon^{\mu\nu\rho\sigma} \epsilon^{\alpha\beta\gamma\delta}  R_{\mu\nu\alpha\beta} R_{\rho\sigma \gamma\delta}\, \CA(\amCB) \cr
&&&\;   +{i\ov 48\pi}   \int d^4 x\sqrt{g} \, \epsilon^{\mu\nu\rho\sigma} {R_{\mu\nu\alpha}}^\beta {R_{\rho\sigma \beta}}^\alpha\,  \CB(\amCB)~.
\eea
At constant values of the extended CB parameters, this evaluates to:
\be\label{Sgrav}
S_{\rm grav} = 2\pi i \Big( \chi\, \CA(\amCB)   + \sigma\, \CB(\amCB)\Big)~,
\ee
where $\chi$ and $\sigma$ are the topological Euler characteristic and the signature of $\CM_4$, respectively. This gives the famous contribution: 
\be\label{Sgrav AB}
e^{-S_{\rm grav}} =  \bA(\amCB)^\chi  \bB(\amCB)^\sigma~, \qquad \qquad \bA(\amCB) \equiv e^{-2\pi i \CA(\amCB)}~,\qquad
\bB(\amCB) \equiv e^{-2\pi i \CB(\amCB)}~.
\ee
The prepotential $\CF$ and the gravitational couplings $\CA$ and $\CB$ can be determined from the Seiberg-Witten geometry of the 5d theory on a circle, in principle, or else from an explicit instanton counting computation on the $\Omega$-background -- for the rank-one 5d SCFTs with $E_n$ symmetry, this was discussed at length in \cite{Closset:2021lhd}.

On general grounds, the prepotential $\CF$ suffers from branch-cut ambiguities:
\be\label{F ambiguities}
 \CF(\amCB) \sim \CF(\amCB) + {n_2 \ov 2} \amCB^2 + n_1 \amCB + {n_0\ov 2}~, \qquad n_0, n_1, n_2 \in \Z~.
\ee
Such shifts are incurred, in particular, when performing large gauge transformations along the 5d circle. It follows that the exponentiated action $\exp{(-S_{\rm flux})}$ is singled-valued if and only if the intersection pairing is even, so that $(\m, \m)\in 2\Z$ for any integer-quantized flux $\m$, which is true if $\CM_4$ is spin. More generally, we need to modify the quantization condition on our fluxes, so that $A^I$ describe spin$^c$ connections rather than $U(1)$ gauge fields. For $U(1)_I$ bundles, we would have $\m^I_k\in \Z$, while more generally we may choose:
\be\label{choose epsI}
{1\ov 2\pi} F^I =\sum_k \left( \varepsilon^I \k_k+  \m_k^I\right) [\S_k]~.
\ee
Here, $\k$ was defined in \eqref{def c1k}, $\varepsilon^I \in \half \Z$, and $\m_k^I\in \Z$. The parameters $\varepsilon^I$ must be carefully chosen  depending on the theory so that it be well-defined on $\CM_4$, as we will discuss in more detail in section~\ref{subsubsec:spincharge} below. They are the infrared analogue of the extended DW-twist parameter $\varepsilon$ introduced in section~\ref{subec: 4d hyper} for the hypermultiplet.  The spin$^c$  connections $A^I$ can be formally viewed as connections on the ill-defined line bundles
\be
 \CL_I=  \CK^{\varepsilon_I}  \otimes  L_{I}~,
\ee
where $L_{I}$ is a $U(1)$ line bundle with first Chern class $\m^I$.  
The necessity of introducing spin$^c$ connections arises from the fact that our 4d $\CN=2$ KK theories generally contain spinors even after the standard DW twist -- in the infrared description on the Coulomb branch, these arise as massive BPS particles coupled to the low-energy (background and dynamical) photons, which can have arbitrary (twisted) spin. We will give the precise condition on $\varepsilon^I$ in section \ref{subsec:hs states and GV} below. For now,  we claim that the $\varepsilon^I$'s can always be chosen so that the low-energy theory is well-defined; in particular, choosing these parameters correctly will render $e^{-S_{\rm TFT}(\amCB)}$  fully gauge-invariant, single-valued and locally holomorphic in $\amCB$.%
\footnote{Here, holomorphy is a formal consequence of supersymmetry since anti-holomorphic terms are $\CQ$-exact.}

\medskip
\noindent
Let us now define the `flux operators':
\be
\Flux_{I,J}(\amCB)= \exp\left( -2 \pi i  {\d^2 \CF(\amCB) \ov \d \amCB_I \d \amCB_J}  \right)~,
\ee
which are meromorphic functions on the ECB parameters $\amCB_I$. Such insertions can be understood as local operators in the twisted infrared theory.  Alternatively, we consider the insertion of \eqref{Sflat gen 4d} for specific fluxes, which can be viewed as the top-dimensional topological descendant of $\CF(\amCB)$, viewed itself as a local operator (at least formally).   Using the topological invariance, we can localise $F^I\wedge F^J$ to have support at a point on $\CM_4$, giving rise to the local insertion:%
\footnote{What we call the `flux operator' has been denoted `the  $C$ coupling' in recent works \protect\cite{Manschot:2021qqe, Aspman:2022sfj}. The flux operator insertion can be also be interpreted as a contact term localised at the intersection of the 2-cycles carrying the flux~\protect\cite{toappearKMMTZ}.}
\be
e^{-S_{\rm flux}}= \prod_{I,J} \Flux_{I,J}(\amCB)^{\half (\m^I+\varepsilon^I \mK, \m^J+\varepsilon^J \mK)}~.
%= 
% \exp\left( - \pi i  \sum_{I, J}  (\m^I+\varepsilon^I \mK, \m^J+\varepsilon^J \mK)  {\d \CF(\amCB) \ov \d \amCB^I \d \amCB^J}  \right)~.
\ee
Using the fact that  $(\k,\k)=2\chi +3\sigma$,
 it is convenient to factorise these contributions as:
\be\label{Sflux Pi G}
e^{-S_{\rm flux}}= Z^{\rm flux}_{\CM_4}(\amCB; \varepsilon)_\m \; \bG(\amCB; \varepsilon)^{2\chi +3\sigma}~,
\ee
where we defined:
\be\label{flux op gen}
 Z^{\rm flux}_{\CM_4}(\amCB; \varepsilon)_\m 
 \equiv   \prod_{I,J} \Flux_{I,J}(\amCB)^{\half (\m^I+\varepsilon^I \mK, \m^J)} 
 =  \exp\left( - \pi i  \sum_{I, J}  (\m^I+2\varepsilon^I \mK, \m^J)  {\d^2 \CF(\amCB) \ov \d \amCB^I \d \amCB^J}  \right)~,
\ee
and:
\be\label{flux K mixing}
\bG(\amCB; \varepsilon)\equiv e^{-2 \pi i \CG(\amCB; \varepsilon)}~, \qquad \quad
\CG(\amCB; \varepsilon)\equiv   \half  \sum_{I, J} \varepsilon^I  \varepsilon^J  {\d^2 \CF(\amCB) \ov \d \amCB^I \d \amCB^J}~.
\ee
  The full exponentiated topological field theory action \eqref{STFT 5d} evaluated on the CB then gives us the `CB partition function' on $\CM_4$ with gauge and flavor fluxes $\m$:
\be\label{ZM4 m geom}
Z_{\CM_4\times S^1}(\amCB; \varepsilon)_\m =  Z_{\CM_4}^{\rm  geom}(\amCB; \varepsilon) \,  Z^{\rm flux}_{\CM_4}(\amCB; \varepsilon)_\m~.
\ee
This object is really the {\it holomorphic integrand} that will enter the $U$-plane integral of the 4d $\CN=2$ KK theories, as discussed in the introduction. 
 Here, we conjecture that the two factors in \eqref{ZM4 m geom} are separately well-defined on any K\"ahler manifold (this is clearly true when $\CM_4$ is spin, but not so obvious in the non-spin case).
  Consider first the ``flux operator'' contribution \eqref{flux op gen}. The $\varepsilon^I$ parameters should be such that  \eqref{flux op gen} is single-valued. A sufficient set of conditions would be
\bea
 \half(\m^I, \m^J)+ \varepsilon^I (\k, \m^J) \in &
 \; \Z  \quad \text{if}\; I=J~,\\
 \varepsilon^I (\k, \m^J)+ \varepsilon^J (\k, \m^I)  \in & \; \Z  \quad \text{if}\; I\neq J~,
  \eea
  for any $\m^I_k\in \Z$, but this is much too strong in general. Instead, the correct condition on the $\varepsilon^I$'s will depend on the 5d BPS spectrum of the field theory (see section~\ref{subsubsec:spincharge}.

The ``geometrical" factor in \eqref{ZM4 m geom} has contributions from the ordinary gravitational couplings \eqref{Sgrav AB} and from \eqref{flux K mixing}, which is dictated by our choice of (background) spin$^c$ connections. It is given by: 
\be\label{ZM4geom}
Z_{\CM_4}^{\rm geom}(\amCB; \varepsilon) = \bA(\amCB)^{\chi}\bB(\amCB)^{\sigma}\bG(\amCB;\varepsilon)^{2\chi+3\sigma}~.
\ee
The $\bA$ and $\bB$ couplings are given in terms of the low-energy Seiberg-Witten geometry as~\cite{Witten:1995gf, Moore:1997pc, Losev:1997tp}:
\be\label{AB IR expect}
  \bA= \alpha  \left( \det_{ij} {d U_i\ov d a^j}\right)^\half~, \qquad\qquad
\bB =\beta \left( \Delta^{\rm phys}\right)^{1\ov 8}~,
\ee
with $\alpha, \beta$ some numerical constants. Here, $U_i(\amCB)$ are the gauge-invariant $U$-parameters, which parametrise the Coulomb branch of the 4d $\CN=2$ KK theory, and $\Delta^{\rm phys}$ is the so-called physical discriminant \cite{Shapere:2008zf} of the Seiberg-Witten fibration  (see \cite{Closset:2021lhd} for a recent discussion). The gravitational couplings can also be extracted from the Nekrasov partition function (see section~\ref{sec:NekGlue}), as discussed  in \cite{Manschot:2019pog, Closset:2021lhd, John:2022yql}. Our conjecture is then that the branch cuts ambiguities in $\bA^\chi \bB^\sigma$, that would generally arise from the expressions  \eqref{AB IR expect}, are precisely cancelled by the third factor $\bG^{2\chi+3\sigma}$ in \eqref{ZM4geom}.%
\footnote{If the K\"ahler manifold $\CM_4$ is spin, we have $\chi\in 4\Z$ and  $\sigma\in 16\Z$. Then the  $\bG$ factor is well-defined by itself, and it can be reabsorbed into the flux operator.}

\paragraph{The free hypermultiplet.}
 Let us consider the 5d hypermultiplet on $\CM_4\times S^1$ coupled to a single $U(1)$ vector multiplet with charge $1$, whose partition function we computed in the previous section. In the present CB approach, we simply need to know the effective prepotential and gravitational couplings for the free hypermultiplet. They are given by:
 \be\label{F prepot hyper}
 \CF=-{1\ov (2\pi i)^3} \trilog(Q)~, \qquad \quad
 \cA= 0~, \qquad \quad \cB= -{1\ov 16\pi i}\log(1-Q)~,
 \ee
with $Q\equiv e^{2\pi i a}$. These gravitational couplings will be further discussed from the perspective of the $\Omega$-background in the next section. We then have:
\be
\bA=1~, \qquad\qquad \bB= (1- Q)^{1\ov 8}~.
\ee
 The non-trivial physical discriminant $\Delta^{\rm phys}=(1-Q)$ encodes the singularity on the (extended) Coulomb branch at $Q=1$, where the hypermultiplet becomes massless. 
Taking the extended topological twist with $\varepsilon =-\half +\delta$, and some background flux $\m$, we also have
\bea
&\bG= (1-Q)^{-{\varepsilon^2\ov 2}}= (1-Q)^{-{1\ov 8}} (1-Q)^{-\half \delta (\delta-1)}~, \qquad
&  Z^{\rm flux}_{\CM_4}  = (1-Q)^{-\half (\m+2\varepsilon \k, \m)}~.
\eea
Then, the formula \eqref{ZM4 m geom} gives us:
\be\label{Zhyper M4 S1 from IR}
Z^{\CH}_{\CM_4\times S^1}(a; \varepsilon)_\m = {\left(1\ov 1-Q\right)}^{\chi_h+\half (\m+\delta \k -\k, \m +\delta\k)}~, 
\ee
in perfect agreement with \eqref{Z Hyper M4xS1 gen eps}.

\subsection{KK theory on $\CM_5$: the fibering operator}
We now consider the non-trivial fibration $S^1\rightarrow \CM_5\rightarrow \CM_4$. From the 4d point of view, all 5d fields decompose in KK towers and there is always a distinguished $U(1)_{\rm KK}$  symmetry in 4d corresponding to the momentum along the fifth direction. A non-trivial fibration of the circle amounts to introducing background fluxes for the KK symmetry on $\CM_4$:
\be
 \int_{\S_k} c_1(\CL_{\rm KK}) ={1\ov 2\pi} \int_{\S_k} \FKK = \sum_l {\bf I}_{kl} \p_l~.
\ee
On the CB of the infrared topologically-twisted 4d $\CN=2$ KK theory, the non-trivial fibration of the fifth direction over $\CM_4$ is then encoded in a `flux operator' for $U(1)_{\rm KK}$, which we call the {\it fibering operator}.
 The expression for the latter is easily determined by dimensional analysis. Reinstating dimensions, the mass parameter for $U(1)_{\rm KK}$ is really $ \mu_{\rm KK}=1/\beta$, so that $\CF_{\rm 4d} =\mu_{\rm KK}^2 \CF$ and one finds:
\be
{\d^2 \CF_{\rm 4d} \ov \d \mu_{\rm KK}^2}  = 2\left(1-\amCB^I {\d\ov \d \amCB^I}+ \half \amCB^I\amCB^J {\d^2\ov \d \amCB^I\d \amCB^J} \right)\CF(\amCB)~,
\ee
and:
\be
{\d^2 \CF_{\rm 4d}\ov \d \mu_{\rm KK}\d (\mu_{\rm KK}\amCB^I)}   = \left(1-\amCB^J {\d\ov \d \amCB^J} \right){\d \CF\ov  \d\amCB^I}~.
\ee
For a principal circle bundle with first Chern numbers $\p_k$, we then write down the fibering operator:
\be\label{F full general}
\h \FiberOp_\p(\amCB; \varepsilon) \equiv \FiberOp(\amCB)^{\half (\p, \p)} \, \prod_I \FiberOpK_I(\amCB)^{(\p, \m^I+ \varepsilon^I \mK)}~,  
\ee
where we defined:
\be\label{F fibering def 1}
 \FiberOp(\amCB) \equiv \exp\left(-4\pi i \left(1-\amCB^I {\d\ov \d \amCB^I}+ \half \amCB^I\amCB^J {\d^2\ov \d \amCB^I\d \amCB^J} \right)\CF(\amCB) \right)~,
\ee
and:
\be\label{F fibering def 2}
 \FiberOpK_I(\amCB) \equiv 
  \exp\left(-2\pi i  \left(1-\amCB^J {\d\ov \d \amCB^J}\right){\d \CF \ov  \d\amCB^I}  \right)~.
\ee
The functions \eqref{F fibering def 1} and \eqref{F fibering def 2} are entirely determined by the exact effective prepotential of the 4D $\CN=2$ KK theory, and they are unaffected by the ambiguities \eqref{F ambiguities}. Moreover, while $\FiberOp(\amCB)^\half$ and $ \FiberOpK_I(\amCB)^\half$ suffer from branch-cut ambiguities in general, the product \eqref{F full general} is expected to be unambiguous. This is exactly like in the case of the flavor flux operators discussed above. For spin manifolds, the intersection form is even and the  factors in \eqref{F full general} are individually well-defined, while on a non-spin $\CM_4$ we again conjecture that the fibering operator \eqref{F full general} remains well-defined once the parameters $\varepsilon^I$ are correctly chosen. 

\paragraph{The $\CM_5$ partition function and gauge invariance.} Putting all the contributions together, we arrive at the full $\CM_5$ partition function  at fixed values of the (gauge and flavor) $U(1)_I$ vector multiplets. We have:
\bea\label{ZM5 full}
Z_{\CM_5}(\amCB; \varepsilon)_\m = Z_{\CM_4}^{\rm geom}(\amCB; \varepsilon)\,  Z_{\CM_4}^{\rm flux}(\amCB; \varepsilon)_\m \,  \h \FiberOp_\p(\amCB; \varepsilon)_\m~,
\eea
with:
\bea
&Z_{\CM_4}^{\rm geom}(\amCB; \varepsilon) = \bA(\amCB)^{\chi}\bB(\amCB)^{\sigma}\bG(\amCB;\varepsilon)^{2\chi+3\sigma}~,\cr
& Z_{\CM_4}^{\rm flux}(\amCB; \varepsilon) =  \Flux(\amCB)^{\half (\m+2\varepsilon \mK, \m)}~, \cr
& \h \FiberOp_\p(\amCB; \varepsilon)_\m =  \FiberOpK(\amCB)^{(\p, \m+ \varepsilon \mK)}\,  \FiberOp(\amCB)^{\half (\p, \p)}~.
\eea
Here we suppressed the $I, J$ indices.%
\footnote{
We will also often omit the $\varepsilon$ from the notation, from now on, to avoid clutter.}
 Importantly,   the partition function \eqref{ZM5 full} is fully gauge invariant.  Consider the large gauge transformations along $U(1)_I$:
\be\label{gauge transfo general}
\amCB_J \rightarrow \amCB_J+ \delta_{IJ}~, \qquad
\m_J \rightarrow \m_J+ \delta_{IJ} \p~,
\ee
which we denote by the shorthand $(\amCB, \m) \rightarrow (\amCB+ \delta_I, \m+ \delta_I \p)$. 
 Gauge invariance implies that:
\be\label{ZM5 gauge inv}
Z_{\CM_5}(\amCB+\delta_I)_{\m+ \delta_I \p} = Z_{\CM_5}(\amCB)_\m~. 
\ee
This is indeed the case. To check this, note that $Z_{\CM_4}^{\rm geom}(\amCB)$ is invariant by itself, and that we have the following large gauge transformations of the building blocks:
\bea
&\Flux_{J,K}(\amCB+ \delta_I)& =&\;  \; \Flux_{J,K}(\amCB)~,\cr
& \FiberOpK_J(\amCB+ \delta_I)& =&\;\; \Flux_{I,J}(\amCB)^{-1}  \, \FiberOpK_J(\amCB)~, \cr
& \FiberOp(\amCB+ \delta_I)&=&\; \;  \Flux_{I,I}(\amCB)\,  \FiberOpK_I(\amCB)^{-2} \, \FiberOp(\amCB)~.
\eea
% The property \eqref{ZM5 gauge inv}  directly follows. 

\medskip
\noindent
{\bf Matching the one-loop computation.} 
Consider the free hypermultiplet coupled to a $U(1)$ vector multiplet. By an application of the general formulas \eqref{F fibering def 1}-\eqref{F fibering def 2},  using the hypermultiplet prepotential \eqref{F prepot hyper}, we find:
\be
\FiberOpK = \FiberOpK^\CH(a)~, \qquad \qquad \FiberOp =\FiberOp^\CH(a)~,
\ee
in terms of the meromorphic functions introduced in \eqref{H K and F hyper}, so that:
\bea\label{hyper fiber full}
\h\FiberOp_\p(a; \varepsilon)_\m =
\FiberOp^\CH_\p(a)_\m \equiv \exp\Bigg( -{(\p,\p)\ov 4\pi^2}\trilog(e^{2\pi i a})
-{(\p, \p) a -(\p, \m+\varepsilon \k)\ov 2\pi i} \dilog(e^{2\pi i a}) \\
\qquad\qquad \qquad
- {a((\p,\p)a - 2(\p, \m+\varepsilon \k))\ov 2} \log(1-e^{2\pi i a}) 
\Bigg)~.
\eea
By multiplying with \eqref{Zhyper M4 S1 from IR}, we obtain the full partition function of a free hypermultiplet on $\CM_5$. 
 This matches precisely with the direct one-loop computation of section~\ref{sec:hyperOneloopReg}.

%%%%
\subsection{Higher-spin state contributions}\label{subsec:hs states and GV}
The prepotential of many five-dimensional superconformal field theories compactified on $S^1$ admits an expansion in terms of 5d BPS states:%
\footnote{Here we ignore some possible `classical' terms, which would contribute additional factors to the CB partition function.}
\be\label{CF expansion GV}
\CF= -{1\ov (2\pi i)^3}\sum_{\bbeta}\sum_{j_l, j_r} c_0^{(j_l, j_r)} N_{j_l, j_r}^\bbeta\, \trilog(Q^\bbeta)~.
\ee
Here, in keeping with common notation, we denote by $\bbeta_I\equiv q_I$ the charges under the $U(1)^{r+r_F}$ symmetry on the extended Coulomb branch, with
 $Q^\bbeta\equiv \prod_I Q_I^{\bbeta_I}$ and $Q_I\equiv e^{2\pi i \amCB^I}$,  
and with the universal coefficients $c_0^{j_l, j_r}$ as in \eqref{cAB0 coefs 1}, namely
\be
c_0^{(j_l, j_r)}  =(-1)^{2j_l+2j_r} (2j_l+1)(2j_r+1)~.
\ee
In the context of geometrical engineering of 5d SCFTs in M-theory on a toric threefold, the theory-dependent non-negative integers $N_{j_l, j_r}^\bbeta$ in \eqref{CF expansion GV}  are the refined Gopakumar-Vafa invariants~\cite{Hollowood:2003cv, Iqbal:2007ii}, as we will review momentarily. The expansion \eqref{CF expansion GV 2} can be written  simply as:
 \be\label{CF expansion GV 2}
\CF= -{1\ov (2\pi i)^3}\sum_{\bbeta} d_\bbeta\, \trilog(Q^\bbeta)~,
\ee
 with
 \be
 d_\bbeta\equiv \sum_{j_l, j_r} (-1)^{2j_l+2j_r} (2j_l+1)(2j_r+1)  N_{j_l, j_r}^\bbeta~,
 \ee
 the effective number of 5d BPS states of charge $\bbeta$. Given the expression \eqref{CF expansion GV 2} for the prepotential, we can directly compute the CB fibering operator in terms of the hypermultiplet result \eqref{hyper fiber full}, at least formally, as a product over the charge lattice:
\be 
\FiberOp_\p(\amCB)_\m =\prod_\bbeta \left[ \FiberOp^\CH_\p(\bbeta(\amCB))_{\bbeta(\m)} \right]^{d_\bbeta}~. 
\ee
Thus, the higher-spin contributions are the same as for $d_\bbeta$ hypermultiplets, in perfect agreement with the second line of \eqref{ZM5jljr}. 

One can similarly expand the flux operators. To obtain the full CB partition function,  we should also consider the contribution of higher-spin particles to the gravitational couplings $\CA$ and $\CB$.  In section~\ref{subsec:GVexpNonEqHS} below, we will show that:
\bea\label{CA and CB refined GV expansion}
& \CA = {1\ov 2\pi i}\sum_{\bbeta}\sum_{j_l, j_r} c_\CA^{(j_l, j_r)} N_{j_l, j_r}^\bbeta\,\log(1- Q^\bbeta)~,\\
& \CB = {1\ov 2\pi i}\sum_{\bbeta}\sum_{j_l, j_r} c_\CB^{(j_l, j_r)} N_{j_l, j_r}^\bbeta\,\log(1- Q^\bbeta)~,
\eea
when expanding in terms of the refined GV invariants, with the coefficients $c_{\CA, \CB}^{(j_l, j_r)}$ given in \eqref{cAB0 coefs 1}. One then easily checks that the CB partition function on $\CM_5$ can be written entirely in terms of the refined GV invariants of the 5d theory, as:
\bea\label{ZM5 full product}
Z_{\CM_5}(\amCB)_\m  =\prod_{\bbeta}\prod_{j_l, j_r} \left[ Z_{\CM_5}^{(j_l,j_r)}(\bbeta(\amCB))_{\bbeta(\m)} \right]^{N_{j_l, j_r}^\bbeta}~, 
\eea
using the explicit expression \eqref{ZM5jljr}. The expression \eqref{ZM5 full product} is the partition function that we would obtain by combining the localisation results of section~\ref{sec:oneloopdet} with the assumption that the full partition function can be obtained as a product over the 5d BPS states.  What we have just shown is that this factorisation is consistent with the low-energy approach of the present section. In fact, the factorisation \eqref{ZM5 full product}   is  simply equivalent to the expansions~\eqref{CF expansion GV} and \eqref{CA and CB refined GV expansion} of the low-energy effective couplings. 

\subsubsection{Spin/charge constraints on the 5d BPS spectrum}\label{subsubsec:spincharge}
To conclude this section, let us mention an important constraint that arises when trying to put a general 5d SCFT on our supersymmetric $\CM_5$, for a generic choice of our geometric background. Namely, every BPS particle of spin $(j_l, j_r)$ and charge $\bbeta$, at any point on the 5d Coulomb branch, should be coupled consistently to the base manifold $\CM_4$, {\it at the same time}. Given the CB (gauge and flavor) symmetry $\prod_I U(1)_I$, we need to choose the $\varepsilon$ parameters $\varepsilon^I$, which define the spin$^c$ connections as in \eqref{choose epsI}, in such a way that 
\be\label{spincharge const}
\half + j_l +j_r+ \bbeta(\varepsilon)  \in \Z~, \qquad \forall \, j_l, j_r, \bbeta \;\, \text{with}\; \, N_{j_l, j_r}^\bbeta\neq 0~.
\ee
Here, $ \bbeta(\varepsilon)\equiv  q_I \varepsilon^I$ is the $\varepsilon$ parameter of this particular BPS particle. For any fixed $j_l, j_r, q_I$, this condition is equivalent to the requirement that the vector bundle \eqref{def Vnetc} be well-defined on any K\"ahler manifold $\CM_4$. (Of course, if $\CM_4$ is spin, then this condition is not necessary.)  

Note that, once we fix $\varepsilon^I$, the condition \eqref{spincharge const} only holds if the spin and  electric charges of the BPS states are appropriately correlated (mod 2). The theories for which this holds obey a ``spin/charge'' relation, which is somewhat reminiscent of the 3d spin/charge relation discussed in  \cite{Seiberg:2016rsg} for strongly-coupled electrons; this spin/charge relation  for 4d $\CN=2$ theories was also discussed in~\cite{Cordova:2018acb}.

\section{Fibering operators from gluing Nekrasov partition functions}\label{sec:NekGlue}

In this section, we give a complementary perspective on the CB partition function  \eqref{ZM5 full}, including the fibering operator, by building up $\CM_5$ as a toric gluing of $\C^2\times S^1$ patches, in the case when the base $\CM_4$ is a toric four-manifold. We can then obtain the CB partition function $Z_{\CM_5}$ as an appropriate gluing of 5d Nekrasov partition functions, generalising well-known results for the five-sphere \cite{Lockhart:2012vp, Kim:2012qf, Imamura:2012efi, Qiu:2013pta}.

\subsection{Nekrasov partition functions and refined topological strings}
Partition functions of 4d $\CN=2$ field theories on toric four-manifolds can be computed in terms of the partition functions on toric patches $\C^2$ \cite{Nekrasov:2003vi}, and similarly for the 5d uplift. On each patch, one considers the so-called Nekrasov partition function on $\C^2\times S^1$ with the $\Omega$-background, which is obtained by the identification
\be\label{def OmBack5}
    (z_1, z_2, x_5) \sim (e^{2\pi i \tau_1} \, z_1, e^{2\pi i \tau_2} \, z_2, x_5 +\beta )~,  
\ee
where $(z_1, z_2, x_5)$ are the $\C^2\times S^1$ coordinates, and we also introduced the dimensionless $\Omega$-deformation parameters:
\be
\tau_1= \beta \epsilon_1~, \qquad \qquad
\tau_2= \beta \epsilon_2~,
\ee
not to be confused with the gauge couplings. 
The $\Omega$-background is a $U(1)^2$-equivariant deformation of the topological twist which effectively compactifies the non-compact $\C^2$, with a finite `volume' $1/(\tau_1 \tau_2)$. Using topological invariance, one can equivalently consider a background geometry $D^2_{\tau_1}\times D^2_{\tau_2}\times S^1$, where $D^2_{\tau_{1,2}}$ are elongated cigars fibered over $S^1$ according to \eqref{def OmBack5}. Formally, we can assign the following Euler characteristic and signature to the $\Omega$-deformed $\C^2$ geometry \cite{Nakajima:2003uh}:
\be
\chi(\C^2) = \tau_1 \tau_2~, \qquad \quad 
\sigma(\C^2)= {\tau_1^2+\tau_2^2\ov 3}~.
\ee
Similarly, the first Chern class of the canonical line bundle over $\C^2$ is formally given by:
\be\label{cK tau12}
c_1(\CK_{\C^2}) = \tau_1+\tau_2~.
\ee
Note that we have $c_1(\CK)^2= 2\chi+3\sigma= (\tau_1+\tau_2)^2$. The partition function of a 5d $\CN=1$ theory on $\C^2\times S^1$ is known as the (K-theoretic) Nekrasov partition function \cite{Nakajima:2005fg, Gottsche:2006bm}, and it will be  denoted by:
\be
    Z_{\bC^2\times S^1}(\amCB,\tau_1, \tau_2)~.
\ee
Here, the CB parameters $\amCB^I$ arise as Dirichlet boundary conditions for the $U(1)_I$ vector multiplets at infinity. Whenever we have a four-dimensional gauge-theory interpretation, the Nekrasov partition function admits an expansion in some instanton counting parameter $ \frak{q}= e^{2\pi i \tau_{\rm UV}}$, according to:
\be
    Z_{\bC^2\times S^1}(\amCB,\tau_1, \tau_2) = Z^{\rm cl}_{\bC^2\times S^1}(\amCB,\tau_1, \tau_2) Z^{\rm pert}_{\bC^2\times S^1}(\amCB,\tau_1, \tau_2) \left( 1 + \sum_k \frak{q}^k Z^{\rm Nek}_k(\amCB,\tau_1,\tau_2) \right)~.
\ee
See {\it e.g.} \cite{Nakajima:2003uh, Taki:2007dh, Tachikawa:2014dja} for reviews of instanton counting, and \cite{Hayashi:2017jze, Kim:2019uqw, Kim:2020hhh} for some more recent advances. When considering the Donaldson-Witten twist, we are interested in the non-equivariant limit $\tau_{1,2}\rightarrow 0$. In that limit, the partition function diverges in a way which precisely encodes the low-energy couplings $\CF$, $\CA$ and $\CB$ of the CB theory, namely \cite{Nekrasov:2003vi, Nakajima:2003pg}:
\be \label{NonEquivariantLimit}
   % \hspace{-0.5cm}
    \log Z_{\bC^2\times S^1}(\amCB,\tau_1, \tau_2) \approx -{2\pi i \ov \tau_1\tau_2} \left( \CF(\amCB) + (\tau_1 + \tau_2)H(\amCB) + \tau_1 \tau_2 \CA(\amCB) + {\tau_1^2 + \tau_2^2 \ov 3}\CB(\amCB)\right)~.
\ee
The term $H(\amCB)$ in \eqref{NonEquivariantLimit} is allowed by dimensional analysis, but it does not represent an additional effective coupling. In fact, for the $U(1)^2$-equivariant DW twist, we must have $H(\amCB)=0$ because there are no supergravity background fields that could contribute to this coupling (see {\it e.g.} \cite{Katz:2020ewz}). More generally, $H$ is fully determined in terms of $\CF$ by the choice of background $U(1)$ gauge fields, as we will see momentarily.

\subsubsection{Nekrasov partition functions for the extended topological twist}
When patching together Nekrasov partition functions into compact four- or five-manifolds, we will have to be careful about whether the base $\CM_4$ is spin or not. In general, we should consider the possibility of an extended DW twist on $\C^2\times S^1$, with parameters $\varepsilon^I$. We propose that this corresponds to twisting the background gauge fields at infinity according to:
\be
\amCB^I \rightarrow \amCB^I + \varepsilon^I (\tau_1+\tau_2)~,
\ee
in agreement with the identification \eqref{cK tau12}. 
Namely, the Nekrasov partition function for the extended DW twist is simply given by:
\be\label{ZC2 with eps}
Z_{\bC^2\times S^1}(\amCB,\tau_1, \tau_2; \varepsilon) = Z_{\bC^2\times S^1}(\amCB+ \varepsilon (\tau_1+\tau_2),\tau_1, \tau_2)~.
\ee
Hence, the non-equivariant limit of the partition function reads:
\bea    \label{NonEquivariantLimitExtendedTwist}
 &\log Z_{\bC^2\times S^1}(\amCB,\tau_1, \tau_2; \varepsilon) \approx\\
 &
  -{2\pi i \ov \tau_1\tau_2} \left( \CF(\amCB) + (\tau_1 + \tau_2)H(\amCB; \varepsilon) + \tau_1 \tau_2 \CA(\amCB) + {\tau_1^2 + \tau_2^2 \ov 3}\CB(\amCB) + (\tau_1+\tau_2)^2 \CG(\amCB; \varepsilon) \right)~,
\eea
with: 
\be\label{defH and CG eps}
H(\amCB; \varepsilon)= \varepsilon^I {\d \CF\ov \d \amCB^I}~, \qquad\qquad
\CG(\amCB; \varepsilon)= \half \varepsilon^I  \varepsilon^J  {\d^2 \CF(\amCB) \ov \d \amCB^I \d \amCB^J}~.
\ee
This parameterisation of the non-equivariant limit naturally parallels the discussion of section~\ref{subsec:fluxop} for the CB effective couplings, with $\CG$ being exactly as in \eqref{flux K mixing}.

\subsubsection{Gluing transformations in the non-equivariant limit}
We wish to glue together Nekrasov partition functions from different patches to obtain the CB partition function of a compact five-manifold $\CM_5$. The most general gluing rules between two patches, for our purposes,  are:%
\footnote{More general gluings could be considered (similarly to the 3d computations in \protect\cite{Closset:2018ghr}), but this would go beyond the class of principal circle bundles that we consider in this paper.}
\be\label{new vars for Znek}
\tau_i \rightarrow \t \tau_i\equiv {\h\tau_i\ov \gamma}~,\qquad \qquad \amCB\rightarrow \t \amCB \equiv {\amCB+  \h\n\ov \gamma}~,
\ee
where we defined:
\be
\h \tau_i \equiv \alpha_i \tau_1 + \beta_i \tau_2~, \qquad \gamma\equiv \gamma_1 \tau_1 + \gamma_2 \tau_2 + 1~,
\ee
for some integers $\alpha_i$, $\beta_i$, $\gamma_i$ (with $i=1,2$), and:
  \be
  \h\n  \equiv \h \n_1 \h\tau_1 + \h \n_2 \h\tau_2~.
  \ee 
  The parameters $\h\n_i$ allow us to introduce background fluxes. 
 To  recover the DW twist on $\CM_5$, we need to consider the non-equivariant limit of the Nekrasov partition function in the variables \eqref{new vars for Znek}. In the limit $\tau_i \rightarrow 0$ and using   the ansatz \eqref{NonEquivariantLimit},  one finds:
\bea\label{general noneq lim}
 &\log Z_{\bC^2\times S^1}(\t\amCB,\t\tau_1, \t\tau_2) \approx\\
 &
  -{2\pi i \ov \h\tau_1\h \tau_2} \Bigg( \CF   + (\h\tau_1 + \h\tau_2)H + 2(\gamma-1)\left(\CF - \amCB^I  \d_I\CF\right)+\h \tau_1 \h\tau_2 \CA(\amCB) + {\h\tau_1^2 + \h\tau_2^2 \ov 3}\CB(\amCB)  \\
&\qquad + (\gamma-1)^2 \left(\CF - \amCB^I  \d_I\CF+\half  \amCB^I\amCB^J  \d_I\d_J\CF \right) +(\gamma-1)(\h \tau_1 +\h\tau_2) \left( H - \amCB^I \d_I H \right) \\
&\qquad  + \h \n^I \left( \gamma \d_I \CF + (\h\tau_1+\h \tau_2) \d_I H -(\gamma-1) \amCB^J \d_I \d_J\CF  \right) + \half \h \n^I \h \n^J \d_I \d_J\CF  \Bigg)~,
\eea
where we used the notation $\d_I=  {\d \ov \d {\amCB^I}}$.  When considering the extended topological twist as in \eqref{ZC2 with eps}, the general non-equivariant limit is obtained from \eqref{general noneq lim} through the substitution:
\be \label{recovering extended twist}
H \rightarrow \varepsilon^I \d_I \CF~, \qquad\qquad \CA \rightarrow \CA + 2 \CG~,
\qquad \qquad \CB\rightarrow \CB+ 3\CG~,
\ee
with $\CG$ defined in \eqref{defH and CG eps}. 

\subsubsection{Refined topological string partition function and refined GV invariants}
We are particularly interested in the 5d SCFTs that can be engineered at canonical singularities in M-theory \cite{Morrison:1996xf, Intriligator:1997pq}; see {\it e.g.} \cite{Jefferson:2018irk, Apruzzi:2019opn,  Apruzzi:2019enx, Closset:2018bjz, Closset:2020scj} for recent studies. Then, the Coulomb-branch low-energy effective theory on the $\Omega$-background is obtained by considering the low-energy limit of M-theory on:
\be
\C^2  \times S^1 \times \t \MG~,
\ee
with the $\Omega$-background turned on along $\C^2$. Here, $\t \MG$ denotes the crepant resolution of a threefold canonical singularity $\MG$. Let us further choose $\MG$ and its resolution to be toric. Then, the Nekrasov partition function of the five-dimensional theory can be computed using the refined topological vertex formalism \cite{Aganagic:2003db, Iqbal:2007ii}. 

In the geometric-engineering picture, the various Coulomb-branch parameters of the 5d theory are now  K\"ahler parameters of the crepant resolution $\t\MG$. 
%; see {\it e.g.} \cite{Closset:2018bjz} for a detailed discussion. 
%
In keeping with standard notation, we use the fugacities $q$, $t=p^{-1}$ and $Q$, defined as:
\be 
    q\equiv e^{2\pi i\tau_1}~, \qquad\qquad  p= t^{-1}\equiv  e^{2\pi i \tau_2}~,\qquad\qquad
    Q^\bbeta\equiv  e^{2 \pi i \int_\bbeta (B + i  J)}= e^{2\pi i \bbeta(\amCB)}~,\
\ee
where  $\bbeta \cong [\CC] \in H_2(\t \MG, \Z)$  denote the homology class of any effective curve in $\t \MG$, and $B + i  J$ is the complexified K\"ahler form in Type IIA string theory.

The Nekrasov partition function of the 5d theory is expected to be equivalent to the refined topological string partition function for the threefold $\t\MG$ \cite{Hollowood:2003cv, Iqbal:2007ii}.  In the M-theory approach, the 5d BPS states arise as M2-branes wrapped over curves. One can then write the Nekrasov partition function 
 as a product over these BPS states, of electric charge $\bbeta$ and spin $(j_l, j_r)$ \cite{Iqbal:2007ii, Lockhart:2012vp}:
\be\label{refined top string gen}
Z_{\C^2\times S^1}(\amCB, \tau_1, \tau_2) = \prod_\bbeta \prod_{j_l, j_r=0}^\infty  \left[  {\bf Z}_{\C^2 \times S^1}^{j_l, j_r}(Q^\bbeta, q, p)\right]^{N^\beta_{j_l, j_r}}~,
\ee
where the non-negative integers $N^\beta_{j_l, j_r}$ are the refined Gopakumar-Vafa invariants.
Here,  the higher-spin particles contribute as:
\be\label{Zjljr Omegaback}
  {\bf Z}_{\C^2 \times S^1}^{j_l, j_r}(Q, q, p) \equiv   \prod_{m_l=- j_l}^{j_l}   \prod_{m_r=- j_r}^{j_r}  (Q  q^{\half +  m_r+ m_l} p^{\half+m_r- m_l}; q, p)_\infty^{(-1)^{1+2j_l+2j_r}}~,
\ee
which is written in terms of the double-Pochhammer symbol:
\be\label{defqqPsym}
    \left(x;q,p\right)_{\infty} \equiv \prod_{j,k = 0}^{\infty} (1-xq^j p^k)~.
\ee
Note that the definition \eqref{defqqPsym} is only valid for ${\rm Im}(\tau_i)>0$. This can be analytically continued to  $|q|\neq 1$, $|p|\neq 1$ \cite{2003math......6164N}, which gives us the formal identities:
\be \label{doublePocc continuation}
     (x;q^{-1},p)_{\infty} = (x q;q,p)^{-1}_{\infty}~,\qquad 
          (x;q,p^{-1})_{\infty} = (x p; q,p)^{-1}_{\infty}~.
\ee
 The expression \eqref{refined top string gen} gives us the Nekrasov partition function for the ordinary $\Omega$-deformed DW twist, and we can also obtain the extended DW twist expression by the substitution $Q \rightarrow Q (qp)^\varepsilon$. For completeness, let us also mention that the unrefined topological string limit corresponds to setting $t=p^{-1}= q$, giving us:
 \be\label{ top string gen}
Z_{\rm top}(Q, q)=\prod_\bbeta \prod_{j_l=0}^\infty  \t  {\bf Z}_{\rm top}^{j_l}(Q^\bbeta, q)^{N^\beta_{j_l}}~,
\ee
with:
\be
\t {\bf Z}_{\rm top}^{j_l}(Q, q)  =   \prod_{m_l=- j_l}^{j_l}  \prod_{k=1}^\infty  \left[\left(1-Q  q^{k + 2m_l}  \right)^{k}\right]^{(-1)^{2 j_l}}~,
\ee
in terms of the unrefined GV invariants  $N^\beta_{j_l} \equiv  \sum_{j_r} (-1)^{2j_r} (2 j_r+1) N^\beta_{j_l, j_r}$. The (refined) GV invariants have to be computed explicitly, for any given toric threefold $\t\MG$, for instance using the (refined) topological vertex formalism \cite{Iqbal:2007ii}.

\subsubsection{Quantum trilogarithm and GV expansion in the non-equivariant limit}
\label{subsec:GVexpNonEqHS}
It is interesting to consider the non-equivariant limit of the expression \eqref{refined top string gen}. We find it useful to introduce the `quantum trilogarithm' defined as:
\be \label{quantum trilog def}
   \trilog (x;q,p) \equiv  -\log(x;q,p)_{\infty} = \sum_{n=1}^{\infty} {x^n \ov n} {1 \ov (1-q^n)(1-p^n)}~.
\ee
In the small-$\tau_i$ limit, it admits an asymptotic expansion:
\be \label{Quantum Trilog Bernoulli}
    \trilog(x; q,p) = \sum_{n,m =0}^{\infty} {(-1)^{n+m} \ov n!\, m!} B_{n} B_{m} (2\pi i \tau_1)^{n-1}(2\pi i \tau_2)^{m-1} {\rm Li}_{3-n-m}(x)~,
\ee
where $B_n$ are the Bernoulli numbers.%
\footnote{With the convention that $B_0=1$ and $B_1=\half$.}
We then have:
\be \label{Quantum Trilog Expansion}
    \trilog (x;q,p) \approx {1 \ov (2 \pi i)^2 \tau_1 \tau_2 } \trilog(x) - {1 \ov 4 \pi i} {\tau_1 + \tau_2 \ov \tau_1 \tau_2} \dilog(x) - {1 \ov 12}\left(3+ {\tau_1 \ov \tau_2} + {\tau_2 \ov \tau_1}\right) \log(1-x)~.
\ee
For a massive hypermultiplet with the extended DW twist, we have
\be
\log Z_{\C^2\times S^1}^\CH(\amCB, \tau_1, \tau_2; \varepsilon) =   \trilog (Q (qp)^{\half +\varepsilon} ;q,p)~.
\ee
Setting $\varepsilon=0$ for simplicity,  the $\tau_i\rightarrow 0$ limit reads:
\be
\log Z_{\C^2\times S^1}^\CH(\amCB, \tau_1, \tau_2) \approx -{2\pi i \ov  \tau_1 \tau_2 }\left(
-{1 \ov (2 \pi i)^3} \trilog(Q)   - \left({\tau_1^2+\tau_2^2\ov 3} \right) {1 \ov 16 \pi i}\log(1-Q)\right)~,
\ee
from which we can read off \eqref{F prepot hyper}. More generally, for the equivariant DW twist ($\varepsilon=0$), we have the refined GV expansion:
\be
\log  Z_{\C^2\times S^1} = \sum_\bbeta\sum_{j_l, j_r} N_{j_l, j_r}^\bbeta \log  {\bf Z}_{\C^2 \times S^1}^{j_l, j_r}(Q^\bbeta, q, p)~,
\ee
 with:
 \be
 \log {\bf Z}_{\C^2 \times S^1}^{j_l, j_r}(Q, q, p) =(-1)^{2j_l +2j_r}  \sum_{m_l=- j_l}^{j_l}   \sum_{m_r=- j_r}^{j_r}  \trilog(Q  q^{\half +  m_r+ m_l} p^{\half+m_r- m_l}; q, p)~.
 \ee
By taking the small-$\tau_i$ limit and comparing to \eqref{NonEquivariantLimit}, one can extract the contribution of a spin-$(j_l, j_r)$ particle (of unit electric charge, $\bbeta=1$) to the low-energy effective couplings. By a straightforward computation, one finds:
\bea
&\CF^{j_l, j_r}= -{c_0^{(j_l, j_r)}\ov (2\pi i)^3}\trilog(Q)~, &&\\
&\CA^{j_l, j_r} = {1\ov 2\pi i} c_\CA^{(j_l, j_r)}  \,\log(1- Q)~, \qquad
&& \CB^{j_l, j_r} = {1\ov 2\pi i} c_\CB^{(j_l, j_r)}  \,\log(1- Q)~,
\eea
with the coefficients:
\bea    \label{HigherSpin Coefficients}
& c_0^{(j_l, j_r)} =(-1)^{2j_l+2j_r} (2j_l+1)(2j_r+1)~,\\
& c_\CA^{(j_l, j_r)} = (-1)^{2j_l+2j_r} (2j_l+1)(2j_r+1)  {j_r(j_r+1)- j_l(j_l+1) \ov 3}~, \\
& c_\CB^{(j_l, j_r)} = (-1)^{2j_l+2j_r} (2j_l+1)(2j_r+1)\left(-{1\ov 8}+{j_l(j_l+1)+ j_r(j_r+1) \ov 2}\right)~, 
\eea 
exactly as anticipated in section~\ref{subsec:hs states and GV}, and in perfect agreement with the index  computation of section~\ref{subsec:HSM5}. Note also that, for the extended topological twist, one has the additional terms $H$ and $\cG$ in \eqref{NonEquivariantLimitExtendedTwist}, namely:
\bea
    H^{j_l,j_r} = -{c_0^{(j_l, j_r)}\ov (2\pi i)^2}\varepsilon\, \dilog(Q)~, \qquad \qquad \cG^{j_l,j_r} = {c_0^{(j_l, j_r)}\ov 4\pi i}\varepsilon^2\log(1-Q)~,
\eea
according to \eqref{defH and CG eps}.

%%%%%%%%%%%%%%%%%%%%%%%%%%%%
\subsection{Gluing rules for circle fibrations}
Let us now consider the explicit gluing of Nekrasov partition functions to obtain the circle-fibered five-manifold:
\be
S^1 \rightarrow \CM_5 \rightarrow \CM_4~,
\ee
where $\CM_4$ is a toric four-manifold. 
 For definiteness, we will mostly focus on the  case when $\CM_4$ is one of the five toric Fano surfaces, $\mathbb{P}^2$, $\bF_0\cong \bP^1\times \bP^1$, or $dP_n$ (the blow-up of $\bP^2$ at $n$ points) with $n \leq 3$, whose Euler characteristic and signature are:
\bea    \label{toric Kahler 4-manifolds}
\begin{tabular}{|c||c|c|c|c|c|}
\hline
     & $\bP^2$ & $\bF_0$ & $dP_1$ & $dP_2$ & $dP_3$  \\ \hline 
    $\chi$ & 3 &  4 & 4 & 5 & 6 \\
    $\sigma$ & 1 & 0 & 0 & $-1$ & $-2$ \\ \hline
\end{tabular}
\eea
Let us first review the case of a trivial fibration, $\CM_5 = \CM_4\times S^1$, before considering the case of a non-trivial fibration.

%
%In this section we review the proposal of \cite{Lockhart:2012vp} for the partition function of $S^5$, expressed in terms of Nekrasov partition functions. We rederive the gluing of the various factors of the instanton partition function from the five-sphere metric, based on the `gluing' of the three coordinate patches of the base four-manifold $\bP^2$.
%
%We further apply this logic lens spaces $L(p;1,1,1)$, which are circle fibrations over $\bP^2$ with non-trivial Chern numbers. Moreover, we consider non-trivial fibrations over $\bP^1 \times \bP^1$, for which we propose similar gluings of the Nekrasov partition function. K\"ahler metrics for the higher del Pezzo surfaces are not known, but given the 5d fibering operator we are able to propose similar gluings for fibrations over the toric $dP_n$ surfaces.

%%%%%%%%%%%%%%%%5
\subsubsection{The $\cM_4 \times S^1$ partition function}

%Let us first consider the partition function on $\cM_4 \times S^1$, with $\cM_4$ a toric K\"ahler 4-manifolds. 

It was conjectured in \cite{Nekrasov:2003vi} that the $\Omega$-deformed Coulomb-branch partition function on a toric manifold $\mathcal{M}_4$ can be obtained by gluing Nekrasov partition functions for each fixed point of the toric action. The full partition function is then obtained, in principle, by a particular contour integral over the CB parameters, together with a sum over fluxes, both of which one should determine by a more careful analysis. This approach was further developed in \cite{Bawane:2014uka, Bershtein:2015xfa, Bershtein:2016mxz}, and generalised to 5d theories on $\CM_4\times S^1$ in \cite{Hosseini:2018uzp}. The full partition function then reads:
\be \label{M4xS1 conjecture}
    {\bf Z}_{\CM_4 \times S^1} = \sum_{\mathfrak{n}_l} \oint d\hspace{0.7pt}\amCB\prod_{l=1}^{\chi(\CM_4)} Z_{\bC^2\times S^1} (\amCB+\tau_1^{(l)}\mathfrak{n}_l + \tau_2^{(l)}\mathfrak{n}_{l+1}, \tau_1^{(l)}, \tau_2^{(l)})~,
\ee
with $\frak{n}_l$ being fluxes associated with the toric divisors $D_l\subset \CM_4$, corresponding to a line bundle:
\be
L= \CO(-\sum_l \n_l D_l)~,
\ee
over $\CM_4$. Note that there are $\chi(\CM_4)$ toric divisors, with 2 linear relations amongst them. The previously defined $U(1)_I$ background fluxes $\m^I$ are then given by:
\be \label{definition fluxes}
  \sum_{k=1}^{\chi-2}  \m_k^I [\S_k]  = - \sum_{l=1}^{\chi} \frak{n}_l^I [D_l]~.
\ee
The equivariant parameters $\tau^{(l)}_i$ are linear combinations of $\tau_{1,2}$, which we shall comment on momentarily. The non-equivariant limit of the integrand of \eqref{M4xS1 conjecture} can be obtained by a direct computation using \eqref{general noneq lim} (with $\gamma=1$), wherein all divergent pieces cancel out between patches, leaving us with a finite quantity. One finds:
\bea  \label{NonEquivariantLimitFullZ M4*S1}
    \log Z_{\cM_4\times S^1}(\amCB) \approx -2\pi i & \Bigg(\chi \cA(\amCB) +  \sigma \cB(\amCB) + \left(\sum  D_l \right) \cdot\left(\sum \frak{n}^I_l D_l \right) {\d H(\amCB) \ov \d \amCB^I} + \\
    & \hspace*{2.5cm}  +\frac{1}{2}\left(\sum \frak{n}^I_l D_l \right) \cdot \left(\sum \frak{n}^J_l D_l \right){\d^2\mathcal{F}(\amCB) \ov \d \amCB^I \d \amCB^J}   \Bigg)~.
\eea
For the DW twist, we have $H=0$ and therefore:
\be
  Z_{\cM_4\times S^1}(\amCB) = \bA(\amCB)^\chi \bB(\amCB)^\sigma  \Flux(\amCB)^{\half (\m, \m)}~,
\ee
in terms of the quantities defined in section~\ref{subsec:fluxop}.  
This reproduces and generalises the results of \cite{Bawane:2014uka, Hosseini:2018uzp, Bonelli:2020xps}.
 More generally, as  explained at length in previous sections, we should consider the extended DW twist with $\varepsilon \neq 0$, in which case we should substitute \eqref{recovering extended twist} into \eqref{NonEquivariantLimitFullZ M4*S1}. Then, using the fact that $\CK_{\CM_4}\cong \CO(-\sum_l D_l)$, we find:
\be
  Z_{\cM_4\times S^1}(\amCB; \varepsilon) = \bA(\amCB)^\chi \bB(\amCB)^\sigma \bG(\amCB; \varepsilon)^{2\chi + 3\sigma}  \Flux(\amCB)^{\half (\m+2 \varepsilon \k, \m)}~,
\ee
which exactly reproduces the formula \eqref{ZM4 m geom}.  Next, let us explain how the Nekrasov partition functions have been glued together.

%%%%%%%%%%%%%%%%%%%%%%%%%%%%%%%%%%%%%%%
\medskip
\noindent {\bf Equivariant parameters.}  The patch-dependent equivariant parameters $\tau^{(l)}$ in \eqref{M4xS1 conjecture} are determined by the toric data, as follows (see \textit{e.g.} \cite{Hosseini:2018uzp} for a more detailed discussion). A compact toric surface $\cM_4$ is described by a set of vectors $\vv{n}_l \in \mathbb{Z}^2$, with $l= 1, \ldots, d$, which we order such that $\vv{n}_l$ and $\vv{n}_{l+1}$ are adjacent (with $\vv{n}_{d+1}\equiv \vv{n}_1$). Each such vector is associated to a non-compact divisor $D_l$.

Each pair of vectors $(\vv{n}_l, \vv{n}_{l+1})$ defines a two-dimensional cone $\sigma_l$, to which we can associate an affine variety $V_{\sigma_l}$. The construction is based on the dual cone $\hat{\sigma}_l$ generated by the primitive integer vectors $\vv{m}_{l}$ and $\vv{m}_{l+1}$, which are orthogonal to $\vv{n}_{l+1}$ and $\vv{n}_l$, respectively, and point inwards inside $\sigma_l$. The set of holomorphic functions on $V_{\sigma_l}$ is given by monomials $z_1^{\mu_1}z_2^{\mu_2}$, for all $\vv{\mu} \in \hat{\sigma_l}$. Then, since $V_{\sigma_l}\cong \C^2$ by assumption that $\CM_4$ be smooth, the local coordinates on $V_{\sigma_l}$ can be chosen as:
\be \label{toric M4 local coords}
\rho_1^{(l)} = z_1^{m_{l,1}}z_2^{m_{l,2}}~, \qquad \rho_2^{(l)} = z_1^{m_{l+1,1}}z_2^{m_{l+1,2}}~.
\ee
The toric variety $\cM_4$ is obtained by gluing together the affine varieties $V_{\sigma_l}$, by identifying dense open subsets associated with the common vectors spanning the neighbouring cones $\sigma_l$. 
%We will implement a similar logic in the next section, where we consider the five-sphere and its coordinate patches.

Due to the $\Omega$-background, the $\cM_4 \times S^1$ partition function only receives contribution from the $\chi(\CM_4)$ fixed points of the toric action. There is then a single contribution from each chart $V_{\sigma_l}$ of $\cM_4$, as written explicitly in \eqref{M4xS1 conjecture}. Thus, the equivariant parameters will `transform' under the $(\bC^*)^2$ action similarly to \eqref{toric M4 local coords}, leading to:
\be\label{equivParamGenFormula}
    \tau_1^{(l)} = \vv{\tau}\cdot \vv{m}_l~, \qquad \quad \tau_2^{(l)} = \vv{\tau}\cdot \vv{m}_{l+1}~.
\ee
Furthermore, the (background) gauge fluxes $\n$ appearing in \eqref{M4xS1 conjecture} are similarly
local contributions from each patch. Thus, the Coulomb branch VEVs change according to:
\be
    \amCB^{(l)} = \amCB + \tau_1^{(l)} \frak{n}_l + \tau_2^{(l)} \frak{n}_{l+1}~.
\ee
 At this stage, it is natural to wonder how this procedure can be modified to account for a non-trivial $U(1)_{\rm KK}$ flux, leading to the non-trivial fibration $\CM_5$. % We will consider this in the next subsection, after checking in detail some non-trivial fibrations over $\bP^2$ and $\bF_0$.
 % Let us first consider some examples of trivial fibrations $\cM_4 \times S^1$.
Before exploring this, let us briefly consider a couple of examples of toric gluings for $\CM_4\times S^1$.

%%%%%%%%%%%%%%%%%%%%%%%%%%%%%%%%%%%%%%%
\medskip

\noindent  {\bf The $\cM_4 = \bP^2$ case.} The simplest example of a toric K\"ahler four manifold is that of $\bP^2$, for which the toric fan and intersection numbers are:
\bea\label{dP0 glsm}
\begin{tikzpicture}[x=.9cm,y=.9cm,baseline={([yshift=-.5ex]current bounding box.center)}]
%\draw[step=.9cm,gray,very thin] (-1,-1) grid (1,1);
%\draw[ligne] (0,-1)--(1,1)--(-1,0)--(0,-1); 
\draw[->] (0,0)--(1,0);
\draw[->] (0,0)--(0,1);
\draw[->] (0,0)--(-1,-1);
\node[bd] at (1,0)[label=right:{{\scriptsize$D_1$}}]  {}; 
\node[bd] at (0,1)[label=above:{{\scriptsize$D_2$}}]  {}; 
\node[bd] at (-1,-1)[label=left:{{\scriptsize$D_3$}}]  {}; 
\node[bd] at (0,0)  {}; 
\node[bd] at (-1,1)  {}; \node[bd] at (0,-1)  {}; \node[bd] at (-1,0)  {}; 
\node[bd] at (1,1)  {};  \node[bd] at (1,-1)  {}; 
\end{tikzpicture}
\qquad\quad
\begin{tabular}{c|c cc|c}
  & $D_1$  &  $D_2$& $D_3$&$\CK$ \\%heading
 \hline 
$\S$ & $1$ & $1$& $1$                & $-3$  \\
\end{tabular}
\eea
The toric divisors satisfy the linear relations $\S\cong D_1 \cong D_2 \cong D_3$, with $S\cong H$ the hyperplane class. Therefore, given the above triple intersection numbers, we have in our previous notation:
\bea    \label{fluxes P2}
    \m = -(\frak{n}_1 + \frak{n}_2 + \frak{n}_3)~.
\eea
Note also that the canonical divisor is given by $\cK \cong -\sum D_l \cong-3D_1$, and thus the Chern number is ${\bf k}  = -3$. The equivariant parameters \eqref{equivParamGenFormula} are given by \cite{Hosseini:2018uzp}:
\be \label{equivariant tau P2}
   \tau_1^{(l)} = (\tau_1, \, -\tau_1 + \tau_2, \, -\tau_2)\,,  \qquad \quad \tau_2^{(l)} = (\tau_2, \, -\tau_1, \, \tau_1-\tau_2)\,.
\ee

%%%%%%%%%%%%%%%%%%%%%%%%%%%%%%%%%%%%%%%
\medskip

\noindent {The \bf $\cM_4 = \bF_0$ case.}  Consider now the case of $\bF_0 \cong S^1 \times S^1$, with the following toric data:
\bea\label{F0 glsm}
\begin{tikzpicture}[x=.9cm,y=.9cm,baseline={([yshift=-.5ex]current bounding box.center)}]
%\draw[step=.9cm,gray,very thin] (-1,-1) grid (1,1);
%\draw[ligne] (0,-1)--(1,0)--(0,1)--(-1,0)--(0,-1); 
\draw[->] (0,0)--(1,0);
\draw[->] (0,0)--(-1,0);
\draw[->] (0,0)--(0,1);
\draw[->] (0,0)--(0,-1);
\node[bd] at (-1,0)[label=left:{{\scriptsize$D_3$}}]  {}; 
\node[bd] at (0,-1)[label=below:{{\scriptsize$D_4$}}]  {}; 
\node[bd] at (1,0)[label=right:{{\scriptsize$D_1$}}]  {}; 
\node[bd] at (0,1)[label=above:{{\scriptsize$D_2$}}]  {}; 
\node[bd] at (0,0)[label=below:{{}}]  {}; 
\node[bd] at (-1,1)  {};  \node[bd] at (1,1)  {}; 
\node[bd] at (-1,-1)  {};  \node[bd] at (1,-1)  {}; 
\end{tikzpicture}
\qquad\quad
\begin{tabular}{c|c ccc|c} 
  & $D_1$  &  $D_2$& $D_3$& $D_4$& $\CK$ \\
 \hline
$\S_1$ & $0$ & $1$& $0$    & $1$               & $-2$    \\
$\S_2$ & $1$ & $0$& $1$    & $0$                & $-2$    \\
\end{tabular}
\eea
The toric divisors satisfy the linear relations $\S_1\cong D_1 \cong D_3$ and $\S_2\cong D_2 \cong D_4$ and, thus, there are two distinct compact curves $\S_1$ and $\S_2$, corresponding to the two $\bP^1$ factors in $\bF_0\cong \bP^1\times \bP^1$. In this basis, the intersection form reads:
\be
    Q_{\bF_0} = \left( \begin{matrix}
     0 & 1 \\ 1 & 0
    \end{matrix}\right)~.
\ee
Furthermore, the canonical divisor is $\cK = -2D_1 - 2D_2$, leading to the Chern numbers ${\bf k} = (-2,-2)$. Given the above triple intersection numbers, we also have the fluxes:
\bea
     \m = (\m_1, \m_2) = (-\frak{n}_1 - \frak{n}_3, \, -\frak{n}_2 - \frak{n}_4)~.
\eea
Finally, the equivariant parameters are given by \cite{Hosseini:2018uzp}:
\be
   \tau_1^{(l)} = (\tau_1, \, \tau_2,\, - \tau_1,\, -\tau_2)\,, \qquad \quad \tau_2^{(l)} = (\tau_2,\, -\tau_1,\, -\tau_2,\, \tau_1)\,.
\ee

\noindent
{\bf The $dP_n$ cases.}  For completeness, let us also give the equivariant parameters for the remaining toric del Pezzo surfaces. The toric fans for $dP_n$ with $n=1,2,3$ read:
\bea
\begin{tikzpicture}[x=.9cm,y=.9cm,baseline={([yshift=-.5ex]current bounding box.center)}]
%\draw[step=.9cm,gray,very thin] (-1,-1) grid (1,1);
%\draw[ligne] (0,-1)--(1,1)--(0,1)--(-1,0)--(0,-1); 
\draw[->] (0,0)--(1,0);
\draw[->] (0,0)--(0,1);
\draw[->] (0,0)--(-1,-1);
\draw[->] (0,0)--(0,-1);
\node[bd] at (1,0)[label=right:{{\scriptsize$D_1$}}]  {}; 
\node[bd] at (0,1)[label=above:{{\scriptsize$D_2$}}]  {}; 
\node[bd] at (-1,-1)[label=left:{{\scriptsize$D_3$}}]  {}; 
\node[bd] at (0,-1)[label=below:{{\scriptsize$D_4$}}]  {}; 
\node[bd] at (0,0) {}; 
\node[bd] at (1,1) {};
\node[bd] at (-1,1) {}; \node[bd] at (-1,0) {}; \node[bd] at (1,-1) {};
\end{tikzpicture}
\qquad\qquad
\begin{tikzpicture}[x=.9cm,y=.9cm,baseline={([yshift=-.5ex]current bounding box.center)}]
%\draw[step=.9cm,gray,very thin] (-1,-1) grid (1,1);
%\draw[ligne] (0,-1)--(1,0)--(1,1)--(0,1)--(-1,0)--(0,-1); 
\draw[->] (0,0)--(1,0);
\draw[->] (0,0)--(0,1);
\draw[->] (0,0)--(-1,-1);
\draw[->] (0,0)--(0,-1);
\draw[->] (0,0)--(-1,0);
\node[bd] at (1,0)[label=right:{{\scriptsize$D_1$}}]  {}; 
\node[bd] at (0,1)[label=above:{{\scriptsize$D_2$}}]  {};
\node[bd] at (-1,0)[label=left:{{\scriptsize$D_3$}}]  {}; 
\node[bd] at (-1,-1)[label=left:{{\scriptsize$D_4$}}]  {}; 
\node[bd] at (0,-1)[label=below:{{\scriptsize$D_5$}}]  {}; 
\node[bd] at (0,0) {}; 
\node[bd] at (1,1) {};
\node[bd] at (-1,1) {}; \node[bd] at (1,-1) {};
\end{tikzpicture}\qquad\qquad
\begin{tikzpicture}[x=.9cm,y=.9cm,baseline={([yshift=-.5ex]current bounding box.center)}]
%\draw[step=.9cm,gray,very thin] (-1,-1) grid (1,1);
\draw[->] (0,0)--(1,0);
\draw[->] (0,0)--(0,1);
\draw[->] (0,0)--(-1,-1);
\draw[->] (0,0)--(0,-1);
\draw[->] (0,0)--(-1,0);
\draw[->] (0,0)--(1,1);
\node[bd] at (1,0)[label=right:{{\scriptsize$D_1$}}]  {}; 
\node[bd] at (1,1)[label=right:{{\scriptsize$D_2$}}]  {}; 
\node[bd] at (0,1)[label=above:{{\scriptsize$D_3$}}]  {};
\node[bd] at (-1,0)[label=left:{{\scriptsize$D_4$}}]  {}; 
\node[bd] at (-1,-1)[label=left:{{\scriptsize$D_5$}}]  {}; 
\node[bd] at (0,-1)[label=below:{{\scriptsize$D_6$}}]  {}; 
\node[bd] at (0,0) {}; 
\node[bd] at (-1,1) {}; \node[bd] at (1,-1) {};
\end{tikzpicture}
\eea
respectively, and the equivariant parameters are:
\bea\nn
  &  dP_1\,:  && \; \tau_1^{(l)} = (\tau_1, -\tau_1 + \tau_2, - \tau_1, -\tau_2)\,,  && \; \tau_2^{(l)} = (\tau_2, -\tau_1, \tau_1-\tau_2, \tau_1)\,,  \\
  &  dP_2\,:  && \; \tau_1^{(l)} = (\tau_1, \tau_2, -\tau_1 + \tau_2, - \tau_1, -\tau_2)\,, && \; \tau_2^{(l)} = (\tau_2, -\tau_1, -\tau_2, \tau_1-\tau_2, \tau_1)\,, \\
  & dP_3\,: && \; \tau_1^{(l)} = (\tau_1-\tau_2, \tau_1, \tau_2, -\tau_1 + \tau_2, - \tau_1, -\tau_2)\,, &&\; \tau_2^{(l)} = (\tau_2, -\tau_1+\tau_2, -\tau_1, -\tau_2, \tau_1-\tau_2, \tau_1)\,. 
\eea

%%%%%%%%%%%

%As a result, the partition function of a free hypermultiplet on $\CM_4 \times S^1$, for toric-K\"ahler 4-manifolds, reduces to:
%\bea    
%    Z_{\cM_4\times S^1}^{hyper}(a) = \left( {1\ov 1-x}\right)^{\chi_h + {1\ov 2}(\boldsymbol{k} +\m, \m)}~,
%\eea
%which is precisely what we found by direct computation in \eqref{Z Hyper M4xS1}. 

%%%%%%%%%%%%%%%%5
\subsubsection{The $S^5$ and $L(\p; 1)$ partition functions}
Let us now turn to the simplest and most important instance of a circle fibration $\CM_5\rightarrow \CM_4$, which is the five-sphere viewed as a circle fibration over the complex projective plane:
\be
    S^1 \longrightarrow S^5 \longrightarrow \bP^2~.
\ee
The gluing approach was first considered in \cite{Lockhart:2012vp}, but it is worthwhile to discuss the argument in some detail. 
The metric of the round five-sphere can be written as $ds^2 = \sum dz_id\b z_i$, in terms of the coordinates $(z_i) \in \mathbb{C}^3$ subject to the constraint $\sum |z_i| = 1$. Alternatively, we can parametrise the five-sphere using the angles $\theta, \phi \in (0,\pi/2)$ and  $\chi_i \in (0,2\pi)$, as:
\be
    z_1 =  e^{i\chi_1}\sin\theta \cos\phi ~, \qquad
    z_2 =  e^{i\chi_2} \sin\theta  \sin\phi ~, \qquad
    z_3 =  e^{i\chi_3} \cos\theta ~.
\ee
In these coordinates, the $S^5$ metric reads:
\be
    ds^2 = d\theta^2 + \sin^2\theta ~d\phi^2 + \sin^2 \theta \cos^2\phi ~d\chi_1^2 + \sin^2\theta \sin^2\phi ~d\chi_2^2 + \cos^2 \theta ~d\chi_3~ .
\ee
To apply our formalism, we should write this metric in the general form \eqref{dsM5}, namely as $ds^2(S^5) = ds^2(\bP^2) + \left(d\psi + \Cfib\right)^2$, where $ds^2(\bP^2)$ is the Fubini-Study metric on the base $\bP^2$. An important requirement is that the connection $\Cfib$ should be well defined on each coordinate patch on the base. Let us then consider the Fubini-Study metric on each patch and subtract it from the $S^5$ metric, in order to find the connection $\Cfib$ on that patch. The $z_i$ coordinates of the five-sphere descend to coordinates of the projective space $\bP^2$. As such, let us denote by $V_i \cong \C^2$ the patch with coordinates $w_j= z_j/z_i$, for $j\neq i$ and $z_i \neq 0$, and  define the corresponding azimuthal coordinates:
\bea \label{parametrisation}
     \text{Patch}\;\; V_1\,:~\quad & \rho_1^{(1)} = \chi_2 - \chi_1~, \quad  & \rho_2^{(1)} = \chi_3 - \chi_1~,  \\
     \text{Patch}\;\; V_2\,:~\quad &\rho_1^{(2)} = \chi_3 - \chi_2~, \quad  & \rho_2^{(2)} = \chi_1 - \chi_2~, \\
    \text{Patch} \;\; V_3\,:~\quad & \rho_1^{(3)} = \chi_1 - \chi_3~, \quad  & \rho_2^{(3)} = \chi_2 - \chi_3~.
\eea
For each coordinate patch, the coordinate along the $S^1$ fiber will be a linear combination of the $\chi_i$ angles, $\psi = \alpha_i \chi_i$.  With the normalisation $\sum_i \alpha_i=1$, we find the $U(1)_{\rm KK}$ connection in each patch to be:
\bea
    \Cfib^{(1)} =\, & {1 \ov 4}\big(1-4\alpha_2-\cos(2\theta)-2\cos(2\phi)\sin(\theta)^2\big) d\rho_1^{(1)}  + {1 \ov 2} \big(1-2\alpha_3+\cos(2\theta) \big) d\rho_2^{(1)}~, \\
    \Cfib^{(2)} =\, &  {1 \ov 2} \big(1-2\alpha_3'+\cos(2\theta) \big) d\rho_1^{(2)}+ {1 \ov 4}\big(1-4\alpha_1'-\cos(2\theta)+2\cos(2\phi)\sin(\theta)^2\big) d\rho_2^{(2)}~, \\
    \Cfib^{(3)} =\, & {1 \ov 4}\big(1-4\alpha_1''-\cos(2\theta)+2\cos(2\phi)\sin(\theta)^2\big) d\rho_1^{(3)} \\ & +{1 \ov 4}\big(1-4\alpha_2''-\cos(2\theta)-2\cos(2\phi)\sin(\theta)^2\big) d\rho_2^{(3)}~. 
\eea
Let us note that the patch $V_i$ is not defined at $z_i=0$, at which point the differential $d\chi_i$ is ill-defined. Then, imposing continuity for the connection and well-definiteness on every coordinate patch (that is, the absence of `Dirac string' singularities), we should pick the following coordinates along the $S^1$ fiber:
\be
    \psi^{(1)} = \chi_1~, \qquad  \psi^{(2)}  = \chi_2~, \qquad \psi^{(3)}  = \chi_3~.
\ee
In this way, we find the following transformations between angles as we change coordinate patches of the $\bP^2$ base:
\be\label{gluinganglesS5}
   \left( \begin{matrix}
        \rho_1^{(2)}   \\
         \rho_2^{(2)}   \\
         \psi^{(2)}    
    \end{matrix} \right) = \left( \begin{matrix}
        -1 & 1& 0\\
        -1 & 0 & 0 \\
        1 & 0 & 1
    \end{matrix} \right) \left( \begin{matrix}
         \rho_1^{(1)}   \\
         \rho_2^{(1)}   \\
         \psi^{(1)} 
    \end{matrix} \right)~, \qquad 
    \left( \begin{matrix}
        \rho_1^{(3)}   \\
        \rho_2^{(3)}   \\
         \psi^{(3)}   
    \end{matrix} \right) = \left( \begin{matrix}
        0 & -1 & 0\\
        1 & -1 & 0 \\
        0 & 1 & 1
    \end{matrix} \right) \left( \begin{matrix}
         \rho_1^{(1)}   \\
         \rho_2^{(1)}   \\
         \psi^{(1)}  
    \end{matrix} \right) ~.
\ee
In the toric description of $S^5$, the five-sphere is a $T^3$ fibration over a triangle \cite{Lockhart:2012vp}. Moreover,  the $\Omega$-background parameters $\tau_{1}^{(i)}$ and $\tau_{2}^{(i)}$ on each patch $V_l\cong \C^2$ can be interpreted as complex structure parameters  for the tori $T^2\subset T^3$ spanned by the angular coordinates $(\rho_1^{(i)}, \psi_i)$ and $(\rho_2^{(i)}, \psi_i)$, respectively.  The $SL(3,\Z)$ transformation matrices \eqref{gluinganglesS5} then suggest the following gluing rules for the Nekrasov partition functions:
\bea \label{ProposedGluing}
   V_1\, : \qquad  && \tau_1~, \qquad \qquad &  \qquad \tau_2~, \quad  &  \amCB~, \qquad \quad  \\
 V_2\, : \qquad&& \tau_1^* = {-\tau_1+\tau_2 \ov \tau_1+1}~, \qquad &  \tau_2^*={-\tau_1 \ov \tau_1+1}~, \qquad &  \amCB^* = {\amCB \ov \tau_1+1}~, \\
   V_3\, : \qquad && \t\tau_1 = {-\tau_2 \ov \tau_2+1}~, \qquad &  \t\tau_2 = {\tau_1 -\tau_2 \ov \tau_2+1}~, \qquad &  \t{\amCB} = {\amCB \ov \tau_2+1}~, \\
\eea
which generalises \eqref{equivariant tau P2}.

\medskip 

\noindent {\bf Lens spaces.} It is also instructive to consider lens spaces $S^5/\Z_p$, as a simple generalisation of the above. 
 The lens space $L(p; q_1,q_2,q_3)$ can be defined as a quotient of $S^5 \subset \C^3$ by the $\Z_p$ action generated by:
\be
    (z_1, z_2, z_3) \mapsto \left( e^{2\pi i {q_1 \ov p}}z_1, e^{2\pi i {q_2 \ov p}}z_2, e^{2\pi i {q_3 \ov p}}z_3\right)~,
\ee
with $q_i, p \in \Z$ and $q_i$ coprime to $p$. For generic values of $q_i$, however, these five-manifolds are not fibrations over $\bP^2$, but rather over singular quotients of $\bP^2$, as one can see  by considering the induced action on the $\bP^2$ coordinates \eqref{parametrisation}.  In this paper, we restrict our attention to principal circle bundles over four-manifolds, which corresponds to the case $q_i = 1$.  We denote the resulting lens space by  $L(p; 1)$. It is simply a principal circle bundle: 
\be
S^1\rightarrow L(\p; 1)\rightarrow \bP^2~,
\ee
with $\p=p$. 
 We can then derive the gluing rules in the same way as for the round $S^5$. One finds:
%\be
%    Z_{L(p; 1)}(\amCB) = \prod_{l=1}^{\chi(\bP^2)}  Z_{\bC^2\times S^1} \left({\amCB \ov \gamma^{(l)}}\,,\, {\tau_1^{(l)}\ov \gamma^{(l)}}\,,\, {\tau_2^{(l)}\ov \gamma^{(l)}}\right)~,
%\ee
\be\label{ZLp1 gluing}
    Z_{L(\p; 1)}(\amCB)_{\m} = \prod_{l=1}^{\chi(\bP^2)}  Z_{\bC^2\times S^1} \left({\amCB + \tau_1^{(l)}\frak{n}_l + \tau_2^{(l)}\frak{n}_{l+1} \ov \gamma^{(l)}}\,,\, {\tau_1^{(l)}\ov \gamma^{(l)}}\,,\, {\tau_2^{(l)}\ov \gamma^{(l)}}\right)~,
\ee
where $\tau_i^{(l)}$ are the equivariant parameters appearing in the $\bP^2 \times S^1$ gluing \eqref{equivariant tau P2}, while the denominators $\gamma$ are given by: 
\be
    \gamma^{(l)} = \left(1,\, p\tau_1+1,\, p\tau_2+1 \right)~.
\ee
In \eqref{ZLp1 gluing} we also allowed for background fluxes, as in \eqref{M4xS1 conjecture}. In the non-equivariant limit (and turning on $\varepsilon$ as before), this expression reproduces exactly the one expected from \eqref{ZM5 full}.

\subsubsection{Fibrations over toric K\"ahler 4-manifolds}

Having discussed fibrations over $\bP^2$, it is natural to consider the generalisation to any toric $\CM_4$. Here, we first derive the gluing formula for $\CM_4=\bF_0 \cong \bP^1 \times \bP^1$, which has an explicitly known K\"ahler metric. We then conjecture a gluing formula in the general case.

\medskip
%%%%%%%%%%%%%%%%%%%%%%%%%

\noindent
\paragraph{The $\cM_4 = \bF_0$ case.} Consider circle fibrations over $\bF_0$, with Chern numbers $\p = (\p_1,\p_2)$, such that the metric of such a space is given by:
\bea
    ds^2 = \sum_{i=1}^2 \left( d\theta_i^2 + \sin^2 \theta_i d\phi_1^2\right) + \left(d\psi +  { 1\ov 2}\sum_{i=1}^2 \p_i(\pm 1+\cos\theta_i)d\phi_i\right)^2~,
\eea
with $\theta_i \in [0,\pi)$, $\phi_i \in [0,2\pi)$ and $\psi \in [0,2\pi)$.   This space has four coordinate patches, corresponding to the two patches of each of the $\bP^1$ spaces. We proceed as before, by finding the well-defined coordinates on each patch. Defining:
\be
    \gamma^{(l)} = (1,\, \p_1 \tau_1 + 1,\, \p_1 \tau_1 + \p_2 \tau_2 + 1, \, \p_2 \tau_2 + 1)~,
\ee
we then propose that the partition function on non-trivial fibrations over $\bF_0$ is given by:
\be
    Z_{\bF_0^{({\p})}}(\amCB) = \prod_{l=1}^{\chi(\bF_0)}  Z_{\bC^2\times S^1} \left({\amCB + \tau_1^{(l)} \frak{n}_l + \tau_2^{(l)} \frak{n}_{l+1} \ov \gamma^{(l)} }\,,\, {\tau_1^{(l)}\ov \gamma^{(l)}}\,,\, {\tau_2^{(l)}\ov \gamma^{(l)}}\right)~,
\ee
where, as before, $\tau_i^{(l)}$ are the equivariant parameters appearing in the $\bF_0 \times S^1$ case and $\frak{n}_l$ are fluxes associated with the toric divisors. 
%%%%%%%%%%%%%%%%%%%%%%%%%%%%%%%%%%

\medskip

\noindent  {\bf General toric K\"ahler surfaces.} We would like to generalise the above result to any principal circle bundle over a toric K\"ahler surface $\cM_4$. The prescription used for non-trivial fibrations over $\bP^2$ and $\bF_0$ involved the coordinates $(\rho_1^{(l)}, \rho_2^{(l)})$ along the coordinate patches $V_{\sigma_l}$ of the base four-manifold, as well as the coordinate along the fiber $\psi^{(l)}$. As explained in the previous sections, a non-trivial fibration can be viewed as a non-trivial flux for a $U(1)_{\rm KK}$ background symmetry on $\CM_4$. We then propose that the denominators $\gamma^{(l)}$ should be given by:
\be
    \gamma^{(l)} = 1 + \tau_1^{(l)}p_l + \tau_2^{(l)}p_{l+1}~,
\ee
where $p_l$ are $U(1)_{\rm KK}$ fluxes associated with the non-compact toric divisors $D_l$, such that:
\be \label{def Chern number p}
   \sum_{k=1}^{\chi-2}  \p_k [\S_k]=-\  \sum_{l=1}^{\chi } p_l D_l~,
\ee
as in \eqref{definition fluxes}. Thus, the CB partition function on non-trivial fibrations over $\cM_4$ with Chern numbers $\p$ should generalise to:
\be \label{Toric gluing M5}
    {\bf Z}_{\cM_5} = \sum_{\mathfrak{n}_l} \oint d\hspace{0.7pt}\amCB\prod_{l=1}^{\chi(\CM_4)} Z_{\bC^2\times S^1} \left({\amCB+\tau_1^{(l)}\mathfrak{n}_l + \tau_2^{(l)}\mathfrak{n}_{l+1} \ov \gamma^{(l)}}, {\tau_1^{(l)} \ov \gamma^{(l)}}, {\tau_2^{(l)} \ov \gamma^{(l)}}\right)~,
\ee
naturally generalising the Nekrasov conjecture \cite{Nekrasov:2003vi}. (Here, as in the original conjecture, the precise form of the contour integration and of the sum over fluxes remain to be determined.)

Given the factorised integrand $Z_{\CM_5}$ in \eqref{Toric gluing M5} with the non-trivial $\Omega$-background, we can again check our general formalism by taking the non-equivariant limit, generalising the formula \eqref{NonEquivariantLimitFullZ M4*S1}.  
For every toric Fano four-manifold $\CM_4$ in \eqref{toric Kahler 4-manifolds},   using \eqref{general noneq lim}, we find the following expression:
\bea  
   & \log Z_{\cM_5}(\amCB) \approx  \log Z_{\cM_4\times S^1}(\amCB)  -2\pi i \left(\sum D_l \right) \cdot\left(\sum  p_l D_l \right) \left( H(\amCB) - \amCB^I {\d H(\amCB) \ov \d \amCB^I}\right)\\
  &\qquad    +\left(\sum  p_l D_l \right) \cdot\left(\sum  \n_l^I D_l \right) \log \FiberOpK_I(\amCB) + {1\ov 2}\left(\sum  p_l D_l \right) \cdot\left(\sum  p_l D_l \right)\log \FiberOp(\amCB)~,
\eea
where $Z_{\cM_4\times S^1}$ is given by \eqref{NonEquivariantLimitFullZ M4*S1} and $\FiberOp$, $\FiberOpK$ are precisely the quantities defined in \eqref{F fibering def 1} and \eqref{F fibering def 2}, respectively. Then, using the relation \eqref{def Chern number p}, as well as the substitution $H \rightarrow \varepsilon^I \d_I \cF$ for the extended topological twist, we recover the complete master formula \eqref{ZM5 full} for the CB partition function on $\cM_5$.

%%%%%%%%%%%%%%%%%%%%%%%%%
\medskip

\noindent 
We should note that the proposal \eqref{Toric gluing M5} appears slightly different from the results above for  $\mathbb{P}^2$ and $\mathbb{F}_0$, though all the formulas agree perfectly in the non-equivariant limit. For instance,  for $\mathbb{P}^2$, the gluing \eqref{Toric gluing M5} uses:
\be
    \gamma^{(l)}_{\rm new} = \big(1+p_1\tau_1 + p_2 \tau_2,\,  1-(p_2+p_3)\tau_1 +p_2\tau_2,\, 1+p_1\tau_1 - (p_1+p_3)\tau_2\big)~,
\ee
where $\p = -(p_1+p_2+p_3)$, while previously we derived:
\be
    \gamma^{(l)} = \big(1,\, \p \tau_1 + 1,\, \p \tau_2+1\big)~.
\ee
However, setting $p_1 = p_2 = 0$, the two expressions become identical. Similar comments hold true in general. It might be the case that the individual fluxes $p_l$ (and $\n_l$) have an intrinsic meaning on the $\Omega$-deformed $\CM_5$, which we did not explore. Our main motivation, here, was to provide a strong consistency check for our formulas for the fibering operator in the DW-twisted theory, hence our main focus was on the non-equivariant limit.

\subsection{The spin/charge relation for the $E_n$ theories}
Finally, let us further comment on the spin/charge relation implied by the constraint \eqref{spincharge const}. In appendix~\ref{appendix:GVinvs}, we explicitly compute the refined GV invariants of the so-called toric $E_n$ 5d SCFTs, which are rank-one SCFTs that can be deformed to a 5d $\CN=1$ $SU(2)$ gauge theory   with $N_f=n{-}1$ fundamental hypermultiplets. Consider first the $E_1$ theory, corresponding to the $SU(2)_0$ gauge theory. In this case, perturbatively in the 5d gauge-theory limit, we only have the massive W-boson, of spin $(0, \half)$, which satisfies the condition
\be
\half + j_l +j_r+ \bbeta(\varepsilon)  \in \Z~,
\ee
with charge $\bbeta=(0,1)$ in the basis corresponding to the two factors $\bP_b\times \bP_f$ of the local $\bF_0$  geometry in M-theory. Hence we need to have $\varepsilon^{I=2} \mod 1=0$ in this basis. Similarly, looking at the first instanton particle, $\bbeta=(1,0)$, we have $\varepsilon^{I=1} \mod 1=0$. (The SCFT has a symmetry exchanging the two charges, $\bbeta=(m,n)\leftrightarrow (m,n)$.) Hence, by consistency, we should have $\half+j_l+ j_r\in \Z$ for any other particle in the spectrum, of any charge. For instance, we   have the states:
\be  
    N_{j_l, j_r}^{(1,d)} =  \delta_{j_l, 0} \delta_{j_r, d+{1\ov 2}}~,\qquad \qquad d\in \Z_{\geq 0}~,
\ee
which capture the one-instanton correction in the $SU(2)$ gauge-theory interpretation. 
The spin/charge relation should similarly hold for any higher-spin state. This is indeed the case, at least to the order that we have checked it. (See table~\ref{tab: E1 refined GV invariants} in appendix.)

Similarly, consider the local $dP_2$ geometry, corresponding to the $E_2$ theory, which can be obtained from the $\bF_0$ geometry by blowing up a curve $\CC_m$. The M2-brane wrapped on $\CC_m$ gives a hypermultiplet (a particle of spin $(0,0)$), hence we should choose $\varepsilon^{I=3} \mod 1 =\half$, in the natural basis $(\CC^I)=(\bP_b, \bP_f, \CC_m)$. We thus have the constraint that, for any BPS particle of charge $\bbeta=(m,n,p)$ in $E_2$ theory, we should have $\half+j_l+j_r +\half p \in \Z$. This is indeed the case (see appendix~\ref{app:dP2}). Similarly, for the $E_3$ theory that arises at the local $dP_3$ geometry, we have a basis $(\CC^I)=(\bP_b, \bP_f, \CC_{m_1}, \CC_{m_2})$ and we should choose $\varepsilon^{I=3} \mod 1 =\half$ and  $\varepsilon^{I=4} \mod 1 =\half$. Given the charges $\bbeta=(m,n,p,q)$, we then need $\half +j_l+j_r+ \half(p+q)\in \Z$, which is the case (see appendix~\ref{app:dP3}). Another interesting example is the $E_0$ SCFT, realised as the local $\bP^2$ geometry in M-theory \cite{Morrison:1996xf}. It has a single charge, $\bbeta=d$, and no gauge theory interpretation. We should then have $\half +j_l+j_r+\half d\in \Z$, which is indeed the case (see \cite[Table 3]{Huang:2013yta}).

Hence, all these models can be coupled consistently to our $\CM_5$ background with the extended DW twist on $\CM_4$. It would be interesting to understand whether the spin/charge relation must always hold, {\it a priori}, in any 5d SCFT.

%%%%%%%%%%%%%%%%%%%%%%%%%%%%%%%%%%%%%%%%%%%%%%

%%%%%%%%%%%%%%%%%%%%%%%%%%%
%%%%%%%%%%%%%%%%%%%%%%%%%%%
\subsection*{Acknowledgements}
We are grateful to  Johannes Aspman, Michele del Zotto, Elias Furrer, Hee-Cheol Kim, Heeyeon Kim, Jan Manschot, Sakura Sch\"afer-Nameki and Luigi Tizzano
 for interesting discussions, feedback and correspondences.
The work  of CC is supported by a University Research Fellowship 2017, ``Supersymmetric gauge theories across dimensions'', of the Royal Society. CC is also a Birmingham Fellow.
The work of HM is supported by a Royal Society Research Grant for Research Fellows.

%%%%%%%%%%%%%
 %%%%%%%%%%%%%%%%%%%%%%%%%%%%%%%%%%%%%%%%%%%%
 %%%%%%%%%%%%%%%%%%%%%%%%%%%%%%%%%%%%%%%%%%%%

 \appendix

 \section{Geometry and supersymmetry conventions}
In this appendix, we state our conventions for 4d and 5d geometry and we recall various basic facts about 4d $\CN=2$ and 5d $\CN=1$ supersymmetry in flat space. We are working in Euclidean signature throughout.

\subsection{Four-dimensional conventions}\label{app:4dconventions}

\subsubsection{Flat-space conventions}\label{app:subsec: 4d flat}
Consider Euclidean flat space, $\R^4$, with the real Cartesian coordinates $x^\mu$. The rotation group is $SO(4)$. In order to define spinors, we consider its universal cover:
\be
{\rm Spin}(4) \cong SU(2)_l \times SU(2)_r~.
\ee
 In 4d, we mostly deal with  Weyl spinors, $\psi_\alpha$ and  $\t\psi^\alphadot$, which transform in the representation with spin $(j_l, j_r)=(\half,0)$ and $(0, \half)$, respectively, in the standard Wess-and-Bagger notation \cite{Wess:1992cp}. In particular, we raise and lower spinor indices with the tensors $\epsilon^{\alpha\beta}$, $\epsilon_{\alpha\beta}$ and  $\epsilon^{\alphadot\betadot}$, $\epsilon_{\alphadot\betadot}$ such that $\epsilon^{12}= \epsilon^{\dot 1 \dot 2}= \epsilon_{21}=\epsilon_{\dot 2 \dot 1}=1$.   We choose the Euclidean $\sigma$-matrices:%
\footnote{Here $\sigma^i$ are the Pauli matrices and $\mathbf{1}$ is the $2{\times} 2 $ identity matrix. Therefore:
$$\sigma^1= \mat{0& 1\\ 1 & 0}~, \quad
\sigma^2= \mat{0& -i\\ i & 0}~, \quad
\sigma^3= \mat{1& 0\\ 0 & -1}~, \quad
\sigma^4= \mat{i & 0\\0 & i}~.$$.}
 \bea
 \sigma^\mu= (\sigma^i,  i \mathbf{1})~, \qquad\qquad \t \sigma^\mu = (-\sigma^i, i \mathbf{1})~.
 \eea
The Dirac spinor $\Psi$ and the $\gamma$-matrices are then given by:
 \be
\Psi_{\bf a}= \mat{\psi_\alpha \cr \t\chi^\alphadot}~,\qquad \qquad {(\gamma^\mu)_{\bf a}}^{\bf b} = \mat{0 & \sigma^\mu \\  - \t\sigma^\mu & 0}~,
 \ee
 with Dirac indices ${\bf a}, {\bf b}=1, \cdots, 4$. Note that:
 \be
 \sigma^\mu \t\sigma^\nu +\sigma^\nu \t\sigma^\mu =- 2 \delta^{\mu\nu}~,\qquad
 \t\sigma^\mu \sigma^\nu +\t\sigma^\nu \sigma^\mu =- 2 \delta^{\mu\nu}~,
 \ee
 and therefore $\{\gamma^\mu, \gamma^\nu \}= 2 \delta^{\mu\nu}$. One also defines the matrices:
 \be
 \sigma^{\mu\nu} \equiv {1\ov 4} \left(\sigma^\mu \t\sigma^\nu -\sigma^\nu \t\sigma^\mu  \right)~, \qquad
\t \sigma^{\mu\nu} \equiv {1\ov 4} \left(\t\sigma^\mu \sigma^\nu -\t\sigma^\nu \sigma^\mu  \right)~,
 \ee
 which generate $SU(2)_l$ and $SU(2)_r$ rotation on $\psi$ and $\t\chi$, respectively. 
They are  respectively anti-self-dual (ASD) and self-dual (SD):
 \be
 \half \epsilon^{\mu\nu\rho\lambda}\sigma_{\rho \lambda}  =- \sigma^{\mu\nu}~, \qquad
 \half \epsilon^{\mu\nu\rho\lambda}\t \sigma_{\rho \lambda}  =\t\sigma^{\mu\nu}~. 
 \ee
We also extensively use the complex coordinates:
 \be
 z^1= x^1 + i x^2~, \qquad z^2= x^3+ i x^4~,
 \ee
on $\C^2 \cong \R^4$.

\subsubsection{Curved-space conventions}\label{app:subsec: 4d curved}
On $\CM_4$ a Riemannian manifold, with the metric:
\be
ds^2(\CM_4) = g_{\mu\nu} \, dx^\mu dx^\nu  = \delta_{ab}\,  \h e^a_\mu  \h e^b_\nu\,  dx^\mu dx^\nu~,
\ee
we use the notation $x^\mu$ for any set of local coordinates. Here, we also introduced the veirbein $\h e^{a}$, with the frame indices $a=1, \cdots, 4$. On a K\"ahler manifold, we also pick a complex frame $e^a, e^{\b a}$ (with indices $a= 1, 2$):%
\footnote{We use the same $a$ for the distinct indices in $\h e^a$ ($a=1, \cdots, 4$) and $e^a$ ($a=1, 2$). That should cause no confusion.}
\be
e^1= \h e^1+ i \h e^2~, \quad e^2= \h e^3+ i \h e^4~, \quad
e^{\b 1}= \h e^1- i \h e^2~, \quad e^{\b 2}= \h e^3- i \h e^4~.
\ee
We note that the complex structure then takes the form:
\be
{J^a}_b=
 \mat{0& -1& 0&0\\ 1&0&0&0 \\ 0&0&0&-1\\ 0&0&1&0}~,
\ee
in the frame basis $\{\h e^a\}$.
Note also that the fully-antisymmetric Levi-Civita tensor, $\epsilon^{\mu\nu\rho\sigma}$, takes the values:
\be
\epsilon^{1 234}=\epsilon_{1 234}= 1~, \qquad\qquad \epsilon^{1 2 \b 1 \b 2}=  4~,
\quad  \epsilon_{1 2 \b 1 \b 2}=  {1\ov 4}~,
\ee
in the real and complex frames, respectively.
Spinors are locally defined with respect to the frame $\{\h e^{a}\}$, using the conventions of subsection~\ref{app:subsec: 4d flat}. 
The covariant derivative of Weyl spinors reads:
 \be
\nabla_\mu\psi = (\d_\mu- \half \omega_{\mu ab} \sigma^{ab}) \psi~, \qquad\quad
\nabla_\mu\t\chi = (\d_\mu- \half \omega_{\mu ab}\t \sigma^{ab}) \t\chi~.
\ee
Here, the spin connection $\omega_\mu$ is defined in terms of the veirbein $\h e^a_\mu$ and the Levi-Civita connection $\Gamma^\lambda_{\mu\nu}$:
\be
 {\omega_{\mu a}}^b = \h e^\nu_a \h e_\rho^b \, \Gamma_{\mu\nu}^\rho - \h e^b_\nu \d_\mu \h e^\nu_a~.
\ee
This is equivalent to the statement that the covariant derivative of the veirbein vanishes:
\be
\nabla_\mu \h e^a_\nu \equiv \d_\mu \h e^a_\nu + \Gamma_{\mu\nu}^\rho \h e_\rho^a - {\omega_{\mu b}}^a \h e^b_\nu =0~.
\ee
The curvature tensor of the spin connection is given by:
\be
{R_{\mu\nu a}}^b(\omega) = \d_\mu { \omega_{\nu a}}^b- \d_\nu { \omega_{\mu a}}^b+ { \omega_{\nu a}}^c  { \omega_{\mu c}}^b - { \omega_{\mu a}}^c  { \omega_{\nu c}}^b~.
\ee
This is simply the ordinary Riemann tensor with two frame indices 
${R_{\mu\nu a}}^b=\h e^\rho_a  \h e^b_\lambda {R_{\mu\nu\rho}}^{\lambda}$. 
Finally, the Ricci scalar is defined as:
\be\label{Ricci def}
R= \h e_\mu^a \h e_b^\nu\, {R^\mu_{\phantom{\mu}\nu a}}^b(\omega)~.
\ee

 \subsubsection{4d $\CN=2$ supersymmetry}\label{App:subsec:4dSUSY}
Consider 4d $\CN=2$ supersymmetry on $\R^4$ (in Euclidean signature). The supersymmetry algebra reads:
\bea\label{4d susy alg R4}
&\{Q_\alpha^I, \b Q_{\dot \beta J}\}=  2 \sigma^\mu_{\alpha\dot\beta} P_\mu\,  {\delta^{I}}_J~,\cr
&\{Q_\alpha^I, Q_\beta^J\} = 2 \varepsilon_{\alpha\beta} \epsilon^{IJ} \t Z~,\qquad
& \{\t Q_{\dot\alpha I},\t Q_{\dot\beta J}\} =  2 \varepsilon_{\alphadot\betadot} \epsilon_{IJ}  Z~,
\eea
with $I=1,2$  the $SU(2)_R$ index and $Z$ the complex central charge. We denote any $SU(2)_R$ doublet ${\bf 2}$ by  $\varphi^I$, and its conjugate $\b {\bf 2}$ by  $\varphi_I$. The $SU(2)_R$ indices are raised and lowered according to:
 \be
 \varphi_I = \epsilon_{IJ} \varphi^J~, \qquad \varphi^I = - \epsilon^{IJ} \varphi_J~.
 \ee
Note that $\epsilon_{12}= \epsilon^{12}=1$ here, in our convention for $SU(2)_R$. 
Under the maximal $R$-symmetry of the $\CN=2$ supersymmetry algebra, $U(2) \cong SU(2)_R \times U(1)_r$, the supercharges transform as:
\be\label{U2R 4d susy}
\begin{array}{c|c|c|c}
 & SU(2)_R \times U(1)_r &  T_3^{SU(2)_R}   & U(1)_R^{\CN{=}1}  \\
 \hline
 Q^I_\alpha &({\bf 2})_{-1} &\pm \half & - \delta^{I1}\\
 \t Q_{\alphadot J} &({\bf \b 2})_{1} & \mp\half & +\delta_{J1}\\
 \hline
Z & ({\bf 1})_{2} & 0 & 
\end{array}
\ee
The central charge $Z$ breaks the $U(1)_r$ symmetry explicitly.  It is sometimes useful to decompose the $\CN=2$ supersymmetry multiplets with respect to $\CN=1$ multiplets. We will choose the $\CN=1$ subalgebra generated by:
\be\label{Neq1subalg}
 Q=Q^{I=1}~, \qquad \qquad \t Q= \t Q_{J=1}~, 
 \ee
 with the $\CN=1$ $U(1)_R$ symmetry indicated in  \eqref{U2R 4d susy}.

 \medskip
 \noindent
 Let us present the supersymmetry transformations for the vector multiplet and the charged hypermultiplet, for a general supersymmetry transformation: 
\be
  \delta = i \xi_I Q^I + i \t\xi^I \t Q_I~, 
   \ee
 with any constant (commuting) Weyl spinors $\xi$ and $\t \xi$.

\paragraph{The vector multiplet.} 
 The vector multiplet has components:
 \be\label{vector}
 \CV_{\CN{=}2}= \left(A_\mu, \phi, \t\phi, \lambda^I,\t \lambda_I, D_{IJ}\right)~, 
 \ee
in the Wess-Zumino gauge. This includes the auxiliary fields $D_{IJ}=D_{JI}$, which transform as a triplet of $SU(2)_R$.  Note the default positions for the $SU(2)_R$ indices in \eqref{vector}. 
The field $A_\mu$ is a gauge field for some gauge group $G$, and all the other fields transform into the adjoint representation of $\Fg= {\rm Lie}(G)$.
 The Euclidean $\CN=2$ super-Yang-Mills action reads:
 \bea\label{Neq2 SYM}
&\SL_{{\rm  SYM}} &=&\; {1\ov g^2} \tr\Bigg( {1\ov 4}F_{\mu\nu} F^{\mu\nu}+D_\mu \t\phi D^\mu \phi  + i\t\lambda_I \t\sigma^\mu D_\mu \lambda^I + {1\ov 4 } D_{IJ} D^{IJ}\cr
&&&\qquad\quad + \half [\t\phi, \phi]^2
-{i \ov \sqrt2}  \lambda_I [\t\phi ,\lambda^I]+ {i\ov \sqrt2} \t\lambda^I [\phi ,\t\lambda_I] \Bigg)~.
\eea
Here, `$\tr$' denotes the trace in some fundamental representation of $\Fg$, which we normalise as
\be\label{def tr G}
\tr(T^A T^B)= \delta^{AB}~,
\ee
with $T^A$ the generators of $\Fg$ in this representation. 
A more canonical convention would be to take the trace in the adjoint,
\be
\SL_{{\rm  SYM}} = {1\ov g^2 h^\vee} \Tr\Bigg( {1\ov 4}F_{\mu\nu} F^{\mu\nu}+\cdots \Bigg)~,
\ee
with the standard normalisation in terms of the dual Coxeter number $h^\vee$ of $\Fg$,
\be
\Tr(T^A_{\rm adj} T^B_{\rm adj})=h^\vee \delta^{AB}~.
\ee
We use `$\tr$' defined in \eqref{def tr G} throughout this paper, which has the advantage of being valid for any semi-simple $G$, including the $U(1)$ factors, but we could as easily formally substitute $\tr \rightarrow {1\ov h^\vee}\Tr$ for each simple factor. 
The supersymmetry transformations of the vector multiplet are given by
 \bea\label{susy vec offshell Neq2}
 & \delta \phi &=&\; \sqrt{2} \xi_I\lambda^I~, \cr
 &\delta \t\phi  &=&\; \sqrt2 \t\xi^I \t \lambda_I~, \cr
& \delta A_\mu  &=&\; i \xi^I \sigma_\mu\t \lambda_I  -i \t\xi_I \t\sigma_\mu \lambda^I~, \cr
&\delta \lambda^I  &=&\;-  i \xi^I  [\t\phi,\phi]  - i D^{IJ}\xi_J +\sigma^{\mu\nu}\xi^I F_{\mu\nu} +i \sqrt2 \sigma^\mu \t\xi^I D_\mu \phi~, \cr
&\delta \t\lambda_I  &=&\; - i \t\xi_I [\t\phi, \phi] - i D_{IJ} \t\xi^J -  \t\sigma^{\mu\nu} \t\xi_I F_{\mu\nu}+i \sqrt2 \t\sigma^\mu \xi_I D_\mu \t\phi~,\cr
& \delta D^{IJ}  &=&\; -\xi^I \sigma^\mu D_\mu \t\lambda^J -\xi^J\sigma^\mu D_\mu \t\lambda^I - \sqrt2 \xi^I [\t\phi, \lambda^J]- \sqrt2 \xi^J [\t\phi, \lambda^I]\cr
&&&\;  -\t\xi^I\t\sigma^\mu D_\mu \lambda^J -\t\xi^J\t\sigma^\mu D_\mu \lambda^I + \sqrt2 \t\xi^I [\phi, \t\lambda^J]+ \sqrt2 \t\xi^J [\phi, \t\lambda^I]~.
 \eea
 Here, the gauge-covariant derivatives act on any adjoint-valued field $\varphi$ as  $D_\mu \varphi = \d_\mu \varphi - i [A_\mu, \varphi]$.  Note that the supersymmetry transformations \eqref{susy vec offshell Neq2} hold off-shell. 

Upon projection onto the $\CN=1$ subalgebra \eqref{Neq1subalg}, the vector multiplet decomposes into an $\CN=1$ vector multiplet $ \CV_{\CN=1}= (A_\mu, \lambda, \t\lambda, D)_{\CN=1}$ and an $\CN=1$ chiral multiplet $\Phi= (\phi, \psi, F)$, $\t \Phi= (\t\phi, \t\psi, \t F)$, with:
 \be
 \lambda^I = (\psi, \lambda)_{\CN{=}1}~, \quad
\t \lambda_I = (\t\psi, \t\lambda)_{\CN{=}1}~, \quad 
 D_{IJ} = \mat{\sqrt2 i \t F &D+[\t\phi, \phi] \\ D+[\t\phi, \phi] & \sqrt2 i F }_{\CN{=}1}~.
 \ee

\paragraph{The hypermultiplet.}
Consider a hypermultiplet $\CH$ valued in the representation $\FR$ of $\Fg$. The physical field components are:
\be
\CH= \left( q^I~,\, \t q_I~,\, \eta~,\, \t\eta~,\, \chi~,\, \t\chi \right)~.
\ee
Note that the fields $q^I$, $\eta$ and $\t\chi$ are valued in $\FR$, while $\t q_I$, $\t\eta$ and $\chi$ are valued in the conjugate representation, $\b\FR$.
The Lagrangian for the hypermultiplet coupled to a vector multiplet then reads:
\bea\label{H kin 4d}
&\SL_{\CH} &=&\;\; D_\mu \t q_I D^\mu q^I +i \t\eta \t\sigma^\mu D_\mu \eta + i \chi \sigma^\mu D_\mu\t \chi 
+ \t q_I \left( \{ \t\phi, \phi\} {\delta^I}_J + {D^I}_J\right) q^J  
\cr
&&&  + \sqrt2 i \chi \phi \eta - \sqrt2 i  \t\eta \t\phi  \t\chi + \sqrt2 i \t q_I \lambda^I \eta - \sqrt2 i \t\eta \t \lambda_I q^I \cr
&&& +\sqrt2 i \epsilon_{IJ} \chi \lambda^I q^J + \sqrt2 i \epsilon^{IJ} \t q_I \t\lambda_J \t\chi~.
\eea
 Here the trace over gauge indices is left implicit. 
 The on-shell $\CN=2$ supersymmetry transformations for the hypermultiplet read:
\bea\label{onshellhyper 4d R4}
&\delta q^I= \sqrt2 \xi^I \eta+  \sqrt2 \t\xi^I \t\chi~, \cr
&\delta \t q_I= \sqrt2 \xi_I \chi   - \sqrt2 \t \xi_I \t\eta~,  \cr
&\delta \eta =2 i \xi_I \t\phi q^I  -  i \sqrt2 \sigma^\mu\t\xi_I D_\mu q^I~,   \cr
&\delta \chi=2i \xi^I  \t q_I \t\phi +  i \sqrt2 \sigma^\mu \t\xi^I D_\mu \t q_I~, \cr
&\delta \t\eta=  i\sqrt2 \t\sigma^\mu \xi^I D_\mu \t q_I -2i \t\xi^I  \t q_I \phi~, \cr
&\delta \t\chi =i \sqrt2 \t\sigma^\mu \xi_I D_\mu q^I +  2 i \t\xi_I \phi q^I~.
\eea
Focussing on the right-chiral supersymmetries, we have the supersymmetry algebra:
\be 
\{\delta_{\t\xi}, \delta_{\t\xi'}\} \varphi =2i  \sqrt{2} (\t\xi_I \t\xi'^I) \phi  \varphi~,
\ee
with $\phi$ acting on the field $\varphi$ in the appropriate gauge representation ($\FR$ or $\t \FR$). It is satisfied upon imposing the equations of motion for the fermions:
\bea
& i \t \sigma^\mu D_\mu \eta - \sqrt2 i  \t\phi \t\chi  - \sqrt2 i \t\lambda_I q^I=0~, \quad
&i \t\sigma^\mu D_\mu \chi - \sqrt2 i \t\eta  \t\phi + \sqrt2 i \epsilon^{IJ} \t q_I \t\lambda_J=0~, \cr
& i  \sigma^\mu D_\mu \t\eta + \sqrt2 i  \chi \phi+ \sqrt2 i \t q_I \lambda^I=0~, \quad
&i \sigma^\mu D_\mu \t\chi + \sqrt2 i   \phi \eta+ \sqrt2 i \epsilon_{IJ} \lambda^I q^J=0~.
\eea
Note also  that, in terms of the $\CN=1$ subalgebra \eqref{Neq1subalg}, we have two chiral multiplets:
\be
Q= (q, \psi_{Q})_{\CN{=}1}~, \qquad Q' = (q',  \psi_{Q'})_{\CN{=}1}~,
\ee
transforming in the representations $\FR$ and $\b\FR$ of $\Fg$, respectively, 
and their charge conjugate anti-chiral multiplets:
\be
\t Q= (\t q, \t\psi_{Q})_{\CN{=}1}~, \qquad{ \t Q}' = ({\t  Q'}, \t \psi_{Q'})_{\CN{=}1}
\ee
In the $\CN=2$ notation, we have:
\bea
&q^I= (\t q', q)^{\CN{=}1}~, \qquad \qquad
&&\t q_I = (q', \t q)^{\CN{=}1}~, \cr
& \eta= \psi_Q~, \; \;
\t\eta=\t \psi_Q~, 
 \qquad && \chi = \psi_{Q'}~, \; \; 
 \t \chi = \t \psi_{Q'}~.
\eea

\medskip

\noindent \paragraph{Off-shell formulation on K\"ahler Manifolds.} The  $\CN=2$ kinetic term \eqref{H kin 4d} is the sum of the standard $\CN=1$ kinetic term and of the $\CN=1$ superpotential contribution $W = \sqrt2 i q' \phi q$. 

It is well known that an off-shell formulation that realizes all eight supercharges for the hypermultiplet requires an infinite number of fields. Furthermore, the off-shell prescription that is obtained from the $\CN = 1$ language, in terms of the auxiliary fields of $\CN=1$ chiral multiplets, leads to an action that is not `$\mathcal{Q}$-exact' in the usual Donaldson-Witten twist \cite{Labastida:2005zz}. 
%For one localizing supercharge, one way of realising the algebra off-shell is by introducing bosonic auxiliary fields $F^{\check{A}}$, carrying the index $\check{A}$ of a local $SU(2)$ symmetry \cite{Hosomichi:2012ek, Hosomichi:2016flq}. Additionally, auxiliary spinors $\check{\xi}_{\alpha\check{A}}$ transforming in the fundamental representation of this local $SU(2)$ symmetry and satisfy additional ad-hoc constraints, are needed in that approach. 
% In the following, we explain how
 %another,  alternative 

 Here, we give an explicit off-shell realisation of the two specific supersymmetries relevant to the DW twist. (For another approach valid for a single generic supercharge, see {\it e.g.}  \cite{Hosomichi:2012ek, Hosomichi:2016flq}.)
We introduce the spinors $h_{\alpha}, \widetilde{h}_{\alpha}$ as auxiliary bosonic fields such that, for the two supercharges corresponding to the supersymmetry parameters 
\be
 \t \xi_{(1) I}^\alphadot=\delta^{\alphadot 1} \delta_{I 1}~, \qquad\qquad
 \t \xi_{(2) I}^\alphadot=  \delta^{\alphadot 2} \delta_{I 2}~,
 \ee
 we have:
\bea
    &\delta_1 q^I = \sqrt{2}\t\xi^I_{(1)} \t\chi~, & & \delta_2 q^I = \sqrt{2}\t \xi^I_{(2)} \t\chi~, \\
    &\delta_1 \t q_I = -\sqrt{2}\t\xi_I^{(1)} \t\eta~, & & \delta_2 \t q_I = -\sqrt{2}\t \xi_I^{(2)} \t \eta~, \\
    & \delta_1 \eta_{\alpha} = -i\sqrt{2}\left(\sigma^{\mu}\t \xi^{(1)}_I\right)_{\alpha} D_{\mu}q^I + h_{\alpha}~, &&  \delta_2 \eta_{\alpha} = -i\sqrt{2}\left(\sigma^{\mu} \t \xi^{(2)}_I\right)_{\alpha} D_{\mu}q^I~, \\
    & \delta_1 \chi_{\alpha} = i\sqrt{2} \left(\sigma^{\mu} \t \xi_{(1)}^I\right)_{\alpha} D_{\mu} \t q_I~, & & \delta_2 \chi_{\alpha} = i\sqrt{2} \left(\sigma^{\mu} \t \xi_{(2)}^I\right)_{\alpha} D_{\mu} \t q_I + \t h_{\alpha},\\
    & \delta_1 \t \eta^{\dot \alpha} = -2i \t \xi^{I \dot \alpha}_{(1)} \t q_I \phi~, & & \delta_2 \t \eta^{\dot \alpha} = -2i\t \xi^{I \dot \alpha}_{(2)} \t q_I \phi~, \\
    & \delta_1 \t \chi^{\dot \alpha} = 2i\t \xi_{(1)I}^{\dot \alpha} \phi q^I~, && \delta_2 \t \chi^{\dot \alpha} = 2i\t \xi_{(2)I}^{\dot \alpha} \phi q^I~, \\
\eea
with the variations of the auxiliary fields given in terms of the fermion equations of motion as:
\bea
    & \delta_1 h_{\alpha} = 0~, \qquad\qquad \qquad\qquad \delta_2 h_{\alpha} = 2i\kappa \sqrt{2}\phi\eta_{\alpha} + 2i\kappa \left(\sigma^{\mu}D_{\mu}\t \chi \right)_{\alpha} + 2i\kappa\sqrt{2}\epsilon_{IJ}\lambda^{I}_{\alpha}q^J~,  \\
    & \delta_1 \t h_{\alpha} = -2i\kappa \sqrt{2}\chi_{\alpha}\phi - 2i\kappa \left( \sigma^{\mu}D_{\mu}\t \eta \right)_{\alpha}-2i\kappa \sqrt{2}\t q_I \lambda^I_{\alpha}~,   \qquad\qquad\qquad\qquad \delta_2 \t h_{\alpha} = 0~.
\eea
Here $\kappa$ is the scalar constructed from the DW supersymmetry parameters:
\be
    \kappa = \t \xi^I_{(1)}\t \xi_{(2)I}=1~.
\ee
Note that the differential forms $h^{0,1}$ and $\t h^{1,0}$ introduced in section~\ref{subec: 4d hyper} can be expressed in terms of the above auxiliary fields, in the local frame basis, as:
\be
    h^{0,1} = -h_2 e^{\b 1} - h_1 e^{\b 2}~, \qquad \quad \t h^{1,0} = \t h_1 e^1 + \t h_2 e^2~.
\ee
For this choice of supercharges, the kinetic Lagrangian is $\CQ$-exact and can be written as:
\bea
    \SL_{\CH} & = \frac{1}{2}\left(\delta_1+\delta_2\right)\left( -\t h\eta - i\sqrt{2}\chi\sigma^{\mu} \t \xi_I D_{\mu}q^I + i\sqrt{2} \t \eta \t \xi_I \t \phi q^I + 2i \t q_I \t \xi^J \t \lambda^I q_J \right)~,
\eea 
where we defined $\t\xi = \t\xi^{(1)} + \t\xi^{(2)}$.

%%%%%%%%%%%%%%%%%%%%%%%%%%%%%%%%%%%%%%%%%%%%%%%%
  
 \subsection{Five-dimensional conventions}\label{app:5dconventions} 
 \subsubsection{Flat-space conventions}
  We use five-dimensional conventions that naturally reduce to our four-dimensional conventions upon circle reduction along the fifth coordinate. In flat space, we use the coordinates $x^M=(x^\mu, x^5)$, with the index $M=(\mu, 5)$ ($\mu=1,\cdots, 4$).  The five-dimensional spin group is:
\be
  {\rm Spin}(5)\cong {\rm Usp}(4)~.
\ee
The 5d Dirac spinor $\Psi=\Psi_{\bf a}$ transforms in the $\b {\bf 4}$  of ${\rm USp}(4)$, with ${\bf a}=1, \cdots, 4$ the Dirac index. The 5d $\gamma$-matrices, $\gamma^M$, are chosen to be:
  \be
\gamma^\mu= \mat{0 & \sigma^\mu \\  - \t\sigma^\mu & 0}~, \qquad \qquad
\gamma^5= \gamma^1 \gamma^2 \gamma^3 \gamma^4= \mat{\mathbf{1} & 0 \\ 0 & -\mathbf{1}}~,
\ee
with the natural index positions $\gamma^M=( {{\gamma^M}_{\bf a}}^{\bf b})$.
The Dirac representation matrices are: 
\be\label{SigmaMN def}
{(\Sigma^{MN})_{\bf a}}^{\bf b} = {i\ov 4}{ [\gamma^M, \gamma^N]_{\bf a}}^{\bf b}~.
\ee
The Dirac indices are raised with:
\be
\Omega^{{\bf a}{\bf b }}= \mat{\epsilon^{\alpha\beta} & 0 \\ 0 & \epsilon_{\alphadot\betadot}} 
=\mat{0& 1 &0 &0\\ -1&0&0&0\\ 0&0&0&-1\\ 0&0&1&0}~, 
\ee
such that $\Psi^{\bf a}\equiv \Omega^{{\bf a} {\bf b}}\Psi_{\bf b}$.
In these conventions, the 5d Dirac spinor simply reduces to the 4d Dirac spinor upon dimensional reduction along $x^5$. Denoting by $\t \Psi$ the hermitian conjugate to $\Psi$ in Lorentzian signature, we have:
\be
(\Psi_{\bf a})= \mat{\psi_\alpha \\ \t\chi^\alphadot}~, \qquad 
(\t\Psi^{\bf a})= \mat{\chi^\alpha~, \; - \t\psi_\alphadot}~.
\ee
Given two  Dirac spinors $\Psi$ and $\Psi'$, we have the five-dimensional scalar
$\Psi \Psi' \equiv \Psi^{\bf a} \Psi'_{\bf a}= \Omega^{\bf a\bf b}   \Psi_{\bf b} \Psi'_{\bf a} = \psi \psi'+ \t\chi \t\chi'$,
which is a sum of two four-dimensional scalars.

The 5d Dirac representation ({\it i.e.} the $\b{\bf 4}$ of ${\rm Spin}(5)$) is pseudo-real but not real, therefore we cannot impose the Majorana condition. Nonetheless, given a pair of spinors $\Psi^I$ ($I=1,2$), we can impose the Majorana-Weyl (MW) condition:%
\footnote{Here $\t\Psi$ is the Hermitian conjugate of $\Psi$ in Lorentzian signature, which involves conjugation of the $SU(2)$ representation (the index $I$). In 4d notation, we simply have $\Psi^I = \mat{\psi^I \cr -\epsilon^{IJ}\t \psi_J}$ for a 5d Majorana-Weyl spinor.}
\be
\t \Psi^{\bf a}_I = \epsilon_{IJ} \Omega^{{\bf a}{\bf b}} \Psi_{\bf b}^J~.  
\ee
In particular, the five-dimensional supercharge $\CQ=(\CQ^I)$ is a Majorana-Weyl spinor, which reduces to the two distinct Majorana spinors $Q^I$ ($I=1,2$) in 4d.

  \subsubsection{Curved-space conventions}
As explained in section~\ref{subsec:M5 geom}, we are interested in five-manifolds that are fibered over a K\"ahler four-manifold $\CM_4$. In particular, we pick a distinguished `fifth dimension'. 
 We choose the complex frame (with the fifth direction along a real covector $\eta$) and a metric adapted to the fibration structure:
\be\label{5d frame}
(e^{\bf A})= (e^1, e^{\b 1}, e^2, e^{\b 2}, \h e^5\equiv \eta)~, \qquad ds^2= e^1 e^{\b 1}+e^2 e^{\b 2} + \eta^2~.
\ee
We also denote by $x^M$ ($M=1, \cdots, 5$) any local coordinates. The spin connection and curvature tensors are defined exactly as in 4d. 
The covariant derivative on 5d Dirac spinors takes the form:
\be
\nabla_M \Psi  \equiv \big(\d_\mu - {i\ov 2} \omega_{M {\bf AB}} \Sigma^{\bf A\bf B}\big) \Psi~,
\ee
with $\Sigma^{MN}$ defined in \eqref{SigmaMN def}, and $\bf{A}, \bf{B}$ denoting 5d frame indices.

%%%%
 \subsubsection{5d $\CN=1$ supersymmetry}\label{app:5dsusy}
Here we discuss 5d $\CN=1$ supersymmetry in flat space. The five-dimensional supersymmetry algebra reads:
\be\label{5d susy alg R5}
\{\CQ^I_{\bf a}, \CQ^J_{\bf b}\} = 2 \epsilon^{IJ}\left(\gamma^M_{{\bf a}{\bf b}} P_M - i \Omega_{{\bf a}{\bf b}} Z_{\rm 5d}\right)~,
\ee
 with $\CQ^I$ the Majorana-Weyl supercharge, and $Z_{\rm 5d}$ a real central charge. Upon dimensional reduction along the fifth dimension, \eqref{5d susy alg R5} reproduces \eqref{4d susy alg R4} with:
\be
\CQ^I_{\bf a} =\mat{ Q^I_\alpha \\ - \epsilon^{IJ} \t Q^\alphadot_J }~, \qquad Z= P_5+i Z_{\rm 5d}~, \qquad \t Z= P_5-i Z_{\rm 5d}~.
\ee
 We denote the supersymmetry variation by:
 \be
 \delta = -i \zeta_I \CQ^I~.
 \ee
 The commuting-spinor parameter $\zeta_I$ is related to the 4d parameters according to:
 \be
\zeta_I =\mat{-\xi_I \\ \t\xi_I }~.
\ee

 \paragraph{5d $\CN=1$ vector multiplet.}
The five-dimensional vector multiplet consists of a gauge field $A_M$, a real scalar field $\sigma$,  a Majorana-Weyl spinor $\Lambda_I$ ($I=1,2$)  and a triplet of auxiliary scalar fields $D_{IJ}=D_{JI}$:
\be
\CV_{\rm 5d} = \left(A_M~,\, \sigma~,\, \Lambda_I~,\, D^{IJ}\right)~.
\ee
The 5d $\CN=1$ super-Yang-Mills action reads:
\bea\label{SYM 5d flat}
&\SL_{{\rm  SYM}} &=&\;\; {1\ov g_{\rm 5d}^2} \tr\Big( {1\ov 4}F_{MN} F^{MN}+\half D_M \sigma D^M \sigma + {i\ov 2} \Lambda^I \gamma^M D_M \Lambda_I  \cr
&&&\qquad\qquad\;+ {i\ov 2} \Lambda^I [\sigma, \Lambda_I]+ {1\ov 4 } D_{IJ} D^{IJ}\Big)~.
\eea
One easily checks that it reduces to the 4d Lagrangian \eqref{Neq2 SYM} upon dimensional reduction on a circle.
The 5d $\CN=1$ supersymmetry transformations are:
\bea\label{vec5d susy}
& \delta A_M &&=\;  i \zeta_I \gamma_M \Lambda^I~, \cr
& \delta \sigma&&=\; - \zeta_I \Lambda^I~, \cr
& \delta \Lambda_I &&=\;- i \Sigma^{MN} \zeta_I\,  F_{MN} + i \gamma^M \zeta_I D_M \sigma - i D_{IJ} \zeta^J~, \cr
& \delta D^{IJ} &&=\;  \zeta^I \gamma^M D_M \Lambda^J+  \zeta^J \gamma^M D_M \Lambda^I + \zeta^I [\sigma, \Lambda^J] +\zeta^J [\sigma, \Lambda^I]~.
\eea
This reduces to \eqref{susy vec offshell Neq2} in 4d. The dimensional reduction is straightforward to carry out explicitly. The gauge field $A_M$ splits as $(A_\mu, A_5)$, with $A_5$ a real scalar, 
and the 4d complex scalar $\phi, \t\phi$ is given by:
\be
\phi= {1\ov \sqrt{2}} \left( \sigma+ i A_5\right)~, \qquad \quad
\t\phi= {1\ov \sqrt{2}} \left( \sigma- i A_5\right)~.
\ee
The auxiliary fields $D_{IJ}$ are taken to be the same in 5d and 4d, and the  gauginos are related as:
\be
\Lambda_I = \mat{\epsilon_{IJ} \lambda^J \\ \t\lambda_I}~.
\ee
One can check that the SUSY algebra closes up to translations and gauge transformations. Picking any two constant supersymmetry parameters $\xi_I$ and $\zeta_I$ (taken as commuting MW spinors), let us define the following bilinears:
\be
   K^M  = - \xi^I \gamma^{M}\zeta_I~, \qquad\quad
    \kappa = \xi^I \zeta_I~.
\ee
One finds, in particular (see {\it e.g.}  \cite{Hosomichi:2012ek}):
\bea\label{susy transformations vect 5d flat}
  &  \{\delta_{\xi}, \delta_{\zeta} \} \sigma & =&\; -2iK^M D_M \sigma~, \\
  &  \{\delta_{\xi}, \delta_{\zeta} \} A_M & =&\; -2i K^N F_{NM} + 2\kappa D_M \sigma~, \\
 &   \{\delta_{\xi}, \delta_{\zeta} \} \Lambda_{I } & =&\;  -2 i K^M D_M \Lambda_{I } + 2i \kappa \left[ \sigma,\Lambda_{I} \right]~, \\
 &   \{\delta_{\xi}, \delta_{\zeta} \} D_{IJ} & =&\; -2i K^M D_M D_{IJ} +2 i \kappa \left[ \sigma,D_{IJ} \right]~. 
\eea
The supersymmetric Lagrangian \eqref{SYM 5d flat} is `almost' $\CQ$-exact, like in 4d. For any  supersymmetry parameter  $\zeta_I$, we have:
\bea
    \SL_{\rm SYM} = & {1\ov g^2_{5d}}\, {1\ov \zeta^K  \zeta_K} \,\delta \left( -{1\ov 4} \zeta^I \gamma^{MN}\Lambda_I F_{MN} - {i\ov 2}\zeta_J \Lambda_I D^{IJ} - {i\ov 2}\zeta^I \gamma^M\Lambda_I D_{M}\sigma\right) +\\
   & \quad + {1\ov 8g^2_{5d}}\, {\zeta^I \gamma_R \zeta_I \ov \zeta^J  \zeta_J}\, \epsilon^{MNPQR}F_{MN}F_{PQ}~.
\eea
The second line reduces to the instanton density in 4d. 

 %%%%%%%%%%%%%%%%%%%%%%%%%
 \paragraph{5d $\CN=1$  hypermultiplet.}\label{app:5d hyper}
Consider the 5d charged hypermultiplet:
\be
\CH_{\rm 5d} =\left( q^I~,\, \t q_I~,\, \Psi~,\, \t\Psi \right)~.
\ee
The free-field Lagrangian for the hypermultiplet coupled to a vector multiplet reads:
\bea\label{kin hyper 5d R5}
&\SL_{\CH} &=&\;\; D_M \t q_I D^M q^I +i \t\Psi \gamma^M D_M \Psi +\t q_I \left(\sigma^2 {\delta^I}_J + {D^I}_J\right) q^J - i \t\Psi \sigma \Psi 
\cr
&&& - \sqrt2 i \t\Psi \Lambda_I q^I- \sqrt2 i \t q_I \Lambda^I \Psi~.  
\eea
The on-shell supersymmetry transformations are:
\bea\label{hyper5d susy}
& \delta q^I &&=\; \sqrt2 \zeta^I \Psi~, \cr
& \delta \t q_I &&=\; - \sqrt2 \zeta_I \t\Psi~, \cr
& \delta \Psi &&=\; \sqrt2 i \gamma^M\zeta_I D_M q^I + \sqrt2 i \zeta_I \sigma q^I~, \cr
& \delta \t\Psi &&=\; \sqrt2 i \zeta^I \gamma^M D_M \t q_I - \sqrt2 i \zeta^I  \t q_I \sigma~, \cr
\eea
with the fields $(q,\Psi)$ transforming in some representation $\FR$ of the gauge group and the fields $(\t q,\t \Psi)$ in the complex conjugate representation $\b \FR$. In our conventions, expressions of the type $\sigma f$ refer to contractions of the type $\sigma^a T^a_{ij} f^j$, where $T^a_{ij}$ are generators for the representation under which the field $f$ transforms, with indices $i,j$, while $a$ are adjoint representation indices. We use the generators of the representation $\FR$ for all fields of the five-dimensional hypermultiplet. Thus, the covariant derivatives become:
\be
    D_M q^I = \partial_M q^I - i A_M q^I~, \qquad \quad   D_M \t q_I = \partial_M \t q_I + i \t q_I A_M~.
\ee
The kinetic Lagrangian left invariant by the above transformations is:
\bea
    \SL_{\CH} & = D_M \t q_I D^M q^I + i\t \Psi \gamma^M D_M\Psi + \t q_I \sigma\sigma q^I + \t q_I D\indices{^I_J}q^J - i\t \Psi \sigma \Psi \\
    & \qquad + i\sqrt{2}\t \Psi \Lambda^I q_I - i\sqrt{2}\t q_I \Lambda^I \Psi~.
\eea
When reducing to 4d, the Lagrangian \eqref{kin hyper 5d R5} and the variations \eqref{hyper5d susy} reproduce \eqref{H kin 4d} and \eqref{onshellhyper 4d R4}, respectively, with the identification:
\be
\Psi= \mat{-\eta \\ \t \chi}~, \qquad \t\Psi = \mat{\chi~, \t\eta}~.
\ee
The equations of motion for the fermions read:
\bea
    i \gamma^M D_M \Psi -i\sigma \Psi + i\sqrt{2} \Lambda^I q_I = 0 ~, \\
    -i D_M \t\Psi \gamma^M - i\t \Psi\sigma -i\sqrt{2} \t q_I \Lambda^I = 0~.
\eea
It is straightforward to first show that:
\bea
    \{\delta_{\xi},\delta_{\zeta}\} q^I & =-2 iK^M D_M q^I + 2i\kappa \sigma q^I~, \qquad
    \{\delta_{\xi},\delta_{\zeta}\} \t q_I & = -2iK^M D_M \t q_I -2 i\kappa \t q_I \sigma~, 
\eea
and using the equations of motion we then find that:
\bea
    \{\delta_{\xi},\delta_{\zeta}\} \Psi & = -2iK^M D_M \Psi + 2i\kappa \sigma \Psi ~, \qquad
    \{\delta_{\xi},\delta_{\zeta}\} \t \Psi & = -2iK^M D_M \t \Psi -2 i\kappa \t \Psi \sigma~.
\eea
For the off-shell formulation, let us focus on two supercharges with the corresponding Killing spinors being the uplift of the four dimensional spinors $\delta^{\dot \alpha \dot 1}\delta_{I 1}$ and $\delta^{\dot \alpha \dot 2}\delta_{I 2}$. In this setup, the five-dimensional spinors of interest are:
\be \label{5d Killing spinors}
    (\zeta_{(1) I \bf a}) = \left(\begin{matrix}
     0 \\ \delta^{\dot \alpha \dot 1}\delta_{I 1}
    \end{matrix}\right)~, \qquad (\zeta_{(2)I \bf a}) = \left(\begin{matrix}
     0 \\ \delta^{\dot \alpha \dot 2 }\delta_{I 2}
    \end{matrix}\right)~.
\ee
We introduce the auxiliary fields (five-dimensional commuting spinors) $h_{\bf a}, \t h^{\bf a}$, with components $(h_1, h_2, 0, 0)$, such that the SUSY variations become:
\bea\label{susy h off shell R5 -1}
    \delta_1 q^I & = \sqrt{2}\,\zeta_{(1)}^I \Psi~, \\
     \delta_2 q^I & = \sqrt{2}\,\zeta_{(2)}^I \Psi~,\\
     \delta_1 \t q_I & = - \sqrt{2}\,\zeta_{(1)I} \t \Psi~,  \\
      \delta_2 \t q_I & = - \sqrt{2}\,\zeta_{(2)I} \t \Psi~, \\
    \delta_1 \Psi_{\bf a} & = i\sqrt{2} (\gamma^M \zeta_{(1)I})_{\bf a} D_M q^I + i\sqrt{2}\zeta_{(1)I \bf a} \sigma q^I + h_{\bf a}~,\\
    \delta_2 \Psi_{\bf a} & = i\sqrt{2} (\gamma^M \zeta_{(2)I})_{\bf a} D_M q^I + i\sqrt{2}\zeta_{(2)I \bf a} \sigma q^I~,\\
    \delta_1 \t \Psi^{\bf a} & = i\sqrt{2} (\zeta^{I}_{(1)} \gamma^M)^{\bf a}   D_M \t q_I - i\sqrt{2} \zeta^{I \bf a}_{(1)} \t q_I \sigma~,\\
    \delta_2 \t \Psi^{\bf a} & = i\sqrt{2} (\zeta^{I}_{(2)} \gamma^M)^{\bf a}   D_M \t q_I - i\sqrt{2} \zeta^{I \bf a}_{(2)}  \t q_I \sigma + \t h^{\bf a}~,
\eea
together with:
\bea\label{susy h off shell R5 -2}
    \delta_1 h_{\bf a} & = 0~, \\
   \delta_2 h_{\bf a} & \approx ~ 2i\kappa \sigma \Psi_{\bf a}  - 2i\kappa \left(\gamma^M D_M \Psi\right)_{\bf a} - 2i\sqrt{2}\kappa \Lambda^I_{\bf a} q_I~,  \\
   \delta_1 \t h^{\bf a} & \approx -2i\kappa \t \Psi \sigma - 2i\kappa \left(D_M\t \Psi \gamma^M\right)^{\bf a} - 2i\sqrt{2} \kappa \t q_I \Lambda^{I \bf a} ~, \\
   \delta_2 \t h^{\bf a} & = 0~.
\eea
The $\approx$ symbol is used to emphasise that these equations should be interpreted as equalities only when restricted to the non-zero components of $h$ and $\t h$, {\it i.e.} for the indices $\bf a\in \{1,2\}$. For the variations to be correct for any $\bf a$, one needs to introduce some additional terms that can be worked out from $\delta\delta\Psi_{\bf a}$. Note also that we restrict the definitions of $\kappa$ and $K^M$ to the two spinors $\zeta_{(1)}$ and $\zeta_{(2)}$. In particular, for the Killing spinors \eqref{5d Killing spinors} and with our choice of frame, we have
\be
K^M \d_M = \d_5~, \qquad\quad \kappa =1~.
\ee
 The off-shell Lagrangian becomes:
\bea
    \SL_{\CH} & = D_M \t q_I D^M q^I + i\t \Psi \gamma^M D_M\Psi + \t q_I \sigma\sigma q^I + \t q_I D\indices{^I_J}q^J - i\t \Psi \sigma \Psi \\
    & \qquad + i\sqrt{2}\t \Psi \Lambda^I q_I - i\sqrt{2}\t q_I\Lambda^I \Psi + \frac{1}{2}\t h^{\bf a} h_{\bf a}~,
\eea
which can be shown to be $\CQ$-exact, 
\bea
   \SL_{\CH}  & = \frac{1}{2\kappa}(\delta_1+\delta_2) \left( \t h \Psi - i\sqrt{2} \, \t \Psi \gamma^M \zeta_I \, D_M q^I + i\sqrt{2} \, \t \Psi \zeta_I \, \sigma q^I + 2i \, \t q_I \zeta^{J} \Lambda^I q_J \right)~.
\eea
Here we used the notation $\zeta = \zeta_{(1)} + \zeta_{(2)}$.

 %%%%%%%%%%%%%%%%%%%%%%%%%%%%%%%%%%%%%%

 \section{More about one-loop determinants}
The material of this appendix complements the discussion of section~\ref{sec:oneloopdet}.

\subsection{Hypermultiplet mode cancellation}\label{app:hyperModePairing}
 
Let us consider the one-loop determinant for the hypermultiplet on $\CM_4$, as in section~\ref{subec:1loop 4d}. Here, for completeness, we would like to explicitly display the mode cancellation between fermions and bosons. Recall that the perturbative contribution to the partition function on $\cM_4$ comes from:
\be 
{\det\left(\Delta_{\rm fer}\right) \ov \det \left( \Delta_{\rm bos} \right) }  = { \det(\Lop^{0,1}) \ov \det(\Lop^{0,0}) \det(\Lop^{0,2})} ~,
\ee
where $\Lop = -i\sqrt{2}\phi$. A different approach to computing this one-loop determinant is to count the modes of $\Lop$, which will be related to the index of the twisted Dolbeault operator as discussed in the main text. To show this, one starts with a fermionic eigenmode with eigenvalue $M_{\rm fer}$:
\be \label{Fermionic eigenvalue eqn}
    \Delta_{\rm fer}\Psi = M_{\rm fer} \Psi~,
\ee
where $\Psi = (\eta^{0,1}, \t\chi^{\,0,0}, \t \chi^{\,0,2})$ and $\Delta_{\rm fer}$ is given by \eqref{fermionic kinetic operator}. Then, combining the first and third equations obtained from \eqref{Fermionic eigenvalue eqn}, one finds a bosonic mode for $Q^{0,0}$, with the bosonic eigenvalue given by:
\be
    M_{\rm bos}  = M_{\rm fer} \left(-2M_{\rm fer} - 2\sqrt{2}(\phi - 2\t\phi) \right)~.
\ee
In fact, the same mode can also be constructed by starting with the fermionic eigenvalue:
\be
    -M_{\rm fer} - {i\ov \sqrt{2}}(\phi - 2\t\phi)~,
\ee
and thus the two fermionic modes are paired with one bosonic mode. This construction relies on the map:
\be
    \t\chi^{\,0,0} \longrightarrow Q^{0,0}~,
\ee
being independent of $\t\chi^{\,0,2}$. Similarly, the bosonic modes for $Q^{0,2}$ are constructed solely from $\t\chi^{\,0,2}$, with the fermionic eigenvalues:
\be
    M_{\rm fer}~, \qquad M_{\rm fer} - {i\ov 2\sqrt{2}}(2\phi +\t \phi)~,
\ee
being mapped to the bosonic mode with:
\be
    M_{\rm fer} \left( 2M_{\rm fer} + {i\ov \sqrt{2}}(2\phi + \t \phi) \right)~.
\ee
In both constructions, however, there are unpaired modes that correspond to degrees of freedom for $\eta^{0,1}$. From the eigenvalue equation \eqref{Fermionic eigenvalue eqn}, these modes are given by:
\be
    \star \partial_A \star \eta^{0,1} = 0~, \qquad \quad \star \b\partial_A \eta^{0,1} = 0~,
\ee
for the two constructions, respectively. For the reverse map, fermionic modes can be built from bosonic modes in two ways, that is:
\be
    \t \chi^{\,0,0} = c Q^{0,0}~, \qquad \quad \eta^{0,1} = c{2(\sqrt{2}\t\phi + i M_{\rm fer}) \ov 2\phi \t\phi - M_{\rm bos}}\, \b\partial_A Q^{0,0}~, 
\ee
or, alternatively:
\be
    \t \chi^{\,0,2} = cQ^{0,2}~, \qquad \quad \eta^{0,1} = c {\sqrt{2}\t\phi - 4iM_{\rm fer} \ov 2\phi\t\phi - 4M_{\rm bos}} \star \partial_A Q^{0,2}~,
\ee
for some constant $c$. The map is again $1$ (boson) to $2$ (fermions), with the eigenvalues being the same as before. In this case, the pairing is not complete if the two fermionic modes built from these maps are not independent. This occurs for:
\be
    \b \partial_A Q^{0,0} = 0~, \qquad \quad \star \partial_A Q^{0,2} = 0~,
\ee
for the two constructions, respectively. This analysis confirms that the one-loop determinant reduces to \eqref{OneLoopDetFullOps}.
 
%%%%%%%%%%%%%%%%%%%%%%%%

\subsection{Vector multiplet one-loop determinant} \label{appendix: vec one-loop}
In this appendix, we compute the one-loop determinant for a W-boson. (We closely follow a similar computation in  \cite{Closset:2015rna}.) Consider a non-abelian 4d $\cN = 2$ vector multiplet on a K\"ahler manifold $\cM_4$ and introduce the usual BRST ghosts $c, \t c$ and auxiliary fields $b$ valued in the adjoint of $\mathfrak{g}$. Then, the standard BRST transformations read:
\bea
   & s A_{\mu} = D_{\mu} c~, \qquad \quad & &  s \varphi_b = i[c,\varphi_b]~, \qquad \quad & & s \varphi_f = i\{c,\varphi_f \}~, \\
    & sc = {i\ov 2}\{c,c \}~, \qquad \quad & & s\t c = -b~, \qquad \quad & & s b = 0~,
\eea
with $\varphi_{b,f}$ the bosonic and fermionic fields in the $\cN=2$ vector multiplet and $s$ the BRST symmetry generator. It follows that $s$ is nilpotent and, furthermore:
 \bea
    \{s,\delta_i\} = 0~, \qquad i=1,2~,
 \eea
 for the two supersymmetry transformations defined in \eqref{susy vec twisted}. We then define the modified supersymmetry transformations:
 \be
    \delta_i' = \delta_i + s~,
 \ee
 which still satisfy the supersymmetry algebra \eqref{4d vec susy closure}:
 \be
(\delta_i')^2  =0~, \qquad\qquad \{ \delta_1', \delta_2'\} = \{ \delta_1, \delta_2\} = 2  \sqrt2 \delta_{g(\phi)}~,
\ee
where $\delta_{g(\phi)}$ is a gauge transformation with parameter $\phi$. Note that the BRST transformation of the vector multiplet fields is just a gauge transformation, so the Yang-Mills lagrangian is automatically invariant under this action. In the absence of supersymmetry, the ghost and gauge-fixing contributions can be recovered from a $s$-exact term. In our case, we have instead:
\be
    \SL_{\rm gf} = (\delta'_1 + \delta'_2) F_{\rm gf}~,
\ee
for some function $F_{\rm gf}$, which ensures that the action preserves supersymmetry. We choose the following conventions:
\be
    \SL_{\rm gf} = {1\ov 2}(\delta'_1 + \delta'_2)\left( \t c \left(G_{\rm gf} + {1\ov 2}\xi_{\rm gf} b \right) \right) = s \left( \t c \left(G_{\rm gf} + {1\ov 2}\xi_{\rm gf} b \right) \right) + {1\ov 2} \t c\, (\delta_1 + \delta_2) G_{\rm gf}~,
\ee
which, upon integrating out the auxiliary field $b$, reads:
\be
    \SL_{\rm gf} = {1\ov 2\xi_{\rm gf}} G_{\rm gf}^2 - \t c \,(s G_{\rm gf}) + {1\ov 2} \t c \,(\delta_1 + \delta_2) G_{\rm gf} ~.
\ee
The gauge-fixing function will typically include a term of the form:
\be
    G_{\rm gf} \supset D_{\mu} A^{\mu} ~,
\ee
leading to the kinetic terms in the Lagrangian:
\be
    \SL_{\rm gf} = {1\ov 2\xi_{\rm gf}} G_{\rm gf}^2 + D_{\mu}\t c \, D^{\mu} c + \ldots ~.
\ee
We will set the gauge-fixing function to:
\be
    G_{\rm gf} = D_{\mu} A^{\mu} + i \xi_{\rm gf} [\phi, \t \phi]~,
\ee
and work in a background where $\phi$ and $\t \phi$ are constant: $\phi = \hat \phi$, $\t \phi = \tilde{\hat{\phi}}$. In the Feynman gauge $\xi_{\rm gf}=1$, expanding at second order in fluctuations around these constant values leads to diagonal kinetic terms between $A_{\mu}$ and $\phi$, $\t \phi$, while the ghost one-loop determinant cancels completely that of $\phi$ and $\t \phi$, since the Lagrangian contains the terms:
\be
    \SL_{\rm SYM} + \SL_{\rm gf} \supset D_{\mu} \t \phi D^{\mu} \phi + D_{\mu} \t c\, D^{\mu} c~.
\ee
For the kinetic terms of the gauge field, we get contributions from:
\be \label{gauge field kinetic terms}
    \SL_{\rm SYM} + \SL_{\rm gf} \supset {1\ov 4}F_{\mu \nu} F^{\mu \nu} + {1\ov 2}(D_{\mu}A^{\mu})^2 - [A_{\mu},\tilde{\hat{\phi}}] [A^{\mu}, \hat \phi]~, 
\ee
where the last term comes from expanding the $D_{\mu} \t \phi D^{\mu} \phi$ kinetic term. In the local frame, in the complex basis, we have:
\be
   {1\ov 4}F_{\mu \nu} F^{\mu \nu} = -2F_{1 \b1}^2 + 4F_{12}F_{\b1 \b2} - 4F_{1\b2}F_{2\b1} - 2F_{2\b2}^2~,
\ee
with the first two terms in \eqref{gauge field kinetic terms} combining to:
\be
     \SL_{\rm kin} \supset -2 A_{\b j} D_{\mu}D^{\mu} A_j - [A_{\mu},\tilde{\hat{\phi}}] [A^{\mu}, \hat \phi]~,
\ee
where the remaining terms correspond to cubic or higher degree terms in the gauge field, and $j \in \{ 1,2\}$. Note that this is almost identical to the bosonic kinetic term of the free hypermultiplet in \eqref{bosonic kinetic operator}, which can be rewritten in terms of the untwisted fields as:
\be
    \SL_{\rm bos}^{\rm hyper} = -\t q_I \left( D_{\mu}D^{\mu} + 2\phi \t\phi\right) q^I~,
\ee
in the supersymmetric background described by \eqref{SUSYbackground1} and \eqref{SUSYbackground2}.
 
Similarly, for the gauginos before the topological twist, the kinetic terms are encoded in:
\be \label{gauginos kinetic terms}
    \SL_{\rm SYM} \supset i\t\lambda_I \t \sigma^{\mu}D_{\mu} \lambda^I - {i\ov \sqrt{2}}\lambda_I[\tilde{\hat{\phi}},\lambda^I] + {i\ov \sqrt{2}}\t\lambda^I[\hat\phi,\t\lambda_I]~, 
\ee
After topological twisting, we group the gauginos into the formal variables $\t \Psi = (\Lambda^{1,0}, \t\Lambda^{0,0}_{(1)}, \t\Lambda^{2,0})$ and $\Psi = (-\Lambda^{0,1}, \t\Lambda^{0,0}_{(2)},\t\Lambda^{0,2})$, such that the kinetic terms of the Lagrangian can be expressed as:
\be
    \SL_{\rm kin} \supset \star \left(\t\Psi \wedge \star \Delta_{\rm fer} \Psi \right)~,
\ee
where:
\bea
    \Delta_{\rm fer} = \left( \begin{matrix}
     -{i\sqrt{2}\ov 4}\,[\tilde{\hat{\phi}}, \cdot\,] & i\,\b\partial_A & ~{i\ov 2}\star \partial_A~ \\
     ~-i\star\partial_A\star~ & i\sqrt{2}\,[\hat\phi,\cdot\,] \dvol & ~0~ \\
     -{i\ov 2}\star \b \partial_A & 0 & ~{i\sqrt{2}\ov 4}\,[\hat\phi,\cdot\,]~
    \end{matrix}\right)~.
\eea
 Note that this is the same as the kinetic operator for the hypermultiplet in \eqref{fermionic kinetic operator}, up to an irrelevant sign in one of the diagonal terms.

One can now work in the Cartan-Weyl basis $E_{\alpha}$, $H_a$, with $\alpha$ denoting the non-vanishing roots and $H_a$ the Cartan elements, by expanding every field as $\varphi = \sum_a \varphi_a H_a +\sum_\alpha \varphi_{\alpha} E_{\alpha}$. The computation is similar to the 2d argument presented in \cite{Closset:2015rna}. 
From this analysis, we  conclude that, at one-loop, the $W$-bosons enter the topologically twisted partition function on $\CM_4$ exactly as a twisted hypermultiplet with higher-spin $(j_l, j_r)=(0, \half)$, in agreement with the discussion in section~\ref{subsec:HSM5}.

%%%%%%%%
\subsection{Index theorems: review and computations}\label{app:index}
In this appendix, we review the Atiyah-Singer index theorem (see {\it e.g.}~\cite{Nakahara:2003nw} for an introduction aimed at physicists), and we work out some indices that are useful for our purposes.
 %The one-loop determinant for a particle of spin $(j_l,j_r)$ is determined by the index of some Dolbeault-like complex. 
 
 \medskip
 \noindent 
Given $(E,D)$ an elliptic complex over an $n$-dimensional compact manifold $M$ without boundary:
\be
    \ldots \xrightarrow[]{D_{i-2}} \Gamma(M, E_{i-1}) \xrightarrow[]{D_{i-1}} \Gamma(M,E_i) \xrightarrow[]{D_{i}} \Gamma(M, E_{i+1}) \longrightarrow \ldots~,
\ee
the Atiyah-Singer theorem states that the index ({\it i.e.} the Euler characteristic) of this complex is determined as follows:
\be
    {\rm ind} (E,D) = (-1)^{\frac{n(n+1)}{2}}\int_M {\rm ch} \left( \bigoplus_{i}(-1)^i E_i \right) \frac{{\rm Td}\left(TM^{\mathbb{C}}\right)}{e(TM)} ~.
\ee
Here ${\rm Td}$ denotes the Todd class, ${\rm ch}$  the Chern character and $e$ the Euler class.  $TM$ is the tangent bundle of $M$, while $TM^{\mathbb{C}}$ is the complexified version of it. 

Two quantities that appear often in the context of 4-manifolds are the Euler characteristic $\chi$ and the signature $\sigma$. They are the indices of the de Rham complex and of the signature complex, respectively. They are given by:
\be
     \chi = \sum_{i=0}^4 (-1)^i b_i = \int_M e(TM)~, \qquad \quad \sigma= b_2^+-b_2^- = {1\ov 3}\int_M p_1(TM)~,
\ee
where $b_i$ are the Betti numbers of $M$ and $p_1$ is the first Pontryagin class. For complex 4-manifolds, we also have $e(TM) = c_2(TM)$ in terms of the second Chern class $c_2$, where in the latter the tangent bundle $TM$ is viewed as a complex vector bundle. 

%%%%%%

\paragraph{Dolbeault Complex.} Let us consider the Dolbeault complex of a complex four-manifold $M$. The Dolbeault complex is the elliptic complex for the ${\b \d}$ operator, with $E_i = \Omega^{0,i}\equiv \bigwedge\nolimits^{i} T^*M^-$, where $\Omega^{0,1}\cong T^*M^-$ is the anti-holomorphic cotangent bundle spanned by $\{d{\b z}^{\mu}\}$. The analytical index is given by the alternating sum of Hodge numbers:
\be
 {\rm ind}({\b \d}) = \sum_{i=0}^{2} (-1)^i h^{0,i}   = \chi_h~,
\ee
called the holomorphic Euler characteristic $\chi_h$, and the index theorem gives us:
\be \label{index: Dolbeault}
    {\rm ind}({\b \d}) = \int_M {\rm Td}(TM^+)~,
\ee
with $TM^+$ the tangent bundle spanned by $\{\d/\d z^{\mu} \}$. 
%This follows from the splitting principle for the Chern characters.  
This integral evaluates to:
\be
    {\rm ind}({\b\d}) = \int_{M} {1\ov 12}\left(c_1(TM^+)\wedge c_1(TM^+) + c_2(TM^+)\right) = {\chi + \sigma \ov 4}~,
\ee
which gives us the relation $\chi_h= {\chi+\sigma\ov 4}$. 
%This quantity is also relevant in the context of Seiberg-Witten invariants. 
 %Note that for general 4-manifolds (\textit{i.e.} not admitting a complex structure), $\chi_h$ does not have to be an integer. However, this turns out to be the case for manifolds of \emph{Seiberg–Witten simple type.}
%
 One can also consider the tensor product bundles $\Omega^{0,i} \otimes V$, for some holomorphic vector bundle $V$ over $M$, leading to the twisted Dolbeault complex:
\be
    \ldots \xrightarrow[]{{\b\d}_V} \Omega^{0,i-1}(M) \otimes V  \xrightarrow[]{{\b\d}_V} \Omega^{0,i}(M) \otimes V \xrightarrow[]{{\b\d}_V} \Omega^{0,i+1}(M) \otimes V  \xrightarrow[]{{\b\d}_V} \ldots ~.
\ee
Applying the splitting principle for the Chern character, 
\bea
    {\rm ch} \left( \bigoplus_i (-1)^i \Omega^{0,i}(M)\otimes V \right) =  {\rm ch} (V) \sum_i (-1)^i {\rm ch} \left( \Omega^{0,i}(M) \right),
\eea
together with the computation for the Dolbeault complex, the index theorem specialises to the Hirzebruch-Riemann-Roch theorem:
\be
    {\rm ind}({\b\d}_V) = \int_M {\rm Td}(TM^+) {\rm ch}(V) = \chi(M,V)~.
\ee
  Here, $\chi(M,V)$ is also often called the holomorphic Euler characteristic.

%%%%%%

\paragraph{Dirac Complex.}  Consider a spin bundle ${\bf S}$ over an $n$-dimensional ($n$ even) orientable manifold $M$. The spin bundle splits as $S^+ \oplus S^-$. The Dirac operator, defined as $\slashed{D} = i\gamma^{\mu}\nabla_{\mu}$, is in fact an elliptic operator, and it also splits as:
\be
    \slashed{D}  = \left( \begin{matrix} 0 & D^+ \\ D^- & 0  \end{matrix} \right).
\ee
In this case, the analytical index, ${\rm ind}(D^+) = n_+ - n_-$, counts the difference between the numbers of positive and negative chirality modes. 
It can be shown that the topological index can be solely expressed in terms of the $\hat{A}$-roof genus, which is a characteristic class containing only $4j$-forms. Thus, the index vanishes unless $n=0 \mod 4$. For $4$-dimensional manifolds, it reads:
\be
    {\rm ind}(\slashed{D}) = \int_M \hat{A}(TM) = -\frac{1}{24}\int_M p_1(TM) = -\frac{1}{8}\sigma(M)~.
\ee
More generally, spinors can transform in some representation of a group $G$. They are then sections of $S(M) \otimes E$, where $E$ is an associated vector bundle to the principal $G$-bundle over $M$, in an appropriate representation. The twisted Dirac operator $\slashed{D}_E$ now contains both the spin connection and a gauge connection on $E$, while the index theorem gives us:
\be
  {\rm ind}(\slashed{D}_E) = n_+ - n_- = \int_M \hat{A} (TM) {\rm ch}(E)~.
\ee

\paragraph{Relation between the twisted Dirac and Dolbeault complexes.} On a complex surface, we have a formal equivalence between the indices of the twisted Dirac and Dolbeault complexes, by tensoring with the square root of the canonical line bundle, $\CK^\half$:
\be
    {\rm ind}({\b\d}_V)  =  {\rm ind}(\slashed{D}_E)~, 
     \qquad \qquad V= \CK^\half \otimes E~,
\ee
where the index theorem gives us:
\be
    {\rm ind}({\b\d}_V) = -{{\rm rk}(E)\ov 8}\sigma + \int_M {\rm ch}(E)~.
\ee
In particular, in the case of the extended topological twist of the hypermultiplet (as introduced in section~\ref{subec: 4d hyper}), we have $V=\CK^{\varepsilon+\half}\otimes L$, where $L$ is a well-defined $U(1)$ bundle with flux $c_1(L)=\m$, and one finds:
\be
    {\rm ind}\left({\b\d}_{\CK^{\varepsilon+\half}\otimes L}\right) =-{\sigma\ov 8} +\half \varepsilon^2 (2\chi+3\sigma)+\int_M c_1(L)\wedge \Big(c_1(L)+ 2 \varepsilon\, c_1(\CK)\Big)~.
\ee
 Recall that the standard topological twist corresponds to $\varepsilon=0$, while in general, on a non-spin complex surface, we need to consider the extended DW twist with $\varepsilon\neq 0$.

 \medskip
 \noindent
 Let us also collect a few identities that are useful for computing indices. Firstly, the relevant contributions from the $\h A$ and Todd classes read:
 \bea
& \hat{A} (TM) = 1-{1\ov 24} p_1(TM)~, \\
& {\rm Td}(TM^+) = 1+\half c_1(TM^+) + {1\ov 12}\left(c_1(TM^+)^2 + c_2(TM^+)\right)~.
 \eea
 We have the relations $c_1(TM^+)=- c_1(\CK)$ and $c_2(TM^+)= c_2(\Omega^{0,1})$, as well as:
 \be
  \int_M c_1(K)^2= 2\chi+3\sigma~, \qquad\qquad
 \int_M c_2(\Omega^{0,1})=\chi~.
 \ee
 Finally, the Chern character of any holomorphic vector bundle $E$ expands as:
 \be
 {\rm ch}(E)= {\rm rk}(E)+ c_1(E)+ \half c_1(E)^2- c_2(E)~.
 \ee

%%%%%%%%%%%

\paragraph{The higher-spin particle index.} Finally, let us compute the index of the Dolbeault complex $\b \d_V$ twisted by:
\be
V= \CK^{\half +\varepsilon} \otimes S^{2j_l}(S_-) \otimes S^{2j_r}(S_+) \otimes F~,
\ee
over a K\"ahler four-manifold $\cM_4$, where $F$ is some holomorphic vector bundle. This is index  relevant for computing the contribution of a massive 5d particle of spin $(j_l,j_r)$ (see section~\ref{subsec:HSM5}).
  The symmetric powers of the spin bundles $S_\pm$ can be formally expanded as:
\be
  S^{2j_l}(S_-)\cong \CK^{j_l} \otimes S^{2j_l}(\Omega^{0,1})~,\qquad \quad
  S^{2j_r}(S_+)\cong   \oplus_{m_r=-j_r}^{j_r} \CK^{m_r}~.
\ee
Their Chern characters are given by:\footnote{To compute the Chern character of the $k$-symmetric power of any vector bundle $E$ with Chern roots $x_i$, we can use the fact that:
$$ \sum_k {\rm ch}(S^k(E)) t^k = \prod_{i} {1\ov 1 - t e^{x_i}}~.$$}
\bea
& {\rm ch}\left(S^{2j_l}(S_-)\right)= (2j_l+1)\left( 1 + {j_l(j_l+1)\ov 6}  (c_1(\CK)^2- 4 c_2(\Omega^{0,1}))\right)~, \cr
& {\rm ch}\left(S^{2j_r}(S_+)\right)= (2j_r+1)\left( 1 + {j_r(j_r+1)\ov 6}  c_1(\CK)^2\right)~.
\eea
Then, we immediately find:
\bea
   &  {\rm ind}({\b\d}_V)  &=&\; \;{\rm rk}(F) 
    (2j_l+1)(2j_r+1)\Bigg[-{\sigma\ov 8} +\half \varepsilon^2 (2\chi+3\sigma)-{2\ov 3} j_l(j_l+1)\chi  \\
&&&\;\;+ {j_l(j_l+1)+j_r(j_r+1)\ov 6} (2\chi+3\sigma)\Bigg] \\
&&&\; \; +  (2j_l+1)(2j_r+1)\left[\varepsilon \int_M c_1(F)c_1(\CK)+ \int_M {\rm ch}(F)\right]~,
\eea
  which gives us the result \eqref{indexjljrfull}. 
  
%%%%%%%%%%%%%%%%%%%%%%%%%%%
\section{$E_n$ SCFTs and Gopakumar-Vafa invariants}\label{appendix:GVinvs}

Five-dimensional SCFTs are engineered in M-theory compactifications on Calabi-Yau threefold canonical singularities $\MG$. Their partition function in the $\Omega$-background is computed using the refined topological vertex and is fully determined by the numbers $N_{j_l, j_r}^{\bbeta}$ of massive particles of spin $(j_l, j_r)$ for a fixed K\"ahler class $\bbeta \in H_2(\t \MG, \Z)$. The simplest five-dimensional SCFTs are the so-called Seiberg $E_n$ theories, originally studied in \cite{Seiberg:1996bd, Morrison:1996xf, Intriligator:1997pq}, which we shall also focus on here.

In this appendix, we list some of the refined GV invariants for the local del Pezzo and Hirzebruch geometries. We first consider the local $\bF_0$ threefold to streamline the computation. The GV invariants for this can be found in \cite{Iqbal:2007ii}, as well as in \cite{Huang:2013yta}. For this computation, we use the fact that the refined topological string free energy can be expressed as:
\be 
    \boldsymbol{F}(Q,\tau_1,\tau_2) = \sum_{j_l,j_r\geq 0} \sum_{\bbeta} \sum_{w=1}^{\infty} (-1)^{2j_l +2j_r} N_{j_l,j_r}^{\bbeta} \frac{\chi_{j_l}\left((q_1/q_2)^{w \ov 2}\right) \chi_{j_r}\left((q_1q_2)^{w\ov2}\right)}{w\left(q_1^{w \ov 2}-q_1^{-{w\ov2}}\right)\left(q_2^{w\ov2}-q_2^{-{w\ov2}}\right)} Q^{w \bbeta}~,
\ee
where $Q$ are the K\"ahler parameters, $q_k = e^{2\pi i\tau_k}$, for $k=1,2$, and $\chi_j$ are $SU(2)$ characters:
\be
    \chi_j(q) = \frac{q^{2j+1}-q^{-2j-1}}{q-q^{-1}}~.
\ee
Note that this expression corresponds to the ordinary DW twist, setting $\varepsilon = 0$. We will use this result together with the Nekrasov partition functions to compute the GV invariants $N_{j_l,j_r}^{\bbeta}$. We will also comment on the perturbative part of the partition function by using the results obtained from the SW geometry.

Let us also note that the prepotential can be resummed to:
\be
    \CF = -{1\ov (2\pi i)^3} \sum_{\bbeta}d_{\bbeta}\trilog(Q^{\bbeta})~, \qquad \quad  d_{\bbeta} = \sum_{j_l, j_r} c_0^{(j_l, j_r)} N^{\bbeta}_{j_l, j_r}~,
\ee
with $c_0^{(j_l, j_r)} = (-1)^{2(j_l+j_r)} (2j_l+1)(2j_r +1)$. Similarly, the gravitational couplings become:
\be
    \CA = {1\ov 2\pi i}\sum_{\bbeta}d_{\bbeta}^{\CA}\log(1-Q^{\bbeta})~, \qquad \quad \cB = {1\ov 2\pi i}\sum_{\bbeta}d_{\bbeta}^{\CB}\log(1-Q^{\bbeta})~, 
\ee
with:
\be
d_{\bbeta}^{\CA, \CB} = \sum_{j_l, j_r}c_{\CA,\CB}^{(j_l, j_r)}N^{\bbeta}_{j_l, j_r}~.
\ee
Here, $c_\CA^{(j_l, j_r)}$ and $c_\CB^{(j_l, j_r)}$ are given by \eqref{HigherSpin Coefficients} for a particle of spin $(j_l,j_r)$. Restoring the extended topological twist, the factors $H$ and $\cG$ appearing in the non-equivariant limit \eqref{NonEquivariantLimitExtendedTwist} become:
\bea
    H = -{1\ov (2\pi i)^2}\varepsilon\, \sum_{\bbeta}d_{\bbeta}\, \dilog(Q^{\bbeta})~, \qquad \qquad \cG = {1\ov 2\pi i}\varepsilon^2 \sum_{\bbeta}d_{\bbeta} \log(1-Q^{\bbeta})~.
\eea

%%%%%%%%%%%%%%%%%
 
\subsection{Five-dimensional SCFTs: the $E_1$ theory}

In this subsection, we shall focus on the $E_1$ SCFT. We also offer an alternative way of using the fibering operator and the gluings of Nekrasov partition functions, by expressing the result in terms of instanton corrections. 

Recall first the proposal of \cite{Closset:2021lhd} that the gravitational couplings $\cA$ and $\cB$ can be obtained directly from the Seiberg-Witten geometry, in analogy with the four-dimensional prescription of \cite{Moore:1997pc, Losev:1997tp, Witten:1995gf}. Given any rank-one Seiberg-Witten geometry, with CB parameter $U$, we thus have:
\be \label{A and B from SW curve}
    A(U) = \alpha \left({dU \ov d a} \right)~, \qquad \quad B(U) = \beta \left(\Delta^{phys}\right)^{1\ov 8}~,
\ee
with $a$ the VEV of the 5d vector multiplet, $\Delta^{phys}$ the `physical-discriminant' and $\alpha, \beta$ some constant prefactors determined in \cite{Closset:2021lhd}. Let us also point out that, in the conventions of \cite{Closset:2021lhd}, the physical discriminant is equivalent to the discriminant of the Seiberg-Witten curve, up to a numerical prefactor. 

Focusing on the $E_1$ SCFT, which is the UV completion of the 5d $\cN=1$ $SU(2)_0$ gauge theory, its Seiberg-Witten curve reads:
\bea
    g_2(U) & = \frac{1}{12}\Big(U^4 - 8(1+\lambda)U^2 + 16\left(1 - \lambda + \lambda^2\right)  \Big)~, \\
    g_3(U) & = -\frac{1}{216}\Big(U^6 - 12(1+\lambda)U^4 + 24\left(2 +\lambda+2\lambda^2\right)U^2-32\left(2 - 3\lambda - 3\lambda^2 + 2\lambda^3\right) \Big)~,
\eea
with the discriminant given by:
\be
    \Delta(U)  = \lambda^2\Big( U^4 - 8(1+\lambda)U^2 + 16(1-\lambda)^2\Big)~.
\ee
Here $\lambda$ is the inverse gauge coupling, which, in the geometric engineering limit, becomes the instanton counting parameter as it gets identified with the dynamically generated scale $\Lambda^4$ of the 4d $SU(2)$ gauge theory \cite{Katz:1996fh}. The complexified K\"ahler parameters of the IIA geometry are:
\be
    Q_f \equiv Q^2 = e^{2\pi i t_f} = e^{4\pi i a}~, \qquad \quad Q_b = e^{2\pi i t_b} = e^{2\pi i(2a + \mu_0)}~, \qquad Q_b = \lambda Q_f~,
\ee
with the periods $t_{f,b}$ corresponding to $D2$-branes wrapping the $\bP^1$ curves $\CC_{b,f}$ of $\bF_0 = \bP^1 \times \bP^1$. Thus, the K\"ahler classes are labelled by pairs of integers $\bbeta = (d_1, d_2)$ corresponding to the degrees of $(Q_b,Q_f)$. Of course, for this particular geometry, there is a symmetry exchanging the two integers. We will denote by $Q^{\bbeta} = Q_b^{d_1} Q_f^{d_2}$, for $\bbeta = (d_1, d_2)$, which should not be confused with the notation $ Q_f =Q^2= e^{4\pi i a}$.  

As $\lambda$ plays the role of the instanton counting parameter, the $n^{th}$ instanton contribution of the four-dimensional theory is recovered from all particles of K\"ahler classes $(n, d_2)$, for all $d_2 \in \Z$. Note that this gauge-theory point of view is not the most natural one from a geometric perspective, but it serves as a computational tool.

The simplest Gopakumar-Vafa invariants for the $E_1$ theory are:
\be \label{E1 simplest GV invariants}
    N_{j_l, j_r}^{(1,d)} = N_{j_l, j_r}^{(d,1)} = \delta_{j_l, 0} \delta_{j_r, d+{1\ov 2}}~,
\ee
Following the above discussion, the perturbative contribution to the prepotential will be thus determined by the particles in the class $\bbeta = (0,1)$, which amount for a single W-boson, namely a particle of spin $(j_l,j_r) = (0,{1\ov 2})$. This leads to:
\be
    \cF_{\text{pert}} = {1\ov (2\pi i)^3}2\trilog(Q_f)~, \qquad \quad \cA_{\text{pert}} = \cB_{\text{pert}} = -{1\ov 4\pi i}\log(1-Q_f)~.
\ee
Note that, from the Seiberg-Witten geometry computation \eqref{A and B from SW curve}, there is an additional $\log(Q)$ term for both $\cA$ and $\cB$. These terms depend on the choice of quantisation scheme, as pointed out in \cite{Closset:2021lhd, Closset:2018bjz}. We list some of the higher degree invariants in table \ref{tab: E1 refined GV invariants}. %
\renewcommand{\arraystretch}{1}
\begin{table}[h]
\renewcommand{\arraystretch}{1} \small
\begin{center}
\begin{tabular}{|c||c||c|c|c|c|c|c|c|c|c|c|c|c|c|c|c|c|c|c|c|c|}
\hline
$\bbeta$ & $j_l \backslash j_r$ & $0$ & ${1\ov 2}$ & $1$ & ${3\ov 2}$ & $2$ & ${5\ov 2}$ & $3$ & ${7\ov 2}$ & $4$ & ${9\ov 2}$ & $5$ & ${11\ov 2}$ & $6$ & ${13\ov 2}$ & $7$ & ${15\ov 2}$ & $8$ & ${17\ov 2}$ & $9$ & ${19 \ov 2}$  \\
\hline\hline
\multirow{2}{*}{ (2,2) } & 0 & & & & & & 1 & & 1 & & & & & & & & & &  & &\\ \cline{2-22}
& 1/2 & & & & & & & & & 1 & & & & & & & & &  & &\\
\hline\hline
\multirow{3}{*}{ (2,3) } & 0 & & & & & & 1 & & 1 & & 2 & & & & & & & & & &\\ \cline{2-22}
& 1/2 & & & & & & & & & 1 & & 1 & & & & & & & & &\\ \cline{2-22}
& 1 & & & & & & & & & & & & 1 & & & & & & & &\\
\hline\hline
\multirow{4}{*}{ (2,4) } & 0 & & & & & & 1 & & 1 & & 2 & & 2 & & & & & & & & \\ \cline{2-22}
& 1/2 & & & & & & & & & 1 & & 1 & & 2 & & & & & & &\\ \cline{2-22}
& 1 & & & & & & & & & & & & 1 & & 1 & & & & & &\\ \cline{2-22}
& 3/2 & & & & & & & & & & & & & & & 1 & & & & &\\
\hline\hline
\multirow{5}{*}{ (2,5) } & 0 & & & & & & 1 & & 1 & & 2 & & 2 & & 3 & & & & & &\\ \cline{2-22}
& 1/2 & & & & & & & & & 1 & & 1 & & 2 & & 2 & & & & & \\ \cline{2-22}
& 1 & & & & & & & & & & & & 1 & & 1 & & 2 & & & &\\ \cline{2-22}
& 3/2 & & & & & & & & & & & & & & & 1 & & 1 & & & \\ \cline{2-22}
& 2 & & & & & & & & & & & & & & & & & & 1 & & \\
\hline \hline
\multirow{5}{*}{ (3,3) } & 0 & & & & 1 & & 1 & & 3 & & 3 & & 4 & & & & & & & &\\ \cline{2-22}
& $1/2$ & & & & & & & 1 & & 2 & & 3 & & 3 & & 1 & & & & &\\ \cline{2-22}
& 1 & & & & & & & & &  & 1 & & 2 & & 3 & & & & & &\\\cline{2-22}
& $3/2$ & & & & & & & & &  & &  & & 1 & & 1 & & & & &\\ \cline{2-22}
& 2 & & & & & & & & &  & &  & & & &  & 1 & & & &\\
\hline\hline
\multirow{7}{*}{ (3,4) } & 0 & & 1 & & 1 & & 3 & & 4 & & 7 & & 6 & & 7 & & 1 & & 1 & & \\ \cline{2-22}
& 1/2 & & & & & 1 & & 2 & & 4 & & 6 & & 8 & & 7 & & 2 & & & \\ \cline{2-22}
& 1 & & & & & & & & 1 & & 2 & & 5 & & 6 & & 7 & & 1 & & \\ \cline{2-22}
& 3/2 & & & & & & & & & & & 1 & & 2 & & 4 & & 4 & & 1 & \\ \cline{2-22}
& 2 & & & & & & & & & & & & & & 1 & & 2 & & 3 & & \\ \cline{2-22}
& 5/2 & & & & & & & & & & & & & & & & & 1 & & 1 & \\ \cline{2-22}
& 3 & & & & & & & & & & & & & & & & & & & & 1 \\
\hline
\end{tabular}\normalsize
\end{center}
\caption{Refined GV invariants $N_{j_l,j_r}^{\bbeta}$ for local $\bF_0$ geometry.}
\label{tab: E1 refined GV invariants}
\end{table}%%
%%%%%%%%%%%%%%%%%%%%%%%%%%%
\medskip

\noindent \paragraph{Partition function.} Given the above refined GV invariants, let us now consider the CB partition function of the $E_1$ theory on $\CM_5$. Knowing the expressions for $\cF, \cA$ and $\cB$, or, alternatively, for the GV invariants $N_{\bbeta}^{j_l, j_r}$, the result should follow immediately. We will express the result in terms of the instanton counting parameter $\lambda$.

From the topological string partition function -- or, equivalently, from the Nekrasov instanton partition function of the 5d $SU(2)_0$ gauge theory -- we have the following closed-form expression for the flat-space partition function:
\be
    Z^{\text{1-inst}}_{\bC^2 \times S^1}(a,\tau_1, \tau_2) = { qp (1+qp) \ov (1-q)(1-p)(1-Q qp)(1-Q^{-1}qp)}~,
\ee
See \cite{Closset:2021lhd} for our conventions on the instanton partition function. Note that the above result holds in the $\varepsilon = 0$ case. The extended topological twist can be easily recovered by shifting $a \rightarrow a + \varepsilon(\tau_1+\tau_2)$, as explained in \eqref{ZC2 with eps}. We then note that:
\bea    \label{E1 1-inst correction}
    \cF_{\text{1-inst}}(a) = {1\ov (2\pi i)^3} {2Q\ov (1-Q)^2}~, \hspace*{3.5cm} \cr
    \cA_{\text{1-inst}}(a) = {1\ov 2\pi i}{Q(1+ 6Q + Q^2) \ov 2(1-Q)^4} ~, \qquad \cB_{\text{1-inst}}(a)  = {1\ov 2\pi i}{Q(1+10Q+Q^2) \ov 2(1-Q)^4}~,
\eea
while $H^{\text{1-inst}}$ vanishes for the DW twist. Recall that the 1-instanton contribution comes from the states associated to the K\"ahler classes $\bbeta = (1,k)$, for all $k\in \bN$. This is a sum over an infinite number of particles, which is, in principle, difficult to compute as the GV invariants are only known up to some finite order. Such sums were shown to reproduce the 4d instanton corrections in \cite{Katz:1996fh} using an assumption on the growth of the coefficients $d_{\bbeta}$. For the 1-instanton case, no such assumption is needed due to \eqref{E1 simplest GV invariants}, which leads to $d_{(1,k)} = -2(k+1)$, while:
\be
    d_{(1,k)}^{\cA} =-{1\ov 6}(1+k)(1+2k)(3+2k)~, \qquad \quad  d_{(1,k)}^{\cB} = -{1\ov 2}(1+k)(1+4k+2k^2)~.
\ee
As a result, expanding the polylogarithms in the instanton counting parameter, one recovers \eqref{E1 1-inst correction} after summing over all classes $\bbeta = (1,k)$, for $k\in \bN$. Similar expressions to \eqref{E1 1-inst correction} can be obtained for higher instanton corrections from the instanton partition function (see \textit{e.g.}~\cite{Closset:2021lhd} for a recent discussion) and it can be checked that those are reproduced by the GV invariants.

%We would like to check that the $\cM_4 \times S^1$ partition function on the toric K\"ahler 4-manifolds described in \eqref{toric Kahler 4-manifolds} is of the form \eqref{M4xS1 conjecture}. 
%Note that since the perturbative part is only due to a single $(j_l, j_r)$ particle, the result was already checked before. Moreover, we further check the result explicitly up to the 3-instanton correction.

Consider now the five-sphere partition function in the absence of background fluxes, for which we shall analyze explicitly the 1-instanton correction. Using the gluing \eqref{ProposedGluing}, the contribution to the (logarithm of the) five-sphere partition function reads,  in the non-equivariant limit:
\bea
  &  \lim_{\tau_{1,2}\to 0}  \lambda^2 \left( \log Z^{\text{1-inst}}_{\bC^2 \times S^1}(a,\tau_1, \tau_2) + \log Z^{\text{1-inst}}_{\bC^2 \times S^1}(a^*, \tau_1^*, \tau_2^*) + \log Z^{\text{1-inst}}_{\bC^2 \times S^1} (\tilde{a}, \tilde{\tau}_1, \tilde{\tau}_2) \right)= \cr
    &\quad    \lambda^2\left(-{2Q(1+7Q+Q^2) \ov (1-Q)^4} \right) + \lambda^2 \left({1\ov 4\pi^2} {2Q\ov (1-Q)^2} - {Q(1+Q) \ov \pi i (1-Q)^3} a - {Q (1+4Q+Q^2) \ov (1-Q)^4}a^2 \right)~.
\eea
The first term can be reorganised as:
\be
   -2\pi i \Big(\chi(\bP^2) \cA_{\text{1-inst}}(a) + \sigma(\bP^2) \cB_{\text{1-inst}}(a) \Big)~,
\ee
being the contribution to the $\bP^2 \times S^1$ partition function. Moreover, the remaining term is due to the fibering operator, giving us $\half \log \FiberOp(\amCB)$, with $\FiberOp$ as defined in \eqref{F fibering def 1}, as one can easily check using the 1-instanton correction to the prepotential in \eqref{E1 1-inst correction}.  It does agree with the general form of the fibering operator for $\varepsilon=0$. For a generic 5d SCFTs, one needs to consider the extended topological twist  with $\varepsilon \neq 0$, in general, which is easily done as explained in the main text. 
For the $E_1$ theory, we can consistently choose $\varepsilon=0$.

%%%%%%%%%%%%%%%%%
\subsection{Local $dP_2$ geometry}\label{app:dP2}

For the $E_2$ theory, the SW curve in the parametrisation of \cite{Closset:2021lhd} is given by:
\be
    \sqrt{\lambda} \left(1 + \frac{M_1}{w}\right) + t\left(\frac{1}{w}+w-2U\right)  + \sqrt{\lambda}  \, t^2 = 0~.
\ee
The instanton corrections to the prepotential agree with Nekrasov instanton counting results upon identifying $M_1 = -e^{-2\pi i \mu}$, with $\mu$ the five-dimensional mass parameter (see \cite{Closset:2021lhd} for our conventions on the instanton partition function). Inspired by the topological vertex formalism, let us define $Q_m = e^{-2\pi i\mu}e^{-2\pi i a} = e^{-2\pi i t_m}$. Here $t_m = a+\mu$ is the complexified K\"ahler parameter associated to the exceptional curve $\CC_m$, resulting from blowing-up of $\bF_0$ at a single generic point:\footnote{In \cite{Closset:2021lhd}, when viewing $dP_n$ as a blow-up of $\bF_0$ at $n-1$ generic points, these exceptional curves were denoted by $E_i$, for $i=1, \ldots, n-1$.}
\be
    t_m = \int_{\CC_m} (B+iJ)~.
\ee
The perturbative contributions to the prepotential obtained from the Seiberg-Witten curve can be expressed in a compact form as:
\bea
    (2\pi i)^3\mathcal{F}_{\text{pert}} & = 2 \trilog(Q_f) - \trilog\left(Q_f Q_m\right) - \trilog\left(Q_m^{-1} \right) ~.
\eea
In the basis $(Q_b,Q_f,Q_m)$, we find that the prepotential is reproduced by the states:
\bea
    (0,1,0)~: \quad N_{j_l,j_r} & = \delta_{j_l,0}~\delta_{j_r,\frac{1}{2}}~, \\
    (0,0,-1)~: \quad N_{j_l,j_r} & = \delta_{j_l,0}~\delta_{j_r,0}~, \\
    (0,1,1)~: \quad N_{j_l,j_r} & = \delta_{j_l,0}~\delta_{j_r,0}~. 
\eea
Let us note that the only state contributing to the perturbative part of  $\cA$ is the $(0,1,0)$ term, which is in perfect agreement with the SW geometry results, namely:
\bea
    2\pi i\,\cA_{\text{pert}}& = -{1\ov 2} \log(1-Q_f) ~.
\eea
On the other hand, the $\cB$ gravitational correction receives contributions from all of the above states, leading to:
\bea
    2\pi i\,\cB_{\text{pert}} & = -{1\ov 2} \log(1-Q_f) - {1\ov 8} \log\left(1 - Q_f Q_m \right) - {1\ov 8} \log\left(1 - Q_m^{-1} \right)~.
\eea
The `$1$-instanton' GV invariants in the basis $(Q_b, Q_f, Q_m)$ are given by:
\bea
    (1,n,0)~:& \quad N_{j_l,j_r} = \delta_{j_l,0}~\delta_{j_r,n+\frac{1}{2}}~, \\
     (1,n,1)~:& \quad N_{j_l,j_r} = \delta_{j_l,0}~\delta_{j_r,n}~,\\
     (1,n,d)~:& \quad N_{j_l,j_r} = 0~, ~\text{ for }~ d>2~.
\eea
Let us also comment on the limits to $E_1$ (\textit{i.e.} local $\bF_0$) and $\widetilde{E}_1$ (\textit{i.e.} local $dP_1$). From the instanton counting results, one can check that for the $n$-instanton contribution the terms of order $(M_1)^0$ and $(M_1)^n$ reproduce the $E_1$ and $\widetilde{E}_1$ contributions, respectively (and thus, the two geometries are recovered in the $M_1 \rightarrow 0$ and $M_1 \rightarrow \infty$ limits, respectively). In terms of the GV invariants, this means that the $(n,m,0)$ invariants should precisely reproduce the $\bF_0$ invariants listed in table~\ref{tab: E1 refined GV invariants}. We list some of the low-degree GV invariants for the local $dP_2$ that are not $\bF_0$ invariants in table~\ref{tab: dP2 GV invariants}. Note also that the $(n,m,n)$ invariants do coincide with the $dP_1$ invariants computed in \cite{Iqbal:2007ii}.

\begin{table} \small \renewcommand{\arraystretch}{1}
\begin{center}
\begin{tabular}{|c||c||c|c|c|c|c|c|c|c|c|c|c|c|c|c|c|c|c|c|c|c|}
\hline
$\bbeta$ & $j_l \backslash j_r$ & $0$ & $\frac{1}{2}$ & $1$ & $\frac{3}{2}$ & $2$ & $\frac{5}{2}$ & $3$ & $\frac{7}{2}$ & $4$ & $\frac{9}{2}$ & $5$ & $\frac{11}{2}$ & $6$ & $\frac{13}{2}$ & $7$ & $\frac{15}{2}$ & $8$ & $\frac{17}{2}$ & $9$ & $\frac{19}{2}$ \\
\hline\hline
(2,1,1) & 0 & & & & & 1 & & & & & & & & & & & & & & & \\
\hline\hline
\multirow{2}{*}{ (2,2,1) } & 0 & & & & & 1 & & 2 & & & & & & & & & & & & & \\ \cline{2-22}
& 1/2 & & & & & & & & 1 & & & & & & & & & & & & \\
\hline\hline
\multirow{3}{*}{ (2,3,1) } & 0 & & & & & 1 & & 2 & & 3 & & & & & & & & & & & \\ \cline{2-22}
& 1/2 & & & & & & & & 1 & & 2 & & & & & & & & & & \\ \cline{2-22}
& 1 & & & & & & & & & & & 1 & & & & & & & & & \\
\hline\hline
\multirow{4}{*}{ (2,4,1) } & 0 & & & & & 1 & & 2 & & 3 & & 4 & & & & & & & & & \\ \cline{2-22}
& 1/2 & & & & & & & & 1 & & 2 & & 3 & & & & & & & & \\ \cline{2-22}
& 1 & & & & & & & & & & & 1 & & 2 & & & & & & & \\ \cline{2-22}
& 3/2 & & & & & & & & & & & & & & 1 & & & & & & \\
\hline\hline
\multirow{5}{*}{ (2,5,1) } & 0 & & & & & 1 & & 2 & & 3 & & 4 & & 5 & & & & & & & \\ \cline{2-22}
& 1/2 & & & & & & & & 1 & & 2 & & 3 & & 4 & & & & & & \\ \cline{2-22}
& 1 & & & & & & & & & & & 1 & & 2 & & 3 & & & & & \\ \cline{2-22}
& 3/2 & & & & & & & & & & & & & & 1 & & 2 & & & & \\ \cline{2-22}
& 2 & & & & & & & & & & & & & & & & & 1 & & & \\ 
%\hline
%
%\end{tabular}
%\end{center}
%
%%%%%%%%%%%%%%%%%%%%%%%%%%%%%%%%%%%%%%%%%
%
%\renewcommand{\arraystretch}{1}
%\begin{center}
%\begin{tabular}{|c||c||c|c|c|c|c|c|c|c|c|c|c|c|c|c|c|c|c|c|c|c|}
%\hline
%$\bbeta$ & $j_l \backslash j_r$ & $0$ & $\frac{1}{2}$ & $1$ & $\frac{3}{2}$ & $2$ & $\frac{5}{2}$ & $3$ & $\frac{7}{2}$ & $4$ & $\frac{9}{2}$ & $5$ & $\frac{11}{2}$ & $6$ & $\frac{13}{2}$ & $7$ & $\frac{15}{2}$ & $8$ & $\frac{17}{2}$ & $9$ & $\frac{19}{2}$ \\
\hline\hline
(2,2,2) & 0 & & & & & & 1 & & & & & & & & & & & & & & \\
\hline\hline
\multirow{2}{*}{ (2,3,2) } & 0 & & & & & & 1 & & 1 & & & & & & & & & & & & \\ \cline{2-22}
& 1/2 & & & & & & & & & 1 & & & & & & & & & & & \\
\hline\hline
\multirow{3}{*}{ (2,4,2) } & 0 & & & & & & 1 & & 1 & & 2 & & & & & & & & & & \\ \cline{2-22}
& 1/2 & & & & & & & & & 1 & & 1 & & & & & & & & & \\ \cline{2-22}
& 1 & & & & & & & & & & & & 1 & & & & & & & & \\
\hline\hline
\multirow{4}{*}{ (2,5,2) } & 0 & & & & & & 1 & & 1 & & 2 & & 2 & & & & & & & & \\ \cline{2-22}
& 1/2 & & & & & & & & & 1 & & 1 & & 2 & & & & & & & \\ \cline{2-22}
& 1 & & & & & & & & & & & & 1 & & 1 & & & & & & \\ \cline{2-22}
& 3/2 & & & & & & & & & & & & & & & 1 & & & & & \\
\hline\hline
\multirow{5}{*}{ (2,6,2) } & 0 & & & & & & 1 & & 1 & & 2 & & 2 & & 3 & & & & & & \\ \cline{2-22}
& 1/2 & & & & & & & & & 1 & & 1 & & 2 & & 2 & & & & & \\ \cline{2-22}
& 1 & & & & & & & & & & & & 1 & & 1 & & 2 & & & & \\ \cline{2-22}
& 3/2 & & & & & & & & & & & & & & & 1 & & 1 & & & \\ \cline{2-22}
& 2 & & & & & & & & & & & & & & & & & & 1 & & \\
\hline\hline
(3,1,1) & 0 & & & & & & & 1 & & & & & & & & & & & & & \\
\hline\hline
\multirow{3}{*}{ (3,2,1) } & 0 & & & & & 1 & & 2 & & 3 & & & & & & & & & & & \\ \cline{2-22}
& 1/2 & & & & & & & & 1 & & 2 & & & & & & & & & & \\ \cline{2-22}
& 1 & & & & & & & & & & & 1 & & & & & & & & & \\
\hline
    \end{tabular}
\end{center}\normalsize%
\caption{Refined GV invariants $N_{j_l,j_r}^{\bbeta}$ for local $dP_2$ geometry.}
    \label{tab: dP2 GV invariants}
\end{table}

%%%%%%%%%%%%%%%%%%%%%%%%%%%%%%%%%%%%%%%

\subsection{Local $dP_3$ geometry}\label{app:dP3}

For the $E_3$ theory we choose the following parametrisation \cite{Closset:2021lhd}:
\begin{equation}
    \sqrt{\lambda} \left(1 + \frac{M_1}{w}\right) + t\left(\frac{1}{w}+w-2U\right)  + \sqrt{\lambda} \, t^2(1+M_2 w) = 0~.
\end{equation}
As before, the instanton corrections agree with Nekrasov instanton counting results upon identifying $M_i = -e^{-2\pi i \mu_i}$. As for $E_2$, we define $Q_{m_i} = e^{-2\pi i \mu_i}e^{-2\pi i a}$. The perturbative contributions to the prepotential obtained from the Seiberg-Witten curve are:
\bea
    (2\pi i)^3\mathcal{F}_{\rm pert} & = 2 \trilog(Q_f) - \sum_{i=1}^2 \left( \trilog\left(Q_f Q_{m_i}\right) + \trilog\left(Q_{m_i}^{-1} \right) \right)~.
\eea
Let us also note the symmetry in $Q_{m_1} \leftrightarrow Q_{m_2}$. For this reason, we avoid writing down redundant invariants. In the basis $(Q_b,Q_f,Q_{m_1},Q_{m_2})$, the states that reproduce the above prepotential are:
\bea
    (0,1,0,0)~: N_{j_l,j_r} & = \delta_{j_l,0}~\delta_{j_r,\frac{1}{2}}~, \\
    (0,0,-1,0)~: N_{j_l,j_r} & = \delta_{j_l,0}~\delta_{j_r,0}~, \\
    %(0,0,0,-1)~: N_{j_l,j_r} & = \delta_{j_l,0}~\delta_{j_r,0}~, \\
    (0,1,1,0)~: N_{j_l,j_r} & = \delta_{j_l,0}~\delta_{j_r,0}~. %\\
    %(0,1,0,1)~: N_{j_l,j_r} & = \delta_{j_l,0}~\delta_{j_r,0}~. 
\eea
The only state contributing to $\CA$ at this order is the $(0,1,0,0)$ state, in perfect agreement with the SW geometry results. The $\CB$ gravitational correction receives contributions from all of the above states, as was the case in the $E_2$ theory. The only $1$-instanton GV invariants are given by:
\bea
    (1,n,0,0)~:& \quad N_{j_l,j_r} = \delta_{j_l,0}~\delta_{j_r,n+\frac{1}{2}}~, \\
     (1,n,1,0)~:& \quad N_{j_l,j_r} = \delta_{j_l,0}~\delta_{j_r,n}~,\\
     (1,n,1,1)~:& \quad N_{j_l,j_r} = \delta_{j_l,0}~\delta_{j_r,n-\frac{1}{2}}~.
\eea
For the $2$-instanton GV invariants let us first note that some of these will be the same as the $E_1$, $\widetilde{E}_1$ invariants, namely:
\bea
    (2,n,0,0)_{dP_3} \cong (2,n)_{\bF_0}\,, \qquad  (2,n,2,0)_{dP_3} \cong (2,n,0,2)_{dP_3} \cong (2,n)_{dP_1}\,,
\eea
where the subscript indicates the geometry. In fact, all these also appear as $dP_2$ invariants, together with the $(2,n,1,0)_{dP_3} \cong (2,n,0,1)_{dP_3}$ invariants. Furthermore, the GV invariants $(2,n,2,1)_{dP_3}\cong (2,n,1,2)_{dP_3}$ also correspond to the $dP_2$ invariants $(2,n-1,1)_{dP_2}$, while the $(2,n,2,2)_{dP_3}$ invariants correspond to the $(2,n-2)_{\bF_0}$ invariants. We list some of the new invariants in table~\ref{tab: dP3 GV invariants} (again, due to the $Q_{m_1} \leftrightarrow Q_{m_2}$ symmetry we do not write down explicitly some of these).
%%%%%%%%%%%%%%%%%%%%%%%%
\begin{table}[h]
\renewcommand{\arraystretch}{1}\small
\begin{center}
\begin{tabular}{|c||c||c|c|c|c|c|c|c|c|c|c|c|c|c|c|c|c|}
\hline
$\bbeta$& $j_l \backslash j_r$& $0$ & $\frac{1}{2}$ & $1$ & $\frac{3}{2}$ & $2$ & $\frac{5}{2}$ & $3$ & $\frac{7}{2}$ & $4$ & $\frac{9}{2}$ & $5$ & $\frac{11}{2}$ & $6$ & $\frac{13}{2}$ & $7$ & $\frac{15}{2}$  \\
\hline\hline
(2,1,1,1) & 0 & & & & 1 & & & & & & & & & & & & \\\hline\hline
\multirow{2}{*}{ (2,2,1,1) } & 0 & & & & 1 & & 3 & & & & & & & & & &  \\\cline{2-18}
& 1/2 & & & & & & & 1 & & & & & & & & & \\\hline\hline

\multirow{3}{*}{ (2,3,1,1) } & 0 & & & & 1 & & 3 & & 5 & & & & & & & & \\\cline{2-18}
& 1/2 & & & & & & & 1 & & 3 & & & & & & & \\\cline{2-18}
& 1 & & & & & & & & & & 1 & & & & & & \\\hline\hline

\multirow{4}{*}{ (2,4,1,1) } & 0 & & & & 1 & & 3 & & 5 & & 7 & & & & & &\\\cline{2-18}
& 1/2 & & & & & & & 1 & & 3 & & 5 & & & & &  \\\cline{2-18}
& 1 & & & & & & & & & & 1 & & 3 & & & & \\\cline{2-18}
& 3/2 & & & & & & & & & & & & & 1 & & &  \\\hline\hline

\multirow{5}{*}{ (2,5,1,1) } & 0 & & & & 1 & & 3 & & 5 & & 7 & & 9 & & & & \\\cline{2-18}
& 1/2 & & & & & & & 1 & & 3 & & 5 & & 7 & & & \\\cline{2-18}
& 1 & & & & & & & & & & 1 & & 3 & & 5 & & \\\cline{2-18}
& 3/2 & & & & & & & & & & & & & 1 & & 3 & \\\cline{2-18}
& 2 & & & & & & & & & & & & & & & & 1  \\ 
\hline
\end{tabular}
\end{center}\normalsize
\caption{Refined GV invariants $N_{j_l,j_r}^{\bbeta}$ for local $dP_3$ geometry.}
    \label{tab: dP3 GV invariants}
\end{table}

 %%%%%%%%%%%%%%%%%%%%%%%%%%%%%%%%%%%%%%%%%%%%
 %%%%%%%%%%%%%%%%%%%%%%%%%%%%%%%%%%%%%%%%%%%%
 %%%%%%%%%%%%%%%%%%%%%%%%%%%%%%%%%%%%%%%%%%%%
  %%%%%%%%%%%%%%%%%%%%%%%%%%%%%%%%%%%%%%%%%%%%
 %%%%%%%%%%%%%%%%%%%%%%%%%%%%%%%%%%%%%%%%%%%%
 %%%%%%%%%%%%%%%%%%%%%%%%%%%%%%%%%%%%%%%%%%%%
  %%%%%%%%%%%%%%%%%%%%%%%%%%%%%%%%%%%%%%%%%%%%
 %%%%%%%%%%%%%%%%%%%%%%%%%%%%%%%%%%%%%%%%%%%%
 %%%%%%%%%%%%%%%%%%%%%%%%%%%%%%%%%%%%%%%%%%%%

%%%%%%%
\bibliographystyle{JHEP}
\bibliography{bib5d}{}

\end{document}